\documentclass{article}%
\usepackage{amssymb}
\usepackage{amsmath}
\usepackage{amsfonts}
\usepackage{graphicx}%
\providecommand{\U}[1]{\protect\rule{.1in}{.1in}}

\numberwithin{equation}{section}
\begin{document}

\title{HOMOGENEOUS AND NONLINEAR GENERALIZED MASTER\\EQUATIONS ACCOUNTING FOR INITIAL CORRELATIONS}
\author{Victor F. Los
\and Institute of Magnetism, Nat. Acad. of Sci. of Ukraine,
\and 36-b Vernadsky Blvd., 03142 Kiev, Ukraine}
\maketitle

\begin{abstract}
Deriving the kinetic (irreversible) equations from the reversible microscopic
dynamics of the many-body systems remains one of the principal tasks of
statistical physics. One of the key assumptions used to deriving the Boltzmann
equation is the Bogoliubov principle of weakening of initial correlations
which implies that on a sufficiently large time scale all initial correlations
(existing at the initial instant) are damped. This approach leads to the
evolution equations that are not valid on all time scales and do not allow
considering all the stages of the system evolution, in particular, the
short-term and transient regimes, which are of scientific and practical
interest for considering the ultrafast relaxation and non-Markovian processes,
decoherence phenomena and onset of irreversibility. To take initial
correlations into account, a method, based on the conventional
time-independent projection operator technique, that allows converting the
conventional linear inhomogeneous (containing a source caused by initial
correlations) time-convolution generalized master equation (TC-GME) and
time-convolutionless GME (TCL-GME) into the homogeneous form exactly, is
proposed. This approach results in the exact linear time-convolution and
time-convolutionless homogeneous generalized master equations (TC-HGME and
TCL-HGME) which take the dynamics of initial correlations into account via
modified memory kernels governing the evolution of the relevant part of a
distribution function of a many-particle system. These equations describe the
evolution of the relevant part of a distribution function on all time scales
including the initial stage when the initial correlations matter. The obtained
TC-HGME is applied to the spatially homogeneous dilute gas of classical and
quantum particles. However, to derive the desired nonlinear equations (the
Boltzmann equation in particular) from this actually linear equation, we
should make an additional approximation neglecting the time-retardation of a
one-particle distribution function which restricts the time scale by times
much smaller than the relaxation time for a one-particle distribution
function. To obtain the actually nonlinear evolution equations and avoid any
restrictions on the time scales, we develop a new method, based on using a
time-dependent operator (generally not a projection operator) converting a
distribution function of a total system into the relevant form, that allows
deriving the new exact nonlinear generalized master equations. The derived
inhomogeneous nonlinear GME is a generalization of the linear Nakajima-Zwanzig
TC-GME and can be viewed as an alternative to the BBGKY chain. To include the
initial correlations into consideration, we convert the obtained inhomogeneous
nonlinear GME into the homogeneous form by the method which we used for
conventional linear GMEs. The obtained exact homogeneous nonlinear GME
describes all evolution stages of the system of interest and treats initial
correlations on an equal footing with collisions by means of the modified
memory kernel. As an application, we obtain a new homogeneous nonlinear
equation retaining initial correlations for a one-particle distribution
function of the spatially inhomogeneous nonideal gas of classical particles.
We show that on the kinetic time scale, the time-reversible terms resulting
from initial correlations vanish (if the particle dynamics have the ergodic
mixing property) and this equation can be converted into the Vlasov-Landau and
Boltzmann equations without any additional commonly used approximations..

\end{abstract}

\section{INTRODUCTION}

While the effects of equilibrium statistical mechanics seem to be well
understood and described, nonequilibrium statistical physics does not enjoy
such an opinion. One of the basic tasks of the latter remains deriving the
appropriate evolution equations for the measurable values (statistical
expectations) characterizing a nonequilibrium state of a many-particle system.
It is expected that these equations should generally be evolution equations
converting into kinetic (or other irreversible) equations on some time scale.
The principal questions are how to derive such irreversible equations
rigorously from underlying microscopic (reversible) classical or quantum
dynamical equations and what is the reason for the observable irreversibility.
Several approaches are usually used to address these problems, commonly
starting with the linear Liouville-von-Neumann equation for a distribution
function (statistical operator) of the $N$-particle ($N\gg1$) system under consideration.

One approach leads to the chain of coupled first-order differential equations
for $s$-particle distribution functions $F_{s}(x_{1},\ldots,x_{s},t)$ ($1\leq
s\leq N$) known as the BBGKY hierarchy \cite{Bogoliubov (1946)}. The most
consistent approach to decoupling the BBGKY hierarchy was developed by
Bogoliubov \cite{Bogoliubov (1946)}. He suggested the principle of weakening
of initial correlations, which implies that at a sufficiently large time
$t-t_{0}\gg t_{cor}$ ($t_{cor}$ is the correlation time due to interparticle
interaction), all initial correlations (existing at the initial instant
$t_{0}$) are damped and the time-dependence of multiparticle distribution
functions is consequently determined completely by the time-dependence of a
one-particle distribution function (Bogoliubov's ansatz). This assumes the
existence of the time interval (time scale)
\begin{equation}
t_{cor}\ll t-t_{0}\ll t_{rel} \label{0}%
\end{equation}
(where $t_{rel}$ is the relaxation time for a one-particle distribution
function) and leads to an approximate conversion (valid only on the large time
scale indicated above) of the inhomogeneous equation (including two-particle
correlations) for a one-particle distribution function (of the BBGKY
hierarchy) into a homogeneous equation. The idea behind this is that the
perturbation expansions of the homogeneous equations are much more effective
than that for the inhomogeneous ones (like the equations of the BBGKY chain),
where the expansions of the functions, rather than equations, are involved.
The Bogoliubov ansatz allows thus to avoid the so-called "secular terms", i.e
the terms growing with time, which emerge after integration of the BBGKY chain
inhomogeneous equation for $s$-particle distribution function $F_{s}%
(x_{1},\ldots,x_{s},t)$ ($1\leq s<N$) over time and make the perturbation
expansion of these functions ineffective on the time scale of interest in
kinetic theory. The Bogoliubov principle of weakening of initial correlations
provides a solid mathematical foundation to the Boltzmann hypothesis of the
molecular chaos (Stosszablansatz) which leads to the coarse-grained (smoothed)
over the time interval (\ref{0}) evolution of the one-particle distribution
function (described by the famous Boltzmann kinetic equation).

Using this approach, Bogoliubov successfully derived the kinetic (Markovian)
equations for a one-particle distribution function, the classical and quantum
Boltzmann equations in particular, which describe the time-evolution with the
characteristic time $t_{rel}$ but is unsuitable for taking the initial
evolution stage $t_{0}\leqslant t\leqslant t_{cor}$ into account.

The Bogoliubov principle of weakening of initial correlations is introduced as
the boundary (initial) condition for the BBGKY chain. For example, in
classical physics, it can be written as
\begin{align}
\lim\limits_{t-t_{0}\rightarrow\infty}U(t,t_{0})\left[  F_{N}(x_{1}%
,\ldots,x_{N},t_{0})-f_{r}(x_{1},\ldots,x_{N},t_{0})\right]   &
=0,\nonumber\\
f_{r}(x_{1},\ldots,x_{N},t)  &  =%
{\displaystyle\prod\limits_{i=1}^{N}}
F_{1}(x_{i},t), \label{0a}%
\end{align}
where $F_{N}(x_{1},\ldots,x_{N},t_{0})$ is an $N$-particle distribution
function at $t=t_{0}$, $x_{i}=(\mathbf{x}_{i},\mathbf{p}_{i})$ is the
coordinate of the $i$th particle in the phase space, $f_{r}(x_{1},\ldots
,x_{N},t)$ is the relevant distribution function with no correlations (the
general definition of the relevant distribution function will be given later),
and $U(t,t_{0})$ is the operator determining the evolution of all coordinates
$x_{i}$ in time. The first inequality in (\ref{0}) implies that all
correlations between the particles vanish for $t-t_{0}\gg t_{cor}$, and the
second inequality in (\ref{0}), $t-t_{0}\ll t_{rel}$, allows neglecting the
time retardation and replacing $U(t,t_{0})f_{r}(t_{0})$ with $U(t,t_{0}%
)f_{r}(t)$ (in the first approximation in the small density of particles $n$).
Thus, decoupling the BBGKY chain is nonlinear procedure (\ref{0a}), which
results in a closed nonlinear equation (like the Boltzmann equation) for a
one-particle distribution function $F_{1}(x_{i},t)$ and introduces a
nonlinearity into the process originally described by the linear reversible
Liouville-von-Neumann equation. It is important to note, that shifting $t_{0}$
to $-\infty$ in the Bogoliubov limiting procedure (\ref{0a}) (in order to get
rid of dealing with a singled out instant of time, when a distribution
function $F_{N}(t)$ is replaced with its relevant part) leads to an
irreversibility of the relaxation process.

But we note only the small-scale initial correlations with $t_{cor}\ll
t_{rel}$ can in fact be neglected on the kinetic time scale. Trying to obtain
the collision integral in the kinetic equation with collisions of three
particles (in the two-dimensional case) or more particles (the terms of the
second and higher orders in $n$), we should handle the divergencies, in
particular, resulting from the large-scale initial correlations with
$t_{cor}\gtrsim t_{rel}$ (see, e.g., \cite{Dorfman and Cohen (1967)},
\cite{Ferziger and Kaper (1972)}). Hence, the correct expansion of the
collision integral in powers of $n$ and consideration of kinetic fluctuations
are only possible if we account for initial correlations with $t_{cor}\gtrsim
t_{rel}$ and $l_{cor}\gtrsim l$, where $l_{cor}=\overline{v}t_{cor}$,
$l=\overline{v}t_{rel}$, $\overline{v}$ is the characteristic (mean) particle
velocity, and $l$ is the particle mean-free path \cite{Klimontovich (1982)}.
Moreover, there can be correlations which do not damp, e.g. due the
conservation laws, such as the collective plasma excitations; also, the
initial quantum correlations, caused by the particle statistics, do not damp
with time.

Attempts to obtain the Boltzmann equation rigorously (especially in quantum
physics) continue. This is important from both the conceptual and practical
standpoints. For example, following \cite{Lanford (1975)}, where the classical
Boltzmann equation was rigorously obtained for a system of hard spheres at
very low densities and for short times, the authors of \cite{Benedetto et al
(2004)} analogously (but not rigorously) obtained the quantum Boltzmann
equation but using the factorized initial condition, i.e., the absence of
initial correlations or random phase approximation (RPA) at $t=t_{0}$, which
is doubtful in principle \cite{van Kampen (2004)}.

On the other hand, we note that, as to the best of our knowledge, there is not
yet a way to take initial correlations into account exactly when dealing with
the BBGKY chain.

In another approach, leading to the so called generalized master equations
(GMEs), an $N$-particle distribution function (or statistical operator) is
divided into the essential (relevant) and inessential (irrelevant) parts using
time-independent projection operators $P$ and $Q$
\begin{align}
F_{N}(x_{1},\ldots,x_{N},t)  &  =f_{r}(x_{1},\ldots,x_{N},t)+f_{i}%
(x_{1},\ldots,x_{N},t),\nonumber\\
f_{r}(x_{1},\ldots,x_{N},t)  &  =PF_{N}(x_{1},\ldots,x_{N},t),f_{i}%
(x_{1},\ldots,x_{N},t)=QF_{N}(x_{1},\ldots,x_{N},t),\nonumber\\
P+Q  &  =1. \label{0b}%
\end{align}
We note, that the relevant and irrelevant parts depend on coordinates (and/or
momenta) of all $N$ particles in contrast to the reduced distribution
functions (like $F_{1}(x_{i})$). The relevant part, which is usually of main
interest, is, as a rule, a vacuum (slowly changing) part of the distribution
function (statistical operator), i.e., the part with no correlations (like
$f_{r}(x_{1},\ldots,x_{N},t)$ in (\ref{0a})). Applying the projectors $P$ and
$Q$ to the Liouville-von-Neumann equation, we can obtain the time-convolution
(non-Markovian) GME (TC-GME) \cite{Nakajima (1958)}, \cite{Zwanzig (1960)},
\cite{Prigogine (1962)} and time-convolutionless (time-local) GME (TCL-GME)
\cite{Shibata1 (1977)}, \cite{Shibata2 (1980)}, which are the exact
inhomogeneous equation for the relevant part of a distribution function with a
source (irrelevant part) containing all many-particle correlations at the
initial instant $t_{0}$. We stress that the time-independent (linear)
projection operators $P$ and $Q$ used in this approach commute with the
time-derivative operator in the Liouville-von-Neumann equation for the
distribution function (statistical operator) of the whole system. Therefore,
procedure (\ref{0b}) is linear and transforms the linear Liouville-von-Neumann
equation into the linear GMEs. Bogoliubov's principle of weakening of initial
correlations or simply the factorized initial conditions (RPA) are commonly
used to exclude initial many-particle correlations (a source), which results
in an approximate closed linear homogeneous GMEs for the relevant part of the
distribution function (statistical operator). The latters are then used to
obtain the linear evolution (master) equations for an interesting reduced
distribution function (especially like that for an open subsystem interacting
with a thermal bath).

There is also an approach based on the Kadanoff-Baym equations for the
two-time correlations functions \cite{Kadanoff (1962)}. The original
Kadanoff-Baym equations do not contain initial correlations and follow from
the exact equations for correlation functions under the assumption of
Bogoliubov's principle of weakening of initial correlations. In the framework
of this approach, several attempts were made to include into account the
initial correlations. A review of these works can be found in \cite{Zubarev
(1997)}.

All kinetic equations which follow from that principle are unable to describe
the initial stage of the evolution, $t_{0}\leqslant t\leqslant t_{cor}$.
Therefore, the influence of initial correlations, which can be important for,
e.g., the ultrafast relaxation processes, is not described by these equations.
Moreover, the time-scale $t\lesssim t_{cor}$ may be of significant interest,
because we can expect that on this time-scale the transition from the
short-term \textquotedblright non-kinetic\textquotedblright\ evolution of a
multi-particle system to the long-term \textquotedblright
kinetic\textquotedblright\ behavior occurs.

In particular, what we have said above means that the proper handling of the
initial correlations is crucial in obtaining the closed evolution equations.
Neglecting the initial correlations results in the evolution equations which
are not valid on all time scales (but, e.g., only on the time scale (\ref{0}))
and do not allow consecutive considering the all stages of the system of
interest evolution, in particular, the stage when a system switches from the
initial short-term (reversible) evolution (when initial correlations play a
role) to the long-term (irreversible) kinetic regime. The outlined problem is
common for both principal methods of deriving the kinetic equations: the BBGKY
hierarchy and the GMEs. Moreover, the initial correlations can be important
for considering the ultrafast relaxation and non-Markovian processes and also
for describing the decoherence phenomenon (see, e.g., \cite{Petruccione
(2005)}).

A progress achieved in the research on dynamics of the many-particle systems
makes it possible, in principle, to study the entire evolution process of
systems in statistical mechanics on any time scale and to determine the
conditions under which a system evolves towards the equilibrium state. To do
that, the dynamics of initial correlations should be included into
consideration, and our goal is to obtain new evolution equations exactly
accounting the initial correlations.

To take the initial correlations into account, we first suggest a method,
based on the conventional time-independent projection operator technique
(specified by (\ref{0b})), that allows converting the linear inhomogeneous
TC-GME and TCL-GME into the homogeneous form exactly \cite{Los (2001), Los
(2005)}. This approach leads to the exact linear time-convolution and
time-convolutionless homogeneous GMEs (TC-HGME and TCL-HGME), which take the
dynamics of initial correlations into account via modified memory kernels
governing the evolution of the relevant part of the distribution function
(statistical operator) of a many-particle system. These equations describe the
evolution (influenced by initial correlations) of the relevant part of a
distribution function on all time scales including the initial stage
$t-t_{0}\lesssim t_{cor}$, when the transition from the reversible evolution
to the irreversible kinetic behavior (resulting from the stochastic
instability of the system dynamics) is expected to occur.

We analyze how the obtained linear TC-HGME works for a spatially homogeneous
dilute gas of classical and quantum particles. But to obtain the desired
nonlinear equations (the Boltzmann equation in particular) describing the
evolution of a classical or quantum gas of particles from the linear TC-HGME,
we should make an additional approximation neglecting the time retardation of
the one-particle distribution function (statistical operator). Hence, in this
approach, we are not restricted by the first inequality in (\ref{0}),
$t_{cor}\ll t-t_{0}$, but still need the second inequality. The obtained (in
the linear approximation in the gas density) new homogeneous equations for a
one-particle distribution function (statistical operator) include initial
correlation caused by interaction between particles (as well as the quantum
initial correlations in the quantum case) and describe the evolution process
on the time scale which includes the initial stage $t-t_{0}\lesssim t_{cor}$.
We show that on the appropriate time scale these equations convert to the
classical and quantum Boltzmann equations if all initial correlations (due to
interparticle interactions) vanish on this time scale.

We note, the obtained linear homogeneous GMEs, when applied to an open quantum
system interacting with a thermal bath, are not subject to the time interval
restriction $t-t_{0}\ll t_{rel}$ (here $t_{rel}$ is the relaxation time due to
the system-bath interaction) and describe all stages of a system of interest
evolution including the initial stage $t-t_{0}\lesssim t_{cor}$.

We see, that, strictly speaking, it is impossible to obtain a nonlinear master
equation (like the Boltzmann equation) in the framework of the linear
time-independent projection operator formalism without additional
approximations (like discussed above neglection of time retardation which is
justified by the second inequality in (\ref{0})). Thus, a nonlinear equation
for a density matrix is often simply postulated in the form of a Lindblad-type
master equation whose generator depends parametrically on this density matrix
(see, e.g., \cite{Breuer and Petruccione (2002)}).\ 

To overcome the mentioned restriction, we suggest an approach allowing to
obtain a new class of evolution equations for the relevant part of a
distribution function (statistical operator) that suitable for deriving both
the nonlinear and linear evolution equations for the reduced distribution
functions (statistical operators) of interest and that can hold on any time
scale \cite{Los (2009)}. The class contains exact nonlinear\textbf{\ }%
inhomogeneous and homogeneous GMEs deduced from the Liouville-von-Neumann
equation using a (generally) nonlinear time-dependent operator $P(t)$ that
converts the many-particle distribution function (statistical operator) into
the appropriate relevant form. The time-dependent operator $P(t)$ is not
generally a projection operator. The nonlinear inhomogeneous GME can be
considered as an alternative to the BBGKY chain. In order to include the
initial correlations into consideration, this equation is converted into
nonlinear homogeneous GME by the method used for the linear GMEs. In the case
of a time-independent linear projection operator $P$ (\ref{0b}), the obtained
equations reduce to the conventional linear GME and obtained TC-HGME. In
contrast to these linear equations, the obtained nonlinear equations are more
general and therefore more convenient, in particular, for obtaining
Boltzmann-like equations and studying the spatially inhomogeneous nonideal gas
of particles for which the spatial nonlocality and time retardation are
essential (see, e.g., \cite{Klimontovich (1982)}).

We apply the derived nonlinear GMEs to the spatially inhomogeneous nonideal
diluted gas of classical particles. In particular, from the obtained
homogeneous nonlinear GME in the linear approximation in the density
parameter, we obtain a new homogeneous nonlinear equation for a one-particle
distribution function that holds on all time scales (no restrictions like
(\ref{0}) are needed) and takes the influence of the dynamics of initial
correlations into account in all stages of the evolution process. In this
approximation (linear in $n$), the evolution of collisions and initial
correlations (in the memory kernel) is governed by the exact two-particle
propagator, and the obtained equation is closed in this sense . If the gas
dynamics have the necessary property (ergodic mixing flow in the phase space),
then all initial correlations vanish at the kinetic time scale, and this
equation can be converted into the irreversible nonlinear Boltzmann equation
with a self-consistent nonlinear Vlasov term. No additional commonly used
assumption, like special initial conditions (factorized ones) or a principle
(weakening of initial correlations), is needed.

\section{TIME-CONVOLUTION HOMOGENEOUS GME}

We start with the Liouville-von-Neumann equation for a distribution function
(statistical operator) $F_{N}(t)$ of $N$ ($N\gg1$) particles
\begin{equation}
\frac{\partial}{\partial t}F_{N}(t)=LF_{N}(t)\text{.} \label{1}%
\end{equation}
Here, $F_{N}(t)$ is a symmetric function of $N$ variables $x_{i}%
=(\mathbf{r}_{i},\mathbf{p}_{i})$ $(i=1,....,N)$ representing the coordinates
and momenta of the particles (classical physics) or a statistical operator
(quantum mechanics), which satisfy the normalization conditions
\begin{equation}
\int dx_{1}\ldots\int dx_{N}F_{N}(x_{1},\ldots x_{N},t)=1,TrF_{N}=1,
\label{1'}%
\end{equation}
$L$ is the Liouville operator acting on $F_{N}(t)$ in the case of classical
physics as
\begin{equation}
LF_{N}(t)=\{H,F_{N}(t)\}_{P}=\sum_{i=1}^{N}\{\frac{\partial H}{\partial
\mathbf{r}_{i}}\frac{\partial F_{N}(t)}{\partial\mathbf{p}_{i}}-\frac{\partial
H}{\partial\mathbf{p}_{i}}\frac{\partial F_{N}(t)}{\partial\mathbf{r}_{i}}\},
\label{1''}%
\end{equation}
where $\{H,F_{N}(t)\}_{P}$ is the Poisson bracket and $H$ is the Hamilton
function for the system under consideration, while in the quantum physics case
$L$ is the superoperator acting on a statistical operator as
\begin{equation}
LF_{N}(t)=\frac{1}{i\hbar}\left[  H,F_{N}(t)\right]  , \label{1'''}%
\end{equation}
where $[,]$ is a commutator and $H$ is the Hamiltonian. Note, that for
simplicity we consider the case of a closed isolated system when $H$ does not
depend on time $t$.

The formal solution to Eq. (\ref{1}) is
\begin{equation}
F_{N}(t)=e^{L(t-t_{0})}F_{N}(t_{0}), \label{1a}%
\end{equation}
where $F_{N}(t_{0})$ is a distribution function (statistical operator) at an
initial instant $t_{0}$, when the initial condition for the
Liouville-von-Neumann equation (\ref{1}) should be set.

Employing the projection operator technique \cite{Nakajima (1958)},
\cite{Zwanzig (1960)}, \cite{Prigogine (1962)}, and applying the projection
operators $P=P^{2}$ and $Q=Q^{2}=1-P$ to Eq. (\ref{1}), it is easy to obtain
the equations for the relevant $f_{r}(t)=PF_{N}(t)$ and irrelevant $f_{i\text{
}}(t)=QF_{N}(t)$ (see (\ref{0b})) parts of $F_{N}(t)$
\begin{align}
\frac{\partial}{\partial t}f_{r}(t)  &  =PL[f_{r}(t)+f_{i}(t)],\label{2}\\
\frac{\partial}{\partial t}f_{i}(t)  &  =QL[f_{r}(t)+f_{i}(t)]. \label{3}%
\end{align}

A formal solution to Eq. (\ref{3}) has the form
\begin{equation}
f_{i}(t)=\int\limits_{t_{0}}^{t}\exp[QL(t-t_{1})]QLf_{r}(t_{1})dt_{1}%
+\exp[QL(t-t_{0})]f_{i}(t_{0}). \label{4}%
\end{equation}
Inserting this solution into (\ref{2}), we obtain the conventional exact
time-convolution generalized master equation (TC-GME) known as the
Nakajima-Zwanzig equation for the relevant part of the distribution function
(statistical operator)
\begin{align}
\frac{\partial}{\partial t}f_{r}(t)  &  =PLf_{r}(t)+\int\limits_{t_{0}}%
^{t}PL\exp[QL(t-t_{1})]QLf_{r}(t_{1})dt_{1}\nonumber\\
&  +PL\exp[QL(t-t_{0})]f_{i}(t_{0}). \label{5}%
\end{align}

It is important to stress that $f_{r}(t)$ and $f_{i}(t_{0})$ are the basic
quantities we are dealing with in Eq. (\ref{5}). All functions of dynamical
variables, the average values of which we can calculate with the help of
$f_{r}(t)$ by multiplying equation (\ref{5}) with the corresponding functions
(operators) from the right and integrating over the relevant variables (taking
a trace), are dependent only on the variables which are not projected off by
$P$ ($P$ integrates off all excessive variables in $F_{N}(t) $). Therefore, if
we represent $f_{r}(t)$ and $f_{i}(t_{0})$ in (\ref{5}) as $f_{r}%
(t)=PF_{N}(t)$ and $f_{i\text{ }}(t_{0})=QF_{N}(t_{0})$, correspondingly, then
the projection operators $P$ and $Q$ in these expressions act only on
$F_{N}(t)$ but not on the functions of dynamical variables of interest to the
right of them. This is the essence of the reduced description method, when, in
order to calculate the average values of the functions dependent on a much
smaller number of variables than the whole distribution function $F_{N}(t)$,
we actually need only the reduced distribution function (density matrix) like
the one-particle distribution functions determining the relevant distribution
function $f_{r}(t)$ in (\ref{0a}).

Serving as a basis for many applications, Eq. (\ref{5}), nevertheless,
contains the undesirable and in general non-negligible inhomogeneous term (the
last term in the right hand side of (\ref{5})), which depends (via $f_{i\text{
}}(t_{0})$) on the same large number of variables as the distribution function
$F_{N}(t_{0})$ at the initial instant ($f_{i\text{ }}(t_{0})$ is not composed
of the reduced distribution functions). Therefore, Eq. (\ref{5}) does not
provide for a complete reduced description of a multiparticle system in terms
of relevant (reduced) distribution function. Applying Bogoliubov's principle
of weakening of initial correlations (allowing to eliminate the influence of
$f_{i}(t_{0})$ on the large enough time scale $t-t_{0}\gg t_{cor}$) or using a
factorized initial condition (RPA),when $f_{i\text{ }}(t_{0})=QF_{N}(t_{0}%
)=0$, one can achieve the above-mentioned goal and obtain the homogeneous GME
for $f_{r}(t)$, i.e. Eq. (\ref{5}) with no initial condition term. However,
obtained in such a way homogeneous GME is either approximate and valid only on
a large enough time scale (when all initial correlations vanish) or applicable
only for a rather artificial (actually unreal, as pointed in \cite{van Kampen
(2004)}) initial conditions (no correlations at an initial instant of time).

In order to obtain a homogeneous equation for the relevant part of the
distribution function (statistical operator) we try to transfer the
inhomogeneous initial correlations (irrelevant) term in the right hand side of
Eq. (\ref{5}) to the (super)operator (kernel) acting on the relevant part
$f_{r}(t)$. To achieve this goal we suggest to present the initial
(irrelevant) term $f_{i}(t_{0})=QF_{N}(t_{0})$ as a following exact identity
\begin{align}
f_{i}(t_{0})  &  =[QF_{N}(t_{0})]F_{N}^{-1}(t_{0})e^{-L(t-t_{0})}%
(P+Q)e^{L(t-t_{0})}F_{N}(t_{0})\nonumber\\
&  =C_{0}\exp[-L(t-t_{0})][f_{r}(t)+f_{i}(t)]. \label{6}%
\end{align}
We assume here that the inverse operator $F_{N}^{-1}(t_{0})=$ $[f_{r}%
(t_{0})+f_{i}(t_{0})]^{-1}$ exists (see below), use that $P+Q=1$ and that the
backward propagator of the system is $\exp[-L(t-t_{0})]$. In (\ref{6}) the
initial correlation parameter $C_{0}$ is introduced
\begin{align}
C_{0}  &  =[QF_{N}(t_{0})]F_{N}{}^{-1}(t_{0})=f_{i}(t_{0})[f_{r}(t_{0}%
)+f_{i}(t_{0})]^{-1}\nonumber\\
&  =f_{i}(t_{0})f_{r}^{-1}(t_{0})[1+f_{i}(t_{0})f_{r}^{-1}(t_{0}%
)]^{-1}\nonumber\\
&  =(1-C_{0})f_{i}(t_{0})f_{r}^{-1}(t_{0}), \label{7}%
\end{align}
where the projection operator $Q$ acts only on $F_{N}(t_{0})$, which is
indicated by enclosing $QF_{N}(t_{0})$ in brackets.

Thus, the additional identity (\ref{6}) has been obtained by multiplying the
irrelevant part by the unity $F_{N}^{-1}(t_{0})F_{N}(t_{0})$ (which implies
the existence of $F_{N}^{-1}(t_{0})$) and inserting the unities $\exp
[-L(t-t_{0})]\exp[L(t-t_{0})]=1$ and $P+Q=1$. Therefore, neither divergency
(due to possible vanishing of $F_{N}(t_{0})$) nor indetermination of the $0/0
$ type (behaviors of the numerator and denominator in $F_{N}(t_{0}%
)/F_{N}(t_{0})=1$ are similar) can happen. This holds over all further
(identical) manipulations (see below). As it is seen from (\ref{7}), the
correlation parameter is a series in $f_{i}(t_{0})f_{r}^{-1}(t_{0})$ and,
therefore, one may only need a formal existence of the function (operator)
$f_{r}^{-1}(t_{0})$ (see also (\ref{39q})), which is invert to the relevant
distribution function (statistical operator) chosen with the help of the
appropriate projection operator $P$ (generally, it can provide some
restriction on the class of appropriate projectors). The relevant part, which
is mainly of interest, is, as a rule, a vacuum (relatively slowly changing)
part of a distribution function (statistical operator), i.e. the part with no
correlations (e.g., a product of the one-particle distribution functions as in
(\ref{0a})). It seems plausible that the invert relevant part of the
distribution function (statistical operator) defined in a pointed above sense
(uncorrelated part) can always be constructed.The examples of construction of
the appropriate invert distribution function (operator) $f_{r}^{-1}(t_{0})$
will be given below.

As a result of introducing the additional identity (\ref{6}), we have two
equations, (\ref{4}) and (\ref{6}), connecting $f_{i}(t)$ with $f_{i}(t_{0})$.
Finding $f_{i}(t_{0})$ from these equations and inserting it into (\ref{5}),
we obtain the following exact equation for the relevant part of a distribution
function (statistical operator):
\begin{equation}
\frac{\partial}{\partial t}f_{r}(t)=PLR(t-t_{0})f_{r}(t)+\int\limits_{t_{0}%
}^{t}PLR(t-t_{0})\exp[QL(t-t_{1})]QLf_{r}(t_{1})dt_{1}, \label{8}%
\end{equation}
where function $R(t-t_{0})$ is defined as
\begin{align}
R(t-t_{0})  &  =1+C(t-t_{0})=\frac{1}{1-C_{0}(t-t_{0})},\nonumber\\
C(t-t_{0})  &  =e^{QL(t-t_{0})}\frac{1}{1-C_{0}e^{-L(t-t_{0})}e^{QL(t-t_{0})}%
}C_{0}e^{-L(t-t_{0})},\nonumber\\
C_{0}(t-t_{0})  &  =e^{QL(t-t_{0})}C_{0}e^{-L(t-t_{0})}. \label{9}%
\end{align}

We have derived the time-convolution homogeneous generalized master equation
(TC-HGME) (\ref{8}) for the relevant part of the distribution function
(statistical operator). This equation holds in both the classical and quantum
physics cases if the proper redefinition of the symbols is done and all
(super)operators exist (we will address the latter problem below). We have not
removed any information while deriving equation (\ref{8}), and, therefore, it
is exact integro-differential equation which takes into account the initial
correlations and their dynamics via the modification of the (super)operator
(memory kernel) in GME (\ref{5}) acting on the relevant part of the
distribution function (statistical operator) $f_{r}(t)$. The obtained exact
kernel of TC-HGME (\ref{8}) can serve as a starting point for consecutive
perturbation expansions. In many cases such expansions of the homogeneous
equations (like (\ref{8})) have much broader range of validity than that for
the inhomogeneous ones (like (\ref{5})).

The initial correlations are entering the TC-HGME (\ref{8}) through the
function $R(t)$ (\ref{9}). This function represents a sum of the infinite
series of expansion in the properly defined time dependent parameter
$C_{0}(t-t_{0})$ which describes the influence of initial correlations in
time. Presence of the function $R(t)\neq1$ in Eq. (\ref{8}) has led to the
modification of the first term in the right-hand side of the GME (\ref{5}) as
well as the second (collision) term. This modification reflects the influence
of initial correlations on the corresponding processes (flow and relaxation)
with time. The TC-HGME also allows treating the correlations arising from
collisions and initial correlations on an equal footing by expanding the
memory kernel in series in some small parameter.

However, the problem of the existence (convergence) of $R(t-t_{0})$ can be
raised. The function $R(t-t_{0})$ behaves properly at all times. Moreover, the
expansion of the kernel in (\ref{8}) can result in canceling the pole in
function $R(t-t_{0})$ as we will see below. In such cases there is no problem
with the existence of $R(t-t_{0})$.

\section{TIME-CONVOLUTIONLESS HOMOGENEOUS GME}

Now, let us turn to the case of the so called time-convolutionless
(time-local) GME (TCL-GME). It is believed that such TCL-GME can be more
easily solved and be more convenient for studying the non-Markovian processes
than the integral TC-GME (\ref{5}) (see \cite{Breuer and Petruccione (2002)},
\cite{Shibata1 (1977)}, \cite{Shibata2 (1980)} ).

Using the identity $F_{N}(\tau)=\exp[-L(t-\tau)][f_{r}(t)+f_{i}(t)]$, we turn
Eq. (\ref{4}) into the form
\begin{align}
f_{i}(t)  &  =\left[  1+A(t-t_{0})\right]  ^{-1}\left[  -A(t-t_{0}%
)f_{r}(t)+e^{QL(t-t_{0})}f_{i}(t_{0})\right]  ,\nonumber\\
A(t-t_{0})  &  =-\int\limits_{0}^{t-t_{0}}e^{QL\tau}QLPe^{-L\tau}%
d\tau=e^{QL(t-t_{0})}Qe^{-L(t-t_{0})}-Q. \label{10}%
\end{align}
Inserting Eq. (\ref{10}) for $f_{i}(t)$ into equation (\ref{2}), we obtain the
TCL-GME \cite{Shibata1 (1977)}, \cite{Shibata2 (1980)}
\begin{equation}
\frac{\partial}{\partial t}f_{r}(t)=PL\left[  1+A(t-t_{0})\right]
^{-1}\left[  f_{r}(t)+e^{QL(t-t_{0})}f_{i}(t_{0})\right]  , \label{11}%
\end{equation}
which contains the undesirable inhomogeneous initial condition term
$\varpropto f_{i}(t_{0})$.

Solving Eqs. (\ref{10}) and (\ref{6}), we get for $f_{i}(t_{0})$%
\begin{equation}
f_{i}(t_{0})=\chi^{-1}C_{0}e^{-L(t-t_{0})}\left(  1+A(t-t_{0})\right)
^{-1}f_{r}(t), \label{12}%
\end{equation}
where
\begin{equation}
\chi^{-1}=\left[  1-C_{0}e^{-L(t-t_{0})}\left(  1+A(t-t_{0})\right)
^{-1}e^{QL(t-t_{0})}\right]  ^{-1}. \label{13}%
\end{equation}
The inverse operator (\ref{13}), needed for obtaining $f_{i}(t_{0})$ from Eq.
(\ref{12}), is defined by expanding it into the series in the term $\varpropto
C_{0}$. At $t=t_{0}$, $A(t-t_{0})=0$, and there is no singularity at the
initial moment of time if $\left\vert C_{0}\right\vert <1$. From the existence
of $\left(  1+A(t-t_{0})\right)  ^{-1}$\cite{Shibata1 (1977)}, \cite{Shibata2
(1980)}, it follows that $\chi^{-1}$ exists at any time if $\left\vert
C_{0}\right\vert <1$. As for the function $R(t-t_{0})$ (\ref{9}) in TC-HGME
(\ref{8}), an expansion in some small parameter can result in canceling the
pole in (\ref{13}) and, therefore, the condition $\left\vert C_{0}\right\vert
<1$ can be not necessary.

Inserting Eq. (\ref{12}) into Eq. (\ref{11}) and using (\ref{13}), we obtain a
desired time-convolutionless (time-local) homogeneous generalized master
equation (TCL-HGME) for the relevant part of the distribution function
(statistical operator) $f_{r}(t)$
\begin{equation}
\frac{\partial}{\partial t}f_{r}(t)=PL\left[  1+A(t-t_{0})\right]
^{-1}\left\{  1-C_{0}(t-t_{0})\left[  1+A(t-t_{0})\right]  ^{-1}\right\}
^{-1}f_{r}(t). \label{14}%
\end{equation}
Based on the above mentioned arguments, one can expect that all operators
entering equation (\ref{14}) exist.

Thus, using the introduced identity (\ref{6}), we have transferred the
inhomogeneous initial correlations term of equation (\ref{11}) into the
operator acting on the relevant part of the distribution function (statistical
operator). Equation (\ref{14}) is exact homogeneous time-local equation for
the relevant part of a distribution function (statistical operator) which
accounts for initial correlations via the time-dependent parameter of initial
correlations $C_{0}(t-t_{0})$ defined by (\ref{9}). This equation is expected
to work on any time scale and to describe the entire evolution process.
Equation (\ref{14}) can also appear more convenient for studying the
non-Markovian processes than the TC-HGME (\ref{8}).

\section{EVOLUTION EQUATION RETAINING INITIAL CORRELATIONS FOR A HOMOGENEOUS
DILUTE GAS OF CLASSICAL PARTICLES}

In this section we consider a gas of $N$ ($N\gg1$) interacting identical
classical particles. It is supposed that the Hamiltonian of the system can be
split into two parts
\begin{align}
H  &  =H^{0}+H^{^{\prime}},\nonumber\\
H^{0}  &  =\sum\limits_{i=1}^{N}H_{i}^{0},\text{ }H_{i}^{0}=\frac
{\mathbf{p}_{i}^{2}}{2m},\nonumber\\
H^{^{\prime}}  &  =\sum\limits_{i<j=1}^{N}V_{ij},\text{ }V_{ij}=V(|\mathbf{x}%
_{i}-\mathbf{x}_{j}|)\text{ .} \label{15}%
\end{align}
Here, $H^{0}$describes an ideal gas of noninteracting particles having the
momenta $\mathbf{p}_{i}$ and mass $m$, and the Hamiltonian of interaction
between particles $H^{^{\prime}}$ is assumed to be a sum of potentials
$V_{ij}$ dependent only on the difference of coordinates of two particles. We
do not suppose here that the interaction is weak and assume that all usual
necessary requirements to the properties of forces, by which particles
interact with each other, are met. In particular, we suppose that the bound
states are not formed (an interaction is repulsive).

To obtain the evolution equation in this case, we presume that the density of
particles $n=N/V$ ($V$ is the volume of the system) is small enough and use a
perturbation theory with the small parameter
\begin{equation}
\gamma=r_{o}^{3}n\ll1, \label{16}%
\end{equation}
where $r_{0}$ is the effective radius of the interparticle interaction. This
small parameter guarantees that the collisions between particles are the well
separated events and defines the time hierarchy, because $t_{cor}/t_{rel}%
\sim\gamma\ll1$.

First of all, we have to define the projection operator $P$ selecting the
relevant part $f_{r}(x_{1},\ldots,x_{N},t)$ of the $N-$particle distribution
function. It is convenient to deal with the $N-$particle distribution function
defined as $f_{N}(x_{1},\ldots,x_{N},t)=V^{N}F_{N}(x_{1},\ldots,x_{N},t)$,
which satisfies the same Liouville equation (\ref{1}). Then, the required
projection operator is
\begin{equation}
P=\left[  \prod\limits_{i=2}^{N}f_{1}(x_{i})\right]  \frac{1}{V^{N-1}}\int
dx_{2}\cdots\int dx_{N}\text{ .} \label{17}%
\end{equation}
Here, $f_{1}(x_{i})$ is a one-particle distribution function
\begin{equation}
f_{1}(x_{i},t)=V\int dx_{1}\cdots\int dx_{i-1}\int dx_{i+1}\cdots\int
dx_{N}F_{N}(x_{1,\ldots,}x_{N},t)\text{ } \label{18}%
\end{equation}
taken at the initial instant, $f_{1}(x_{i})=f_{1}(x_{i},t_{0})$. In the
definition (\ref{18}), a symmetry of a distribution function relatively to a
permutation of space-phase coordinates $x_{i}=(\mathbf{x}_{i},\mathbf{p}_{i})
$ is taken into account.

In general, the $s$-particle ($s\leq N$) distribution function is defined as
\begin{equation}
f_{s}(x_{1},\ldots,x_{s},t)=V^{s}\int dx_{s+1}\cdots\int dx_{N}F_{N}%
(x_{1},\ldots,x_{N},t)\text{ .} \label{19}%
\end{equation}

From (\ref{1'}) we have the normalization conditions for the reduced
distribution functions $f_{s}$
\begin{equation}
\int dx_{1}\cdots\int dx_{s}f_{s}(x_{1},\ldots,x_{s},t)=V^{s}\text{ .}
\label{20}%
\end{equation}

Applying the projection operator (\ref{17}) to the $N$-particle distribution
function $f_{N}(t)=V^{N}F_{N}(t)$, we obtain the following relevant part
$f_{r}$
\begin{equation}
f_{r}=Pf_{N}(t)=\left[  \prod\limits_{i=2}^{N}f_{1}(x_{i})\right]  f_{1}%
(x_{1},t)\text{ .} \label{21}%
\end{equation}
We suppressed in (\ref{21}) the arguments of the $f_{r}(x_{1,}\ldots,x_{N},t)$
for brevity.

The irrelevant part comprising initial correlations, $f_{i}(t_{0}%
)=(1-P)f_{N}(t_{0})=f_{N}(t_{0})-f_{r}(t_{0})$, can always be represented by
the following cluster expansion in terms of multiparticle correlation
functions (see, for example, \cite{Balescu (1975)})
\begin{equation}
f_{i}(t_{0})=\sum\limits_{i,j=1,i<j}^{N}g_{2}(x_{i},x_{j})\prod
\limits_{k=1,k\neq i,j}^{N-2}f_{1}(x_{k})+\sum\limits_{i,j,k=1,i<j<k}^{N}%
g_{3}(x_{i},x_{j},x_{k})\prod\limits_{l=1,l\neq i,j,k}^{N-3}f_{1}%
(x_{l})+\ldots, \label{22}%
\end{equation}
where $\prod\limits_{k=1,k\neq i,j}^{N-2}f_{1}(x_{k})$ and $\prod
\limits_{l=1,l\neq i,j,k}^{N-3}f_{1}(x_{l})$ stand for the products of $N-2$
and $N-3$ one-particle distribution functions with $k\neq i,j$ and $l\neq
i,j,k$, respectively, whereas $g_{2}(x_{i},x_{j})$ and $g_{3}(x_{i}%
,x_{j},x_{k})$ are the irreducible two-particle and three-particle correlation
functions (further terms in (\ref{22}) are defined in the same way).

In order to find out the rules of acting of the Liouvillian $L$ and the
functions of the Liouvillian, which enter Eq. (\ref{8}), on the relevant
distribution function (\ref{21}), note, that in accordance with the
Hamiltonian (\ref{15}), the Liouvillian $L$ can be presented as
\begin{align}
L  &  =L^{0}+L^{^{\prime}}\text{ ,}\nonumber\\
L^{0}=\sum\limits_{i=1}^{N}L_{i}^{0},  &  L_{i}^{0}=-\mathbf{v}_{i}%
\cdot\mathbf{\nabla}_{i}\text{, }\mathbf{v}_{i}=\frac{\mathbf{p}_{i}}{m}\text{
, }\mathbf{\nabla}_{i}=\frac{\partial}{\partial\mathbf{x}_{i}}\text{
,}\nonumber\\
L^{^{\prime}}  &  =\sum\limits_{i<j=1}^{N}L_{ij}^{^{\prime}}\text{, }%
L_{ij}^{^{\prime}}=(\mathbf{\nabla}_{i}V_{ij})\cdot(\frac{\partial}%
{\partial\mathbf{p}_{i}}-\frac{\partial}{\partial\mathbf{p}_{j}})\text{ .}
\label{23}%
\end{align}

We suppose, as usual, that all functions $\Phi(x_{1},\ldots,x_{N})$, defined
on the phase space, and their derivatives vanish at the boundaries of the
configurational space and at $\mathbf{p}_{i}=\pm\infty$. These properties of
distribution functions and the form of the Liouville operators (\ref{23}) lead
to the following relations
\begin{align}
\int dx_{i}L_{i}^{0}\Phi(x_{i},\ldots,x_{N},t)  &  =0,\nonumber\\
\int dx_{i}\int dx_{j}L_{ij}^{^{\prime}}\Phi(x_{1},\ldots,x_{N},t)  &  =0.
\label{24}%
\end{align}
These equations also follow, in general, from the Liouville-von- Neumann
equation (\ref{1}) and conditions (\ref{1'}).

Let us consider the first term in the right-hand side of Eq. (\ref{8}).
Applying operators (\ref{23}) to the relevant distribution function (\ref{21})
and taking into account (\ref{24}), we have
\begin{align}
PL^{0}f_{r}  &  =\left[  \prod\limits_{i=2}^{N}f_{1}(x_{i})\right]  L_{1}%
^{0}f_{1}(x_{1},t),\nonumber\\
PL^{^{\prime}}f_{r}  &  =\left[  \prod\limits_{i=2}^{N}f_{1}(x_{i})\right]
n\int dx_{2}L_{12}^{^{\prime}}f_{1}(x_{2})f_{1}(x_{1},t)\text{ .} \label{25}%
\end{align}

To calculate the additional term (hereinafter we put $t_{0}=0$ because Eq.
(\ref{8}) is valid for any initial instant $t_{0}$)
\begin{equation}
C(t)=P(L^{0}+L^{^{\prime}})e^{Q(L^{0}+L^{^{\prime}})t}\frac{1}{1-C_{0}%
e^{-(L^{0}+L^{^{\prime}})t}e^{Q(L^{0}+L^{^{\prime}})t}}C_{0}e^{-(L^{0}%
+L^{^{\prime}})t}f_{r}(t)\text{ ,} \label{26}%
\end{equation}
arising due to initial correlations, we use the following relations
\begin{align}
e^{(A+B)t}  &  =e^{At}+\int\limits_{0}^{t}d\theta e^{A(t-\theta)}%
Be^{(A+B)\theta},\label{27}\\
e^{L^{0}t}\Phi(x_{1},\ldots,x_{N},t)  &  =\left[  \prod\limits_{i=1}%
^{N}e^{L_{i}^{0}t}\right]  \Phi(x_{1},\ldots,x_{N},t)\nonumber\\
&  =\Phi(\mathbf{x}_{1}-\mathbf{v}_{1}t,\mathbf{p}_{1},\ldots,\mathbf{x}%
_{N}-\mathbf{v}_{N}t,\mathbf{p}_{N},t)\text{ ,} \label{28}%
\end{align}
where $A$ and $B$ are the arbitrary operators. Equation (\ref{28}) follows
from the definition (\ref{23}) of the Liouvillian $L^{0}$ of the free
propagating particles. We also note, that the parameter of initial
correlations (\ref{7}) $C_{0}=f_{i}(0)/[f_{r}(0)+f_{i}(0)]$ can be expanded
into the power series in $f_{i}(0)/f_{r}(0)$. Using (\ref{21}) and (\ref{22}),
we have
\begin{equation}
\frac{f_{i}(0)}{f_{r}(0)}=\sum\limits_{i<j=1}^{N}\frac{g_{2}(x_{i},x_{j}%
)}{f_{1}(x_{i})f_{1}(x_{j})}+\sum\limits_{i<j<k}\frac{g_{3}(x_{i},x_{j}%
,x_{k})}{f_{1}(x_{i})f_{1}(x_{j})f_{1}(x_{k})}+\ldots\label{29}%
\end{equation}
If we use of (\ref{27}) and expand (\ref{26}) in $PL^{0}$, $L^{^{\prime}}$,
$QL^{^{\prime}}$ and $f_{i}(0)/f_{r}(0)$, then, it is easy to see that
(\ref{26}) can be presented as a sum of products of operators $PL^{0}%
,L^{^{\prime}},PL^{^{\prime}}$, $\exp(L^{0}\theta)$, one-particle distribution
functions and correlations functions (\ref{29}). Applying the projection
operator (\ref{17}) to such products and accounting for (\ref{21}),
(\ref{23}), (\ref{24}), (\ref{28}) and (\ref{29}), we find that (\ref{26}) can
be presented as a series expansion in the density parameter $\gamma$ (\ref{16}).

In what follows, we restrict ourselves to the linear approximation in the
density parameter $\gamma$ (\ref{16}). Also, in order to simplify formulae, a
spatially homogeneous case will be considered. In this case a one-particle
distribution function does not depend on a particle coordinate, $f_{1}%
(x_{j},t)=f_{1}(\mathbf{p}_{j},t)$ ($\int f(\mathbf{p,}t)d\mathbf{p=}1$), and
the multiparticle correlation functions depend only on the differences of
coordinates, e.g., $g_{2}(x_{i},x_{j})=g_{2}(\mathbf{x}_{i}-\mathbf{x}%
_{j},\mathbf{p}_{i},\mathbf{p}_{j})$. A spatial homogeneity allows us to
cancel all terms with $PL^{0}$(such as the first equation (\ref{25})),
because, according to (\ref{23}) and (\ref{24}), all these terms contain a
derivative with respect to a variable $\mathbf{x}_{1}$, which is not under the
integration. The second equation (\ref{25}) also vanishes in a space
homogeneous case, because we consider a potential $V_{ij}$ (\ref{15})
dependent of the particle coordinates difference. Taking also into account,
that the terms containing $PL^{^{\prime}}$ result in the expressions
proportional at least to the first power of $n$, we can replace operators
$Q=1-P$ in (\ref{26}) by the unity and simplify it to
\begin{equation}
C(t)=PL^{^{\prime}}e^{(L^{0}+L^{^{\prime}})t}\frac{1}{1-C_{0}}C_{0}%
e^{-(L^{0}+L^{^{\prime}})t}f_{r}(t)\text{.} \label{30}%
\end{equation}

Expanding formally $\frac{1}{1-C_{0}}$ and $C_{0}$ in the power series in
$f_{i}/f_{r}$, restricting ourselves to the first order in $n$ and then
converting the series, we obtain the following expression for the contribution
of initial correlations (\ref{26})
\begin{equation}
C(t)=\left[  \prod\limits_{i=2}^{N}f_{1}(\mathbf{p}_{i})\right]  n\int
dx_{2}L_{12}^{\prime}e^{L_{12}t}\frac{g_{2}(x_{1},x_{2})}{f_{1}(\mathbf{p}%
_{1})f_{1}(\mathbf{p}_{2})}e^{-L_{12}t}f_{1}(\mathbf{p}_{2})f_{1}%
(\mathbf{p}_{1},t), \label{31}%
\end{equation}
where $L_{12}=L_{12}^{0}+L_{12}^{\prime}$, $L_{12}^{0}=L_{1}^{0}+L_{2}^{0}$,
i.e. $L_{12}$ is the two-particle Liouvillian (\ref{23}), and $f_{1}%
(\mathbf{p}_{i})=f_{1}(\mathbf{p}_{i},t_{0})$. The parameter of initial
correlations $C_{0}(x_{1},\ldots,x_{N})$ in (\ref{30}) under the accepted
linear approximation in $n$ (binary collisions) reduces to the parameter
$C_{0}^{12}(x_{1},x_{2})$, which contains only a two-particle correlation
function $g_{2}(x_{1},x_{2})=g_{2}(\mathbf{x}_{1}-\mathbf{x}_{2}%
,\mathbf{p}_{1},\mathbf{p}_{2})$:
\begin{equation}
\frac{1}{1-C_{0}^{12}(x_{1},x_{2})}C_{0}^{12}(x_{1},x_{2})=\frac{g_{2}%
(x_{1},x_{2})}{f_{1}(\mathbf{p}_{1})f_{1}(\mathbf{p}_{2})}\text{ .} \label{32}%
\end{equation}
Deriving (\ref{31}), we have also taken into account that each additional
integration over $x_{3},\ldots$ adds an additional power of $n$ and,
therefore, in the linear approximation in $n$, all formulae can contain no
more than one integration over the phase space.

Consider now the second (collision) term of the TC-HGME (\ref{8}). Initial
correlations also contribute to this term through the second term of function
$R(t)=1+\left[  R(t)-1\right]  $ (\ref{9}). We start with the conventional
collision term of the TC-GME taking $R(t)=1$. To simplify this term, we note,
that the following relations hold in general
\begin{align}
PL^{0}e^{L^{0}t}Q  &  =0,\nonumber\\
e^{QL^{0}t}Q  &  =e^{L^{0}t}Q\text{ .} \label{33}%
\end{align}
The second equation (\ref{33}) follows from the first one, which in its turn
follows from (\ref{17}), (\ref{20}), (\ref{23}), (\ref{24}), (\ref{27}) and
(\ref{28}). The term in the considered collision integral, which is
proportional to $QL^{0}f_{r}(t)$, gives zero contribution in the case of
spatial homogeneity. Thus, the kinetic term under consideration acquires the
form
\begin{equation}
K(t)=\int\limits_{0}^{t}dt_{1}PL^{^{\prime}}e^{(L^{0}+QL^{^{\prime}})t_{1}%
}QL^{^{\prime}}f_{r}(t-t_{1})\text{ .} \label{34}%
\end{equation}

Using the procedure, described above, we have the collision term (\ref{34}) in
the accepted linear approximation in the density parameter $\gamma$ as
\begin{equation}
K(t)=\left[  \prod\limits_{i=2}^{N}f_{1}(\mathbf{p}_{i})\right]
n\int\limits_{0}^{t}dt_{1}\int dx_{2}L_{12}^{^{\prime}}e^{L_{12}t_{1}}%
L_{12}^{^{\prime}}f_{1}(\mathbf{p}_{2})f_{1}(\mathbf{p}_{1},t-t_{1})\text{ .}
\label{35}%
\end{equation}

In the same way we can consider the contribution of initial correlations to
the collision integral. Accounting for the second term in $R(t)$ (\ref{9}) in
the kinetic term of equation (\ref{8}) leads to the modification of the
collision integral (\ref{35}). In the linear approximation in $\gamma$ we have
for the kinetic term of equation (\ref{8}) (instead of (\ref{35}))
\begin{align}
\tilde{K}(t)  &  =\prod\limits_{i=2}^{N}f_{1}(\mathbf{p}_{i})n\int
\limits_{0}^{t}dt_{1}\int dx_{2}L_{12}^{^{\prime}}R_{12}(t)e^{L_{12}t_{1}%
}L_{12}^{^{\prime}}\nonumber\\
&  \times f_{1}(\mathbf{p}_{2})f_{1}(\mathbf{p}_{1},t-t_{1})\text{
,}\label{36}\\
R_{12}(t)  &  =1+e^{L_{12}t}\frac{g_{2}(x_{1},x_{2})}{f_{1}(\mathbf{p}%
_{1})f_{1}(\mathbf{p}_{2})}e^{-L_{12}t}\text{ .} \label{37}%
\end{align}

Now, collecting all obtained formulae (\ref{31}), (\ref{36}) and (\ref{37})
and accounting for the definition of the relevant distribution function
(\ref{21}), we obtain from the TC-HGME (\ref{8}) the following equation for a
one-particle momentum distribution function
\begin{align}
\frac{\partial}{\partial t}f_{1}(\mathbf{p}_{1},t)  &  =n\int dx_{2}%
L_{12}^{^{\prime}}G_{12}(t)f_{1}(\mathbf{p}_{2})f_{1}(\mathbf{p}%
_{1},t)\nonumber\\
&  +n\int dx_{2}\int\limits_{0}^{t}dt_{1}L_{12}^{^{\prime}}[1+G_{12}%
(t)]e^{L_{12}t_{1}}L_{12}^{^{\prime}}f_{1}(\mathbf{p}_{2})f_{1}(\mathbf{p}%
_{1},t-t_{1})\text{ ,} \label{38}%
\end{align}
where
\begin{align}
G_{12}(t)  &  =R_{12}-1\nonumber\\
&  =e^{L_{12}t}\frac{g_{2}(x_{1},x_{2})}{f_{1}(\mathbf{p}_{1})f_{1}%
(\mathbf{p}_{2})}e^{-L_{12}t} \label{38a}%
\end{align}
is the parameter of initial correlations in the linear in $\gamma$
approximation which is determined by the time evolution of a two-particle
correlation function. It is interesting to note that the expression like
(\ref{38a}) appears as an additional term in the (correlation) entropy caused
by two-particle correlation function and obtained by means of
quasi-equilibrium statistical operator (see, e.g. \cite{Ropke (1987)}). This
additional contribution to entropy is essential when a two-particle
correlation function damps slowly.

Equation (\ref{38}) is a new equation for a one-particle distribution function
obtained from TC-HGME (\ref{8}) in the linear approximation in $n$ and which
accounts for both initial correlations and collisions in this approximation
exactly. It has been obtained not using the Bogoliubov principle of weakening
of initial correlations or any other hypothesis. First term in the right-hand
side of this equation (linear in $L_{12}^{^{\prime}}$) is exclusively defined
by initial correlations evolving with time. Initial correlations also modify
the collision integral (second term in the right-hand side of this equation).
It contains, except the term related to an interaction between particles
(determined by the Liouvillian $L_{12}^{^{\prime}}$), also time-dependent
initial correlations determined by the two-particle correlation function
(\ref{38a})). Evolution in time of initial correlations as well as of the
collision term is determined by the exact two-particle propagator $\exp
(L_{12}t)$ only, and in this sense Eq. (\ref{38}) is closed. As it follows
from (\ref{27}), the propagator $\exp(L_{12}t)$ satisfies the integral
equation
\begin{equation}
\exp(L_{12}t)=\exp(L_{12}^{0}t)+\int\limits_{0}^{t}dt_{1}\exp[L_{12}%
^{0}(t-t_{1})]L_{12}^{^{\prime}}\exp(L_{12}t_{1})\text{ ,} \label{39}%
\end{equation}
where $\exp(L_{12}^{0}t)$ is the propagator for noninteracting particles. In
the case of a weak inter-particle interaction, we can iterate the integral
equation (\ref{39}) in $L_{12}^{^{\prime}}$ and obtain equation (\ref{38}) in
the desired approximation in the small interaction parameter (note, that the
correlation function $g_{2}(x_{1},x_{2})$ also depends on an interaction
between particles).

The obtained homogeneous integro-differential equation (\ref{38}) differs from
that which could be found from the conventional GME (\ref{5}) using the
projection operator (\ref{17}) and the principle of weakening of initial
correlations or RPA. In the latter case we would have equation (\ref{38}) in
which the time-dependent correlation function factor $G_{12}(t)$ determining
the dynamics of initial correlations would be set zero. Moreover, such an
equation would be valid only at at the time scale $t\gg t_{cor}$, whereas
equation (\ref{38}) is valid on any time scale.

It should be noted, that there can be the large-scale correlations associated
with, e.g., the quantities that are conserved. These correlations can only
vanish on the time scale of the order of the relaxation time $t_{rel}$ of a
distribution function $f_{1}(\mathbf{p}_{i},t)$. If at $t\gg t_{cor}$
$(t_{cor}\ll t_{rel})$ the correlation function $G_{12}(t)$ was set zero, this
would signify that all\textbf{\ }initial\textbf{\ }correlations vanish on this
time scale, i.e. the large-scale correlations do not matter.

Equation (\ref{38}) describes the entire evolution process and is expected to
switch from initial (reversible) regime into the kinetic (irreversible) one,
automatically. It will be shown below, that Eq. (\ref{38}) converts on the
appropriate time scale into the conventional Boltzmann equation if all
correlations (initial and caused by collisions) damp with time.

\section{CONNECTION TO THE CLASSICAL BOLTZMANN\ EQUATION}

Obtained closed equation (\ref{38}) for a one-particle distribution function
is not a kinetic equation in the conventional sense. It does not describe
generally an irreversible evolution with time because it is time-reversible if
the parameter of initial correlations (\ref{32}) remains unchanged when all
the particle velocities are reversed (see below). Thus, a time-asymmetric
behaviour can be secured by the special choice of an initial condition. If we
suppose that $C_{0}^{12}(\mathbf{x}_{1}-\mathbf{x}_{2},\mathbf{p}%
_{1},\mathbf{p}_{2})=C_{0}^{12}(\mathbf{x}_{1}-\mathbf{x}_{2},-\mathbf{p}%
_{1},-\mathbf{p}_{2})$, i.e., $g_{2}(\mathbf{x}_{1}-\mathbf{x}_{2}%
,\mathbf{p}_{1},\mathbf{p}_{2})=g_{2}(\mathbf{x}_{1}-\mathbf{x}_{2}%
,-\mathbf{p}_{1},-\mathbf{p}_{2})$ and $f_{1}(\mathbf{p}_{i})=f_{1}%
(-\mathbf{p}_{i})$, then equation (\ref{38}) is invariant under the
transformation $t\rightarrow-t,$ because such a transformation results in
changing the sign of all Liouvillians $L_{i}^{0}$ and $L_{ij}^{^{\prime}}$.
Such reversibility is understandable on the very early stage of the system
evolution, when $t\lesssim t_{cor}$ and memory of the initial state has not
been lost. On the time scale $t\gg t_{cor}$ the situation can change.

It can be seen from (\ref{38}) that, in order to enter the kinetic
(irreversible) stage of the evolution, the reversible terms connected with
initial correlations should vanish on some time scale. Moreover,
irreversibility of collision integral may be realized if it is possible to
extend the integral over $t_{1}$ to infinity. Thus, we have to consider the
behaviour of initial correlations and the integral over $t_{1}$ in Eq.
(\ref{38}) in time.

It should be noted that the irreversible evolution can go \textquotedblright
forwards\textquotedblright\ or \textquotedblright backwards\textquotedblright.
The question as to which of these evolutions should be chosen cannot be
answered on the basis of the present consideration of the time-reversible
Liouville equation only, and we consider the evolution in the future $(t>0).$

In the accepted approximation of the binary collisions, the initial
correlations are described by a correlation function $g_{2}(x_{1},x_{2})$
depending on coordinates and momenta of only two particles. The time evolution
of this binary correlation function is governed by the two-particle propagator
(\ref{39}). In the zero approximation in the interaction $L_{12}^{^{\prime}}$,
the action of the propagator $\exp(L_{12}^{0}t)$ on a correlation function
$g_{2}(x_{1},x_{2})$ (and on all functions of coordinates) results in shifting
the coordinates $\mathbf{x}_{1} $and $\mathbf{x}_{2}$ with time according to
(\ref{28})
\begin{equation}
\exp(L_{12}^{0}t)g_{2}(x_{1},x_{2})=g_{2}(\mathbf{r}-\mathbf{g}t,\mathbf{p}%
_{1},\mathbf{p}_{2})\text{ ,} \label{40}%
\end{equation}
where $\mathbf{r}=\mathbf{x}_{1}-\mathbf{x}_{2}$ is the initial distance
between particles 1 and 2 and $\mathbf{g}=\mathbf{v}_{1}-\mathbf{v}_{2}$ is
the relative velocity. Therefore, in this case a binary correlation function
changes with time according to a change of a distance between particles
$|\mathbf{v}_{1}-\mathbf{v}_{2}|t$, which is linear in time. In general, an
evolution of a two-particle correlation function $g_{2}(x_{1},x_{2})$ under
the action of the exact two-particle propagator $\exp(L_{12}t)$ can be found
from the well-known solution to the two-body problem.

It is important to know how a correlation function behaves as the time passes.
If there is a big fraction of a \textquotedblright parallel
motion\textquotedblright\ and a distance between particles (or, more
generally, a distance between points in the phase space) does not change
significantly, a correlation function does not vanish, i.e. the initial
correlations do not disentangle with time. It can be so even in the case of
the ergodic flow in the phase space. If a distance between particles grows
with time and the radius of correlation between particles $r_{cor}$ $\backsim
r_{0}$ is limited, the correlation function $g_{2}(x_{1},x_{2})$ fades with
time. But for existence of a limited relaxation time, i.e. a finite time for
vanishing a correlation function, this function should diminish with time
exponentially. As it is well known, such an exponential disentangling of
correlations with time can be guaranteed by a mixing ergodic flow in the phase
space. The mixing ergodic flow is closely connected with the property of the
dynamical systems known as the local (stochastic) instability
(see,e.g.,\cite{Zaslavsky (1984)}). If we assume that the dynamics of the
system of classical particles under consideration have the property of mixing
flow (and this is usually the case), then the initial correlations weak with
time and disappear after a short correlation time $t_{cor}$. In this case, at
$t\gg t_{cor}$, we can neglect the contribution of initial correlations to
equation (\ref{38}) and put $G_{12}(t)=0$.

This cannot, however, be done if there are the correlations associated with
the large-scale fluctuations such as the collective excitations, for example,
the plasma oscillations. A relaxation time for such fluctuations can be of the
order of the kinetic relaxation time $t_{rel}\gg t_{cor}$, and, therefore,
they are essential on the kinetic and other large time scale regimes. The
conventional kinetic equations do not take into account such large-scale
correlations and do not describe the fluctuations of the distribution
functions. A kinetic theory of fluctuations can be found in, e.g.,
\cite{Klimontovich (1982)}.

Dependence of the initial correlation function $g_{2}(x_{1},x_{2})$ and a
potential of an interparticle interaction $V_{ij}=V(|\mathbf{x}_{i}%
-\mathbf{x}_{j}|)$ on a distance between particles is essential for the
behaviour of integrals over $\mathbf{x}_{2}$ in (\ref{38}) with time. Let us
consider an example, when this dependence is defined as
\begin{align}
g_{2}(x_{1},x_{2})  &  =g_{0}\exp(-\frac{r^{2}}{r_{cor}^{2}})\phi
(\mathbf{p}_{1},\mathbf{p}_{2})\text{ ,}\nonumber\\
V(r)  &  =V_{0}\exp(-\frac{r^{2}}{r_{0}^{2}})\text{,} \label{41}%
\end{align}
where $g_{0}$ and $V_{0}$ are the constant parameters and $\phi(\mathbf{p}%
_{1},\mathbf{p}_{2})$ is a properly normalized function of the particle
momenta. We will estimate the time dependence of the terms in Eq. (\ref{38}),
determined by initial correlations, in the case of a weak interparticle
interaction, when the time evolution is governed by the \textquotedblright
free\textquotedblright\ propagator $\exp(L_{12}^{0}t)$. The correlation
function (\ref{41}) under the action of $\exp(L_{12}^{0}t)$ transforms
according to (\ref{40}) as
\begin{equation}
\exp(L_{12}^{0}t)g_{2}(x_{1},x_{2})=g_{2}(\mathbf{r}-\mathbf{g}t,\mathbf{p}%
_{1},\mathbf{p}_{2})=g_{0}\exp(-\frac{|\mathbf{r}-\mathbf{g}t|^{2}}%
{r_{cor}^{2}})\phi(\mathbf{p}_{1},\mathbf{p}_{2})\text{ .} \label{42}%
\end{equation}
The function (\ref{42}) goes to zero at $t\rightarrow\infty$ and any fixed
distance $\mathbf{r}$ and velocity $\mathbf{g}$.

Using (\ref{41}) and (\ref{42}), the first right-hand side term of Eq.
(\ref{38}) can be presented as
\begin{equation}
n\int d\mathbf{p}_{2}\int d\mathbf{r}\left[  \frac{\partial}{\partial
\mathbf{r}}V_{0}\exp(-\frac{r^{2}}{r_{0}^{2}})\right]  (\frac{\partial
}{\partial\mathbf{p}_{1}}-\frac{\partial}{\partial\mathbf{p}_{2}})\frac
{g_{2}(\mathbf{r}-\mathbf{g}t,\mathbf{p}_{1},\mathbf{p}_{2})}{f_{1}%
(\mathbf{p}_{1})f_{1}(\mathbf{p}_{2})}f_{1}(\mathbf{p}_{2})f_{1}%
(\mathbf{p}_{1},t)\text{ ,} \label{43}%
\end{equation}
where $g_{2}(\mathbf{r}-\mathbf{g}t,\mathbf{p}_{1},\mathbf{p}_{2})$ is defined
by (\ref{42}).

It is easy to see, that the integral over $\mathbf{r}$ (\ref{43}) is not equal
to zero only if $t<t_{cor}$ , where $t_{cor}\backsim r_{cor}/\bar{v}\thicksim
r_{0}/\bar{v}$ and $\bar{v}$ is a typical (mean) particle velocity. At $t\gg
t_{cor}$ the integral (\ref{43}) practically vanishes because of the finite
range $r_{0}$ of an interparticle interaction. Of course, such a behaviour is
possible if a contribution of the \textquotedblright parallel
motion\textquotedblright\ with small $g$ to (\ref{43}) is negligible. The
term, related to initial correlations and contributing to the second
right-hand side term of Eq. (\ref{38}) (collision integral), displays the same
behaviour with time. Thus, in the considered example, the terms with initial
correlations in Eq. (\ref{38}) vanish at $t\gg t_{cor}$ if the particles
dynamics is characterized by the ergodic mixing flow in the phase space.

If the effective interaction between particles vanishes at a distance between
particles $r\gg r_{cor},$ then at $t_{1}\gg t_{cor}$ the action of
$\exp(L_{12}^{0}t_{1})$ on the interaction Liouvillian $L_{12}^{^{\prime}}$
(\ref{23}) under the integral over $t_{1}$ in (\ref{38}) results in increasing
the distance between particles beyond the radius of inter-particle interaction
$r_{0}\backsim r_{cor}$ and, therefore, in vanishing the integrand. As follows
from Eq. (\ref{38}), the initial moment $t=0$ refers to the time just before
the collision. Given that the time between collisions $t_{rel}$, which
determines the time of an essential change of the one-particle distribution
function $f_{1}(\mathbf{p}_{i},t)$, exceeds considerably $t_{cor}$ due to the
condition (\ref{16}), $f_{1}(\mathbf{p}_{1},t-t_{1})$ does not practically
change within the interval $0\leqslant t_{1}\leqslant t_{cor}$. Thus,
$f_{1}(\mathbf{p}_{1},t-t_{1})$ under the integral in (\ref{38}) can be
substituted by $f_{1}(\mathbf{p}_{1},t)$ with the accuracy of $t_{cor}%
/t_{rel}\backsim\gamma\ll1$. Moreover, at $t\gg t_{cor}$ the upper limit of
integral over $t_{1}$ can be extended to infinity. After such a transformation
the collision integral becomes Markovian and irreversible.

Therefore, the transition from the microscopic time scale $0\leqslant
t\leqslant t_{cor}$ to the macroscopic one $t\gg t_{cor}$ is essential for
obtaining an irreversible Markovian kinetic equation. On the macroscopic time
scale the reversible-in-time terms of Eq. (\ref{38}) caused by initial
correlations may vanish (if the particle dynamics have the necessary
properties) and the collision term acquires the desired form. The extension of
the upper limit of the integral in Eq. (\ref{38}) to infinity ($t\rightarrow
\infty$) and the existence of this limit are also essential for obtaining the
irreversible equation from reversible Eq. (\ref{38}). Existence of the
limiting value of such an integral over time (at $t\rightarrow\infty$) was
proved in \cite{Prigogine and Resibois (1961)} in the thermodynamic limit
($V\rightarrow\infty$, $N\rightarrow\infty$, $n=V/N$ is finite) and when the
time scales of collisions and of relaxation are widely separated ($t_{cor}\ll
t_{rel}$).

In order to obtain the desired nonlinear equation for the one-particle
distribution function $f_{1}(\mathbf{p}_{1},t)$, we should make an additional
approximation neglecting the time retardation of the one-particle distribution
function which is possible on the time scale $t\ll t_{rel}$. In this case, we
can approximate the function $f_{1}(\mathbf{p}_{2})=f_{1}(\mathbf{p}_{2},0)$
in Eq. (\ref{38}) by $f_{1}(\mathbf{p}_{2},t)$ with the accuracy of $\gamma
\ll1$ (for more details see below).

Summing up, we can rewrite Eq. (\ref{38}) on the considered macroscopic time
scale (\ref{0}) as
\begin{equation}
\frac{\partial}{\partial t}f_{1}(\mathbf{p}_{1},t)=n\int dx_{2}\int
\limits_{0}^{\infty}dt_{1}L_{12}^{^{\prime}}\exp(L_{12}t)L_{12}^{^{\prime}%
}f_{1}(\mathbf{p}_{1},t)f_{1}(\mathbf{p}_{2},t)\text{ .} \label{45}%
\end{equation}

This equation is Markovian, time-irreversible and describes the relaxation of
the system with the characteristic time $t_{rel}$. Using the two-particle
scattering theory, it can be shown that Eq. (\ref{45}) is equivalent to the
conventional Boltzmann equation (see, e.g., \cite{Balescu (1975)}). In the
case of a classical gas with a weak interparticle interaction, the exact
two-particle propagator $\exp(L_{12}t)$ in (\ref{45}) can be replaced by the
\textquotedblright free\textquotedblright\ propagator $\exp(L_{12}^{0}t) $
according to (\ref{39}). In this case (\ref{45}) is equivalent to the Landau
equation (see, \cite{Balescu (1975)}).

It is worth noting, that we have managed to obtain the Boltzmann equation
(\ref{45}), which is nonlinear as it should be, using the linear theory of
projection operators. Actually, Eq. (\ref{38}), which has been used for
deriving the Boltzmann equation (\ref{45}), is linear with respect to the
distribution function $f_{1}(\mathbf{p}_{1},t)$ but contains a one-particle
distribution function $f_{1}(\mathbf{p}_{2})$ at the initial moment $t=0$
(this follows from using the time-independent projection operator (\ref{17})
resulting in the relevant distribution function (\ref{21})). Change in time of
the momentum distribution function can be presented as $f_{1}(\mathbf{p}%
_{i},t)=f_{1}(\mathbf{p}_{i})+(\partial f_{1}/\partial t)_{0}t+\ldots$, where
the derivative can be estimated as $\partial f_{1}/\partial t\thicksim
f_{1}/t_{rel}$ and the relaxation time $t_{rel}$ is determined by the
collision integral in (\ref{38}). On the time scale (\ref{0}), there is no
difference in Eq. (\ref{38}) between $f_{1}(\mathbf{p}_{2})$ and
$f_{1}(\mathbf{p}_{2},t)$ and that makes this equation nonlinear. Thus, the
second inequality in (\ref{0}), $t\ll t_{rel}$, is essential for obtaining a
nonlinear evolution equation from the linear equation (\ref{8}), which
accounts for initial correlations, i.e. derived with no restriction on the
time scale defined by the first inequality in (\ref{0}), $t\gg t_{cor}$.

The described procedure of obtaining the time-irreversible Eq. (\ref{45}) from
Eq. (\ref{38}) clearly indicates that irreversibility emerges on the
macroscopic time scale (\ref{0}) as a result of the loss with time of the
information about both the initial correlations and correlations that emerge
due to collisions.

\section{EVOLUTION EQUATION RETAINING INITIAL CORRELATIONS FOR A HOMOGENEOUS
DILUTE GAS OF QUANTUM PARTICLES}

Let us now test the general scheme outlined in Section 2 by applying it to the
system of interacting quantum particles. In the quantum physics case the
derived HGMEs hold in the forms obtained above but the Liouville operator $L$
in Eq. (\ref{1}) (and elsewhere) as well as the operator $\exp(Lt)$ are now
the superoperators acting according to (\ref{1'''}) on any operator $A(t)$
(particularly on the statistical operator $F_{N}(t)$ of $N$ interacting
quantum particles) as
\begin{equation}
LA(t)=\frac{1}{i\hbar}[H,A(t)],e^{Lt_{1}}A(t)=e^{\frac{1}{i\hbar}Ht_{1}%
}A(t)e^{-\frac{1}{i\hbar}Ht_{1}}, \label{19q}%
\end{equation}
respectively. Accordingly, Eqs. (\ref{8}) and (\ref{14}) for the relevant part
of the statistical operator $f_{r}(t)=PF_{N}(t)$ hold with the appropriate
redefinitions of the symbols and operations used in these equations. The order
of operators in these equations also matters.

In the space representation, the system of $N$ quantum particles is described
by the density matrix $F_{N}(\mathbf{r}_{1},\ldots,\mathbf{r}_{N}%
,\mathbf{r}_{1}^{^{\prime}},\ldots,\mathbf{r}_{N}^{^{\prime}},t)$. This matrix
should satisfy the symmetry conditions which reflect the statistics of the
particles under consideration, i.e.
\begin{align}
P_{ij}F_{N}  &  =F_{N}P_{ij}=\theta F_{N},\nonumber\\
\theta &  =\pm1, \label{20q}%
\end{align}
where $P_{ij}$ is the operator of transmutation of any two variables
$\mathbf{r}_{i}$and $\mathbf{r}_{j}$, when it stands to the left of $F_{N}$,
or of $\mathbf{r}_{i}^{^{\prime}}$and $\mathbf{r}_{j}^{^{\prime}}$, when it
acts from the write side of $F_{N}$ (this rule holds for any operators acting
on the matrixes). The sign plus is applicable for bosons and minus should be
used for fermions.

We focus on the derivation of the evolution equation for a one-particle
density matrix $f_{1}(\mathbf{r}_{1},\mathbf{r}_{1}^{\prime},t)$, which is
defined according to the following definition of the $s-$particle $(s\leq N)$
density matrix
\begin{equation}
f_{s}(\mathbf{r}_{1},\ldots,\mathbf{r}_{s},\mathbf{r}_{1}^{^{\prime}}%
,\ldots,\mathbf{r}_{s}^{^{\prime}},t)=V^{s}Tr_{(s+1,\ldots,N)}F_{N}%
(\mathbf{r}_{1},\ldots,\mathbf{r}_{N},\mathbf{r}_{1}^{^{\prime}}%
,\ldots,\mathbf{r}_{N}^{^{\prime}},t), \label{21q}%
\end{equation}
where $Tr_{(s+1,\ldots,N)}$ denotes the trace taken over the coordinates of
$N-s$ particles ($s+1,\ldots,N$). From the normalization condition (\ref{1'})
$TrF_{N}=1$ it follows that $\frac{1}{V}Trf_{1}=1$.

It is convenient to introduce the following projection operator (compare with
the definition (\ref{17}))
\begin{equation}
P=\left[  \prod_{i=2}^{N}f_{1}(\mathbf{r}_{i},\mathbf{r}_{i}^{\prime})\right]
\frac{1}{V^{N-1}}Tr_{(2,\ldots,N)}, \label{22q}%
\end{equation}
where $f_{1}(r_{i},r_{i}^{\prime})=f_{1}(r_{i},r_{i}^{\prime},t_{0})$ is the
one-particle density matrix taken at the initial moment of time $t_{0}$.
Therefore, the chosen relevant density matrix under consideration is
\begin{equation}
Pf_{N}(t)=f_{r}(t)=\left[  \prod_{i=2}^{N}f_{1}(\mathbf{r}_{i},\mathbf{r}%
_{i}^{\prime})\right]  f_{1}(\mathbf{r}_{1},\mathbf{r}_{1}^{\prime},t)
\label{23q}%
\end{equation}
in accordance with the definition (\ref{21q}) for the $N$-particle density
matrix $f_{N}=V^{N}F_{N}$. As it is seen from equation (\ref{23q}), the
selected relevant part of the $N$ -particle density matrix does not contain
any correlations. In order to define the irrelevant part of the density matrix
$f_{i}(t_{0})=(1-P)f_{N}(t_{0})$, which enters equations (\ref{8}) and
(\ref{14}) and comprise all initial correlations, we have also to take into
account the quantum correlations arising due to the symmetry condition
(\ref{20q}). These correlations exist independently of the interaction between
particles, which we introduce by means of the following Hamiltonian
\begin{align}
H  &  =H^{0}+H^{^{\prime}},\nonumber\\
H^{0}  &  =\sum_{i=1}^{N}K_{i},\text{ }H^{^{\prime}}=\sum_{i<j=1}^{N}%
\Phi_{i,j},\nonumber\\
K_{i}  &  =-\frac{\hbar^{2}}{2m}\mathbf{\triangledown}_{\mathbf{r}_{i}}%
^{2},\text{ }\Phi_{i,j}=\Phi(\left\vert \mathbf{r}_{i}-\mathbf{r}%
_{j}\right\vert ). \label{24q}%
\end{align}
Thus, $f_{N}(t_{0})=f_{r}(t_{0})+f_{i}(t_{0})$ can be presented in the form of
the cluster expansion with
\begin{equation}
f_{r}(t_{0})=\prod_{i=1}^{N}f_{1}(i) \label{25q}%
\end{equation}
and
\begin{equation}
f_{i}(t_{0})=\sum\limits_{i,j=1,i<j=1}^{N}\widetilde{g}_{2}(i,j)\prod
\limits_{k=1,k\neq i,j}^{N-2}f_{1}(k)+\sum\limits_{i.j.k=1,i<j<k=1}%
^{N}\widetilde{g}_{3}(i,j,k)\prod\limits_{l=1,l\neq i,j,k}^{N-3}%
f_{1}(l)+\ldots, \label{26q}%
\end{equation}
where we have used the short notation $f_{1}(i)=f_{1}(\mathbf{r}%
_{i},\mathbf{r}_{i}^{^{\prime}})$ (i.e. $i=\mathbf{r}_{i},\mathbf{r}%
_{i}^{^{\prime}},i=1,\ldots,N$) and, e.g., $\prod\limits_{k=1,k\neq i,j}%
^{N-2}f_{1}(k)$ stands for the product of $N-2$ one-particle functions
$f_{1}(k)$ with $k$ taking values $1,\ldots,N$ but $k\neq i,j$ (compare with
(\ref{22})). A two-particle $\widetilde{g}_{2}(i,j)$ and a three-particle
$\widetilde{g}_{3}(i,j,k)$ correlation functions are defined as (the
correlation functions for more particles can be written down in the same way)
\begin{align}
\widetilde{g}_{2}(i,j)  &  =\theta P_{ij}f_{1}(i)f_{1}(j)+g_{2}%
(i,j),\nonumber\\
\widetilde{g}_{3}(i,j,k)  &  =(P_{ij}P_{jk}+P_{ij}P_{ik})f_{1}(i)f_{1}%
(j)f_{1}(k)+(\theta P_{ik}+\theta P_{jk})g_{2}(i,j)f_{1}(k)\nonumber\\
&  +(\theta P_{ij}+\theta P_{ik})g_{2}(j,k)f_{1}(i)+(\theta P_{ij}+\theta
P_{jk})g_{2}(i,k)f_{1}(j)\nonumber\\
&  +g_{3}(i,j,k). \label{27q}%
\end{align}
The terms in the right-hand sides of Eqs. (\ref{27q}) with one, two, etc.
permutation operators $P_{ij}$ represent a two-, three-particle, etc. quantum
correlations emerging due to the proper symmetry properties of $f_{N}(t_{0})$
guaranteed by the symmetrization operators, e.g.,
\begin{equation}
f_{2}(i,j)=\theta P_{ij}f_{2}(\mathbf{r}_{i},\mathbf{r}_{j},\mathbf{r}%
_{i}^{^{\prime}},\mathbf{r}_{j}^{\prime}). \label{28q}%
\end{equation}
The irreducible two-, three-, etc. correlation functions, $g_{2}(i,j)$,
$g_{3}(i,j,k)$, etc., are due to the interaction between particles $\Phi
_{i,j}$ and are proportional to an interaction parameter $\varepsilon$ in some
power. Note, that in contrast to the classical physics case (see (\ref{22})),
each correlation function in (\ref{27q}) contains correlations of all possible
orders in $\varepsilon$, e.g. $\widetilde{g}_{2}(i,j)$ includes two-particle
quantum correlations (existing even in the absence of interaction, when
$\varepsilon=0$) and two-particle correlations of the first order in
$\varepsilon$ ($g_{2}(i,j)$). The same is valid for $\widetilde{g}_{3}%
(i,j,k)$, which contains a three-particle correlations of the zero, first and
second ($g_{3}(i,j,k)$) orders in $\varepsilon$.

The following useful relations hold for the permutation operators
\begin{align}
P_{ij}P_{jk}  &  =P_{ik}P_{ij},\nonumber\\
P_{jk}P_{ij}  &  =P_{ij}P_{ik}. \label{29q}%
\end{align}

Let us now derive the evolution equation for a gas of interacting quantum
particles in the lowest (first) order in the small density of particles
parameter $n$. The corresponding dimensionless small parameter is again (see
(\ref{16}))
\begin{equation}
\gamma=r_{0}^{3}n\thicksim t_{cor}/t_{rel}\ll1. \label{29q'}%
\end{equation}
We will use the TC-HGME given by Eqs.(\ref{8}) and (\ref{9}) with appropriate
redefinitions of the symbols, as it was mentioned above.

The Liouville superoperator $L=L^{0}+L^{^{\prime}}$, corresponding to the
Hamiltonian (\ref{24q}), is defined by equation (\ref{19q}) as
\begin{align}
L^{0}  &  =\sum_{i=1}^{N}L_{i}^{0},L_{i}^{0}=\frac{1}{i\hbar}(-\frac{\hbar
^{2}}{2m})(\nabla_{\mathbf{r}_{i}}^{2}-\nabla_{\mathbf{r}_{i}^{^{\prime}}}%
^{2}),\nonumber\\
L^{^{\prime}}  &  =\sum_{i<j=1}^{N}L_{ij}^{^{\prime}},L_{ij}^{^{\prime}}%
=\frac{1}{i\hbar}\left[  \Phi(\left\vert \mathbf{r}_{i}-\mathbf{r}%
_{j}\right\vert )-\Phi(|\mathbf{r}_{i}^{^{\prime}}-\mathbf{r}_{j}^{^{\prime}%
}|)\right]  , \label{30q}%
\end{align}
where we have used the abovementioned rule, that an operator standing on the
right side (in the commutator) acts on the primed variables.

Using the definitions (\ref{30q}), one can easily show that the following
relations (similar to the classical physics case) hold
\begin{align}
&  Tr_{(i)}L_{i}^{0}\varphi_{N}(1,\ldots,i,\ldots,N)=0,\nonumber\\
&  Tr_{(i,j)}L_{ij}^{^{\prime}}\varphi_{N}(1,\ldots,i,\ldots,j,\ldots,N)=0,
\label{31q}%
\end{align}
where $\varphi_{N}(1,\ldots,i,\ldots,N)$ is some matrix, defined for the
$N-$particle system, which in the space representation takes the form
$\varphi_{N}(1,\ldots,i,\ldots,N)=\varphi_{N}(\mathbf{r}_{1},\ldots
,\mathbf{r}_{i},\ldots,\mathbf{r}_{N},\mathbf{r}_{1}^{^{\prime}}%
,\ldots,\mathbf{r}_{i}^{^{\prime}},\ldots,\mathbf{r}_{N}^{^{\prime}})$ and
satisfies the necessary boundary conditions.

Thus, using the projection operator (\ref{22q}) and the relevant density
matrix (\ref{23q}), the first two terms in TC-HGME (\ref{8}) $P(L^{0}%
+L^{^{\prime}})f_{r}(t)$ (which coincide with those in TC-GME (\ref{5})) can
be presented as
\begin{align}
PL^{0}f_{r}(t)  &  =\left[  \prod_{i=2}^{N}f_{1}(i)\right]  \frac{1}{i\hbar
}[-\frac{\hbar^{2}}{2m}\left(  \nabla_{\mathbf{r}_{1}}^{2}-\nabla
_{\mathbf{r}_{1}^{^{\prime}}}^{2}\right)  ]f_{1}(\mathbf{r}_{1},\mathbf{r}%
_{1}^{^{\prime}},t),\nonumber\\
PL^{^{\prime}}f_{r}(t)  &  =\left[  \prod_{i=2}^{N}f_{1}(i)\right]  \frac
{n}{i\hbar}\int d\mathbf{r}_{2}[\Phi(\left\vert \mathbf{r}_{1}-\mathbf{r}%
_{2}\right\vert )\nonumber\\
&  -\Phi(|\mathbf{r}_{1}^{^{\prime}}-\mathbf{r}_{2}|)]f_{1}(\mathbf{r}%
_{2},\mathbf{r}_{2})f_{1}(\mathbf{r}_{1},\mathbf{r}_{1}^{^{\prime}},t)\}.
\label{32q}%
\end{align}

To simplify the derivation, we restrict ourselves to the spatially homogeneous
case, when all matrixes are invariant under translations and, in particular,
the matrix $f_{1}(\mathbf{r}_{i},\mathbf{r}_{i}^{^{\prime}})$ should be of the
form
\begin{equation}
f_{1}(\mathbf{r}_{i},\mathbf{r}_{i}^{^{\prime}},t)=f_{1}(\mathbf{r}%
_{i}-\mathbf{r}_{i}^{^{\prime}},t). \label{33q}%
\end{equation}
Then, it is easy to see that for functions (\ref{33q}), $H^{0}$ commutes with
$f_{r}(t)$ and Eqs. (\ref{32q}) vanish, i.e.
\begin{equation}
L^{0}f_{r}(t)=0,PL^{^{\prime}}f_{r}(t)=0. \label{34q}%
\end{equation}
Also, in this spatially homogeneous case, for any matrix $\varphi_{N}%
(1,\ldots,N)$
\begin{equation}
PL^{0}\varphi_{N}(1,\ldots,N)=0. \label{35q}%
\end{equation}

Let us now consider the terms in TC-HGME (\ref{8}) caused by initial
correlations. We note, that in the considered case of identical interacting
quantum particles, it is impossible to simply disregard the inhomogeneous term
in the exact TC-GME (\ref{5}) (assuming that correlations damp with the
distance between particles) due to always existing (at any distances) quantum
initial correlations (see (\ref{27q})). Bogoliubov \cite{Bogoliubov (1946)}
tackled this problem by introducing a special boundary (initial) condition to
the BBGKY chain at $t_{0}\rightarrow-\infty$ implying that at this limiting
moment of time the particles are located at distances beyond the correlation
radius $r_{0}$, which means that correlation functions in (\ref{27q}),
$g_{2}(i,j)$, $g_{3}(i,j,k)$, etc., caused by interaction between particles
(and proportional to some power of $\varepsilon$), vanish, and%

\begin{equation}
\lim\limits_{t_{0}\rightarrow-\infty}f_{i}(t_{0})=\gamma_{N}%
{\displaystyle\prod\limits_{i=1}^{N}}
f_{1}(i), \label{35q'}%
\end{equation}
where
\begin{equation}
\gamma_{N}=(1+\theta P_{1N}+\ldots+\theta P_{N-1,N})\ldots(1+\theta P_{12})
\label{35q''}%
\end{equation}
is the symmetrization operator for $N$ particles. Therefore, in the Bogoliubov
approach, the multi-particle density matrix at $t_{0}\rightarrow-\infty$ is
the properly symmetrized product of the one-particle density matrixes (compare
with (\ref{25q})-(\ref{27q})). As in the classical physics case, this boundary
condition introduces irreversibility into the system evolution and thus allows
obtaining the quantum kinetic equations \cite{Bogoliubov and Gurov (1947)}.
Also, as in the case of a dilute gas of classical particles, in order to
obtain the closed kinetic equation for a one-particle density matrix,
Bogoliubov used the specific form of time-dependence of the multi-particle
density matrix (valid only at $t-t_{0}\gg t_{cor}$), i.e. he assumed that on
this time scale the dependence on time of the multi-particle density matrices
is determined by the time-dependence of a one-particle density matrix.

Here, we do not use any of the abovementioned Bogoliubov's assumptions. Thus,
let us consider the (super)operator (see (\ref{9}))
\begin{equation}
C(t)=R(t)-1=e^{QLt}\frac{1}{1-C_{0}e^{-Lt}e^{QLt}}C_{0}e^{-Lt} \label{36q}%
\end{equation}
determining the influence of initial correlations in Eq. (\ref{8}) on the
evolution process (hereinafter we put $t_{0}=0$, because the HGME is valid for
any initial moment $t_{0}$). In order to expand exponential (super)operators
in the power series, we use the relation (\ref{27}) which is also valid for
any (super)operators $A$ and $B$.

Using (\ref{27}) and expanding (\ref{36q}) in $PL^{0}$ and $QL^{^{\prime}}$,
one can see that all terms with $PL^{0}$ vanish due to Eq. (\ref{35q}) and,
therefore, all $QL$ in (\ref{36q}) can be replaced by $L^{0}+QL^{^{\prime}}$.
Taking into account that the terms containing $PL^{^{\prime}}$ result in the
expressions proportional to at least the first power of $n$ (see,
e.g.,(\ref{32q})) and looking for the lowest (first) order approximation in
$n$ of Eq. (\ref{8}), we can substitute $Q=1-P $ in (\ref{36q}) with the unity
and simplify this expression to
\begin{equation}
C(t)=e^{Lt}\frac{1}{1-C_{0}}C_{0}e^{-Lt}. \label{38q}%
\end{equation}
Using expression (\ref{7}) for $C_{0}$, we have
\begin{equation}
\frac{1}{1-C_{0}}C_{0}=f_{i}(0)f_{r}^{-1}(0). \label{39q}%
\end{equation}
This is a remarkable result showing that in this approximation there is no
problem with the existence (convergence/invertability) of $R(t)$ (\ref{9})
(the same has been shown in Sec. 4 for the classical physics case). In
connection with that it is worth reminding, that if in some part of the
definition space $f_{r}(0)$ goes to zero, so should do $f_{i}(0)$ (at the same
rate or more promptly than the relevant part). It follows from the possibility
of breaking the statistical operator into the uncorrelated ($f_{r}$) and
correlated ($f_{i}$) parts (see (\ref{25q})-(\ref{27q})), which in turn caused
by the fact that if an interaction between particles is switched off
($\varepsilon=0$), the statistical operator should be given by symmetrized
product of one-particle operators. Moreover, in the equations for the relevant
part of statistical operator obtained below, as a result of approximation
(\ref{39q}), the factor $f_{i}(0)f_{r}^{-1}(0)$ is always multiplied by
$f_{r}(0)$ (see (\ref{55q}) below and take into account that in the
approximation under consideration the relevant part changes slowly). This fact
additionally guarantees an absence of any divergencies (in other words, any
manipulations with initial condition term should result in terms behaving like
$f_{i}(0)$ in accordance with the identity (\ref{6})).

Now, it is necessary to define $f_{r}^{-1}(0)$. In the spatially homogeneous
case under consideration, it is convenient to work in the momentum
representation. The relation between the space and momentum representations
for a one-particle density matrix of interest is
\begin{align}
f_{1}(\mathbf{r,r}^{^{\prime}},t)  &  =\frac{1}{(2\pi\hbar)^{3}}\int
f_{1}(\mathbf{p},\mathbf{p}^{^{\prime}},t)e^{i\mathbf{pr}/\hbar}%
e^{-i\mathbf{p}^{^{\prime}}\mathbf{r}^{^{\prime}}/\hbar}d\mathbf{p}%
d\mathbf{p}^{^{\prime}},\nonumber\\
f_{1}(\mathbf{p},\mathbf{p}^{^{\prime}},t)  &  =(2\pi\hbar)^{3}w(\mathbf{p,t}%
)\delta(\mathbf{p}-\mathbf{p}^{^{\prime}}), \label{40q}%
\end{align}
where a function $w(\mathbf{p})$ (with the dimensionality [$p^{-3}$]) is
normalized to the unity and corresponds to a classical momentum distribution
function. According to the definition
\begin{equation}
\int f_{1}(\mathbf{p},\mathbf{p}^{^{\prime\prime}},t)f_{1}^{-1}(\mathbf{p}%
^{^{\prime\prime}},\mathbf{p}^{^{\prime}},t)d\mathbf{p}^{^{\prime\prime}%
}=\delta(\mathbf{p}-\mathbf{p}^{^{\prime}}), \label{41q}%
\end{equation}
we can take for $f_{1}^{-1}(\mathbf{p},\mathbf{p}^{^{\prime}})$ the following
matrix
\begin{equation}
f_{1}^{-1}(\mathbf{p},\mathbf{p}^{^{\prime}},t)=(2\pi\hbar)^{-3}%
w^{-1}(\mathbf{p,t})\delta(\mathbf{p}-\mathbf{p}^{^{\prime}}). \label{42q}%
\end{equation}
Note, that the $\delta$-functions in (\ref{40q})-(\ref{42q}) (related to the
fact that a one-particle distribution function is diagonal in the spatially
homogeneous case) always stay under the integrals or are divided out.

In accordance with (\ref{40q}) and (\ref{42q}), an invert one-particle matrix
in the space representation may be defined as
\begin{align}
f_{1}^{-1}(\mathbf{r,r}^{^{\prime}},t)  &  =\frac{1}{(2\pi\hbar)^{3}}\int
f_{1}^{-1}(\mathbf{p},\mathbf{p}^{^{\prime}},t)e^{i\mathbf{pr}/\hbar
}e^{-i\mathbf{p}^{^{\prime}}\mathbf{r}^{^{\prime}}/\hbar}d\mathbf{p}%
d\mathbf{p}^{^{\prime}}\nonumber\\
&  =\frac{1}{(2\pi\hbar)^{6}}\int w^{-1}(\mathbf{p,t})e^{i\mathbf{p}%
(\mathbf{r}-\mathbf{r}^{^{\prime}})}d\mathbf{p.} \label{43q}%
\end{align}
It satisfies the necessary relation
\begin{equation}
\int f_{1}(\mathbf{r},\mathbf{r}^{^{\prime\prime}},t)f_{1}^{-1}(\mathbf{r}%
^{^{\prime\prime}},\mathbf{r}^{^{\prime}},t)d\mathbf{r}^{^{\prime\prime}%
}=\delta(\mathbf{r}-\mathbf{r}^{^{\prime}}). \label{44q}%
\end{equation}

Now,we can define an operator, invert to the relevant operator (\ref{25q}),
as
\begin{equation}
f_{r}^{-1}(0)=\prod_{i=1}^{N}f_{1}^{-1}(i), \label{45q}%
\end{equation}
because, due to (\ref{41q}) and (\ref{44q}),
\begin{equation}
\lbrack f_{r}(0)f_{r}^{-1}(0)]=%
{\displaystyle\prod\limits_{i=1}^{N}}
I(i). \label{46q}%
\end{equation}
Here and further, $(i)$ stands for $(\mathbf{r}_{i},\mathbf{r}_{i}^{^{\prime}%
})$ or $(\mathbf{p}_{i},\mathbf{p}_{i}^{^{\prime}})$ depending on the
representation used, and $I(i)$ is the unity matrix in the space of
$i$-particle
\begin{align}
I(i)  &  =I(\mathbf{r}_{i},\mathbf{r}_{i}^{^{\prime}})=\delta(\mathbf{r}%
_{i}-\mathbf{r}_{i}^{^{\prime}}),\nonumber\\
I(i)  &  =I(\mathbf{p}_{i},\mathbf{p}_{i}^{^{\prime}})=\delta(\mathbf{p}%
_{i}-\mathbf{p}_{i}^{^{\prime}}) \label{47q}%
\end{align}
in the space and momentum representations, respectively.

Applying the introduced definitions for invert matrixes and using Eq.
(\ref{26q}) for the irrelevant density matrix, the correlation parameter
(\ref{39q}) can be rewritten in the following way
\begin{align}
f_{i}(0)f_{r}^{-1}(0)  &  =\sum_{i,j=1,i<j}^{N}\widetilde{g}(i,j)f_{1}%
^{-1}(i)f_{1}^{-1}(j)\prod_{k=1,k\neq i,j}^{N-2}I(k)\nonumber\\
&  +\sum_{i,j,k=1,i<j<k}^{N}\widetilde{g}(i,j,k)f_{1}^{-1}(i)f_{1}%
^{-1}(j)f_{1}^{-1}(k)\prod_{l=1,l\neq i,j,k}^{N-3}I(l)+\ldots, \label{48q}%
\end{align}
where, e.g., the two-particle part of (\ref{48q}), according to the
expressions for correlation functions (\ref{27q}), is
\begin{equation}
\widetilde{g}(i,j)f_{1}^{-1}(i)f_{1}^{-1}(j)=\theta P_{ij}+g(i,j)f_{1}%
^{-1}(i)f_{1}^{-1}(j), \label{49q}%
\end{equation}
and the matrix elements of the permutation operator $P_{ij}$ are defined in
the space and momentum representations as
\begin{align}
P_{ij}(\mathbf{r}_{i},\mathbf{r}_{j},\mathbf{r}_{i}^{^{\prime}},\mathbf{r}%
_{j}^{^{\prime}})  &  =\delta(\mathbf{r}_{i}-\mathbf{r}_{j}^{^{\prime}}%
)\delta(\mathbf{r}_{j}-\mathbf{r}_{i}^{^{\prime}}),\nonumber\\
P_{ij}(\mathbf{p}_{i},\mathbf{p}_{j},\mathbf{p}_{i}^{^{\prime}},\mathbf{p}%
_{j}^{^{\prime}})  &  =\delta(\mathbf{p}_{i}-\mathbf{p}_{j}^{^{\prime}}%
)\delta(\mathbf{p}_{j}-\mathbf{p}_{i}^{^{\prime}}). \label{49q'}%
\end{align}

Let us consider the additional (to the flow) term in TC-HGME (\ref{8}),
$F(t)=PL^{^{\prime}}C(t)f_{r}(t)$, caused by initial correlations, in the
linear approximation in $n$. Using (\ref{38q}), (\ref{39q}) and (\ref{48q}),
(\ref{49q}), one can obtain in this approximation
\begin{align}
F(t)  &  =PL^{^{\prime}}e^{Lt}\frac{1}{1-C_{0}}C_{0}e^{-Lt}f_{r}(t)\nonumber\\
&  =\left[  \prod_{i=2}^{N}f_{1}(i)\right]  \frac{n}{i\hbar}Tr_{(2)}%
L_{12}^{^{\prime}}e^{-\frac{i}{\hbar}(H_{12}^{0}+H_{12}^{^{\prime}})t}[\theta
P_{12}\nonumber\\
&  +g(1,2)f_{1}^{-1}(1)f_{1}^{-1}(2)]e^{\frac{i}{\hbar}(H_{12}^{0}%
+H_{12}^{^{\prime}})t}f_{1}(2)f_{1}(1,t). \label{50q}%
\end{align}
Deriving (\ref{50q}), we have taken into consideration that each additional
integration over the coordinate or momentum space adds an additional power of
$n$ and, therefore, in the linear approximation in $n$, all formulae can
contain no more than one integration (as in (\ref{50q})). Thus, Eq.
(\ref{50q}) is determined only by the two-particle dynamics described by the
Hamiltonian $H_{12}=H_{12}^{0}+H_{12}^{^{\prime}}$ (and the corresponding
Liouvillian $L_{12}=L_{12}^{0}+L_{12}^{^{\prime}}$). We note, that (\ref{50q})
is different from the corresponding additional (to flow) term (caused by
initial correlations) obtained in Sec. 4 for the classical physics case:
except a two-particle correlation function, Eq. (\ref{50q}) contains a
symmetrization operator $P_{12}$ responsible for quantum correlations.

Taking into account (\ref{49q'}) and invariance of the Hamiltonian $H_{12}$
with regard to renaming the particles, i.e. $H_{12}(\mathbf{p}_{1}%
,\mathbf{p}_{2},\mathbf{p}_{1}^{^{\prime}},\mathbf{p}_{2}^{^{\prime}}%
)=H_{12}(\mathbf{p}_{2},\mathbf{p}_{1},\mathbf{p}_{2}^{^{\prime}}%
,\mathbf{p}_{1}^{^{\prime}})$, it can be shown that in the operator form we
have (see, e.g. \cite{Bogoliubov and Bogoliubov (1984)})
\begin{equation}
P_{12}H_{12}P_{12}^{-1}=H_{12} \label{50q'}%
\end{equation}
\qquad\qquad\ \qquad\qquad\qquad\qquad\qquad\qquad\qquad\qquad\qquad
\qquad\qquad\qquad\qquad\qquad\qquad\qquad\qquad\ \ \ \ \ \ \ \ \ \ \ \qquad
(i.e. a symmetrization operator commutes with the Hamiltonian) and, therefore,
$e^{-\frac{i}{\hbar}(H_{12}^{0}+H_{12}^{^{\prime}})t}P_{12}e^{\frac{i}{\hbar
}(H_{12}^{0}+H_{12}^{^{\prime}})t}=P_{12}$. Now, it is easy to show that the
first term in (\ref{50q}), caused by the symmetrization operator $P_{12}$,
vanishes in the spatially homogeneous case (for non-symmetrized $f_{r}(t)$ it
is demonstrated by Eq. (\ref{34q})). Thus,
\begin{align}
PL_{12}^{^{\prime}}P_{12}f_{r}(t)  &  =n[\prod_{i=2}^{N}f_{1}(i)]\int
d\mathbf{r}_{2}\frac{1}{i\hbar}[\Phi(\left\vert \mathbf{r}_{1}-\mathbf{r}%
_{2}\right\vert )\nonumber\\
-\Phi(\left\vert \mathbf{r}_{1}^{^{\prime}}-\mathbf{r}_{2}\right\vert
)]f_{1}(\mathbf{r}_{1}-\mathbf{r}_{2})f_{1}(\mathbf{r}_{2}-\mathbf{r}%
_{1}^{^{\prime}},t)  &  =0. \label{51q}%
\end{align}
This can be easily proven, if we substitute the integration variable
$\mathbf{r}_{2}$ in (\ref{51q}) by $\mathbf{\rho}=\mathbf{r}_{1}%
-\mathbf{r}_{2}$ and $\mathbf{\rho}^{^{\prime}}=\mathbf{r}_{2}-\mathbf{r}%
_{1}^{^{\prime}}$ in the first and second integrals, respectively, and take
into consideration, that in the adopted approximation (first order in $n$) the
time-dependence of a one-particle density matrix can be disregarded due to the
fact that the change with time is described by the (super)operator which is
already of the first order in $n$ for the spatially homogeneous system.

Hence, the contribution of initial correlations to the flow term of Eq.
(\ref{8}) in the first order in $n$ and for the spatially homogeneous system
is
\begin{equation}
PL^{^{\prime}}C(t)f_{r}(t)=\left[  \prod_{i=2}^{N}f_{1}(i)\right]
nTr_{(2)}L_{12}^{^{\prime}}G_{12}^{q}(t)f_{1}(2)f_{1}(1,t), \label{52q}%
\end{equation}
where
\begin{equation}
G_{12}^{q}(t)=e^{-\frac{i}{\hbar}H_{12}t}g(1,2)f_{1}^{-1}(1)f_{1}%
^{-1}(2)e^{\frac{i}{\hbar}H_{12}t} \label{53q}%
\end{equation}
is a time-dependent quantum two-particle correlation function. This result is
similar to that obtained for the classical physics case (see (\ref{38}) and
(\ref{38a})): the initial correlations lead to the appearance of the
additional, linear in interaction between particles ($H_{12}^{^{\prime}}$),
term which is determined by the evolution in time of a two-particle
correlation function. Moreover, this result is closed in the sense that the
time-evolution is also completely determined by the two-particle dynamics with
the Hamiltonian $H_{12}$.

The collision term of Eq. (\ref{8}) in the adopted approximation can be
obtained in the same way. Accounting for equations (\ref{34q}), (\ref{35q}),
(\ref{37q}), (\ref{38q}), (\ref{39q}), (\ref{48q}), (\ref{49q}), (\ref{50q'})
and remembering that any additional power of $PL^{^{\prime}}$ and any
additional integration (taking \ a trace) over the coordinate or momentum
space adds an additional power of density $n$, it is not difficult to show
that in the linear approximation in $n$ the collision integral in (\ref{8})
reduces to
\begin{align}
K^{q}(t)  &  =\left[  \prod_{i=2}^{N}f_{1}(i)\right]  n%
{\displaystyle\int\limits_{0}^{t}}
dt_{1}Tr_{(2)}L_{12}^{^{\prime}}\left[  1+\theta P_{12}+G_{12}^{q}(t)\right]
e^{L_{12}t_{1}}\nonumber\\
&  \times L_{12}^{^{\prime}}f_{1}(2)f_{1}(1,t-t_{1}). \label{54q}%
\end{align}
This result is different from that obtained for the classical physics case,
$K(t)+\widetilde{K}(t)$, given by Eqs. (\ref{35}), (\ref{36}), due to the
presence of quantum correlations stipulated by a symmetrization operator
$P_{12}$.

Collecting the obtained results (formulae (\ref{52q}) and (\ref{54q})) and
accounting for the definition of the relevant statistical operator
(\ref{23q}), we obtain from the TC-HGME (\ref{8}) the following equation for a
one-particle density matrix in the linear approximation in $n$ for the
spatially homogeneous case
\begin{align}
\frac{\partial}{\partial t}f_{1}(1,t)  &  =nTr_{(2)}L_{12}^{^{\prime}}%
G_{12}^{q}(t)f_{1}(2)f_{1}(1,t)\nonumber\\
&  +n%
{\displaystyle\int\limits_{0}^{t}}
dt_{1}Tr_{(2)}L_{12}^{^{\prime}}\left[  1+\theta P_{12}+G_{12}^{q}(t)\right]
e^{L_{12}t_{1}}L_{12}^{^{\prime}}f_{1}(2)f_{1}(1,t-t_{1}). \label{55q}%
\end{align}

We have obtained a new closed homogeneous integro-differential equation for a
one-particle density matrix retaining initial correlations which is valid on
any time scale (i.e. we can use equation (\ref{55q}) for considering in detail
all stages of evolution of a one-particle density matrix). No approximation
like the Bogoliubov principle of weakening of initial correlations (valid only
on a large time scale $t\gg t_{cor}$) has been used for deriving equation
(\ref{55q}). This equation is exact in the linear approximation in $n$ and,
therefore, accounts for initial correlations and collisions in this
approximation exactly. The first term in the right-hand side of (\ref{55q}),
linear in interaction $L_{12}^{^{\prime}}$, is determined by initial
correlations caused by interaction between particles (this term is absent if
we apply the boundary condition (\ref{35q'})). Initial correlations, both
quantum and due to interaction, also modify the second (collision) term of
(\ref{55q}). Note, that the symmetrization operator $\theta P_{12}$
(accounting for the particle statistics) appears in this term as a result of a
regular procedure outlined above but not as a result of using the boundary
condition (\ref{35q'}), as it is the case when one applies the Bogoliubov
principle of weakening of initial correlations \cite{Bogoliubov (1946)}. Also,
when this principle is applied, the correlation function $G_{12}^{q}(t)$ does
not show up in the second term of the right-hand side of (\ref{55q})).
Evolution with time in (\ref{55q}) is completely determined by the exact
two-particle dynamics, described by the Hamiltonian $H_{12}$ (which is quite
natural for the considered linear in $n$ approximation), and in this sense the
obtained equation is closed.

\section{CONNECTION TO THE QUANTUM BOLTZMANN\ EQUATION}

Equation (\ref{55q}) is not a kinetic equation in the conventional sense
because it is time reversible (if the correlation function $g_{2}%
(i,j)f_{1}^{-1}(i)f_{1}^{-1}(j)$ does not change when the particle velocities
are reversed). Thus, one of the possibilities to secure a time-asymmetric
behaviour can be the special choice of an initial condition (e.g., the
factorized one, like (\ref{35q'})).

It is instructive to consider the transition from Eq. (\ref{55q}) to an
irreversible kinetic equation describing the evolution, say, into the future
($t>0$). As it is seen from (\ref{55q}), in order to enter the kinetic
(irreversible) stage of the evolution, the reversible terms caused by initial
correlation function, $G_{12}^{q}(t)$, should vanish on some time scale. Let
us suppose that this is the case, on the time scale $t\gg t_{cor}$, i.e.
\begin{equation}
G_{12}^{q}(t)=0,t\gg t_{cor}. \label{56q}%
\end{equation}
Here, the correlation time can be estimated as $t_{cor}\thicksim
r_{0}/\overline{v}$, where $r_{0}$ is the radius of inter-particle interaction
and $\overline{v}$ is a typical mean particle velocity. Thus, if the dynamics
of the system of particles under consideration is such (e.g., a mixing ergodic
flow in the phase space) that all\textbf{\ }initial correlations caused by
inter-particle interaction vanish on the time scale $t\gg t_{cor}$, as it is
supposed by condition (\ref{56q}), then Eq. (\ref{55q}) reduces to
\begin{align}
\frac{\partial}{\partial t}f_{1}(1,t)  &  =n%
{\displaystyle\int\limits_{0}^{t}}
dt_{1}Tr_{(2)}L_{12}^{^{\prime}}e^{L_{12}t_{1}}L_{12}^{^{\prime}}(1+\theta
P_{12})f_{r}^{12}(t-t_{1}),\nonumber\\
f_{r}^{12}(t-t_{1})  &  =f_{1}(2)f_{1}(1,t-t_{1}),t\gg t_{cor}, \label{57q}%
\end{align}
where (\ref{50q'}) is also taken into account. Making use of the definitions
(\ref{19q}), this equation can be rewritten as
\begin{align}
\frac{\partial f_{1}(1,t)}{\partial t}  &  =-\frac{n}{\hbar^{2}}%
{\displaystyle\int\limits_{-\infty}^{\infty}}
dE%
{\displaystyle\int\limits_{-\infty}^{\infty}}
dE^{^{\prime}}%
{\displaystyle\int\limits_{0}^{t}}
dt_{1}e^{\frac{1}{i\hbar}(E-E^{^{\prime}})t_{1}}Tr_{(2)}\Phi_{12}%
(E,E^{^{\prime}},t-t_{1}),t\gg t_{cor},\nonumber\\
\Phi_{12}(E,E^{^{\prime}},t-t_{1})  &  =\left[  H_{12}^{^{\prime}}%
,\delta(E-H_{12})\left[  H_{12}^{^{\prime}},(1+\theta P_{12})f_{r}%
^{12}(t-t_{1})\right]  \delta(E^{^{\prime}}-H_{12})\right]  . \label{58q}%
\end{align}
In the adopted first approximation in $n$, we can consider $f_{r}^{12}%
(t-t_{1})$ under the integral in the zero approximation in $n$, in which it
does not change with time. Thus, one can approximate the two-particle relevant
part of a density matrix under the integral in (\ref{58q}) as $f_{r}%
^{12}(t-t_{1})=f_{1}(1,t)f_{1}(2,t)$. This approximation is valid on the time
scale $t\ll t_{rel}$ (see inequality (\ref{0} at $t_{0}=0$)) allowing the
neglection of time retardation and allows to convert the linear equation for
$f_{1}(1,t)$ (\ref{55q}) (or (\ref{58q})) into the nonlinear one. Now, the
integral in (\ref{58q}) over $t_{1}$ can be calculated. Note, that for finite
$t$ equation (\ref{58q}) is time reversible and integral over $t_{1}$ is a
periodic function of $t$ for a finite volume of a system $V$. This equation
becomes time irreversible if the upper limit of integral can be extended to
infinity, $t\rightarrow\infty$, and this limiting value of the integral over
$t_{1}$ exists. However, this limiting value of the integral can evidently
exist only for a large enough volume of the system under consideration. In
other words, in order to exclude the long Poincare's cycles, the thermodynamic
limiting procedure ($V\rightarrow\infty,N\rightarrow\infty,n=\frac{N}{V}$ is
finite) is needed before calculating the integral. The existing of this
integral in the thermodynamic limit was proved in \cite{Prigogine and Resibois
(1961)}. If the mentioned limiting value of the integral exists, then the
integral can be considered in the Abel's sense, i.e.
\begin{align}
&  \lim_{t\rightarrow\infty}%
{\displaystyle\int\limits_{0}^{t}}
dt_{1}e^{\frac{1}{i\hbar}(E-E^{^{\prime}})t_{1}}\nonumber\\
&  =\lim_{\varepsilon\rightarrow0^{+}}\lim_{t\rightarrow\infty}%
{\displaystyle\int\limits_{0}^{t}}
dt_{1}e^{-\epsilon t_{1}}e^{\frac{1}{i\hbar}(E-E^{^{\prime}})t_{1}}%
=-i\hbar\frac{\Pr}{E-E^{^{\prime}}}+\pi\hbar\delta(E-E^{^{\prime}}),
\label{58q'}%
\end{align}
where the order of taking limits matters and $\Pr$ is a symbol of taking a
principal value of an integral.

Thus, we consider the system evolution on the time interval, which satisfies
two inequalities: $t\gg t_{cor}$ (to exclude the influence of initial
correlations) and $t\ll t_{rel}$ (to go from linear Eq. (\ref{55q}) to
nonlinear equation). This is the interval (\ref{0}) which existence is
guaranteed by the condition (\ref{29q'}). Using (\ref{58q'}), Eq. (\ref{58q})
can be rewritten as
\begin{align}
\frac{\partial f_{1}(1,t)}{\partial t}  &  =-\frac{\pi n}{\hbar}\int
dETr_{(2)}\Phi_{12}(E,t),\nonumber\\
\Phi_{12}(E,t)  &  =\left[  H_{12}^{^{\prime}},\delta(E-H_{12})\left[
H_{12}^{^{\prime}},(1+\theta P_{12})f_{1}(1,t)f_{1}(2,t)\right]
\delta(E-H_{12})\right]  . \label{58q'''}%
\end{align}
Obtaining (\ref{58q'''}), we have taken into account that Eq. (\ref{58q})
contains only the diagonal matrix element $\Phi_{12}(E,E^{^{\prime}%
},\mathbf{p}_{1},\mathbf{p}_{2},\mathbf{p}_{1},\mathbf{p}_{2},t)$ as it should
be because of taking a trace over $\mathbf{p}_{2}$ and because the left hand
side of (\ref{58q}) is diagonal relatively to $\mathbf{p}$ and $\mathbf{p}%
^{^{\prime}}$(see (\ref{40q})). The latter property of $Tr_{(2)}\Phi
_{12}(E,t)$ can be also proven explicitly. This diagonal element has the
symmetry property $\Phi_{12}(E,E^{^{\prime}},\mathbf{p}_{1},\mathbf{p}%
_{2},\mathbf{p}_{1},\mathbf{p}_{2},t)=\Phi_{12}(E^{^{\prime}},E,\mathbf{p}%
_{1},\mathbf{p}_{2},\mathbf{p}_{1},\mathbf{p}_{2},t)$, and, therefore, the
imaginary (principal) part of the integral over $E$ and $E^{^{\prime}}$
vanishes, as it follows from (\ref{58q}) and (\ref{58q'}). The mentioned
symmetry of $\Phi_{12}(E,E^{^{\prime}},\mathbf{p}_{1},\mathbf{p}%
_{2},\mathbf{p}_{1},\mathbf{p}_{2},t)$ (\ref{58q}) is a consequence of the
symmetry of matrix elements of $H_{12}^{0}$ and $H_{12}^{^{\prime}}$ (and,
therefore, of $H_{12})$. The latter symmetry follows from the definition
(\ref{24q}) and means that $H_{12}^{0}(\mathbf{p}_{1},\mathbf{p}%
_{2},\mathbf{p}_{1}^{^{\prime}},\mathbf{p}_{2}^{^{\prime}})=H_{12}^{^{0}%
}(\mathbf{p}_{1}^{^{\prime}},\mathbf{p}_{2}^{^{\prime}},\mathbf{p}%
_{1},\mathbf{p}_{2})$ and $H_{12}^{^{\prime}}(\mathbf{p}_{1},\mathbf{p}%
_{2},\mathbf{p}_{1}^{^{\prime}},\mathbf{p}_{2}^{^{\prime}})=H_{12}^{^{\prime}%
}(\mathbf{p}_{1}^{^{\prime}},\mathbf{p}_{2}^{^{\prime}},\mathbf{p}%
_{1},\mathbf{p}_{2})$, where
\begin{align}
H_{12}^{0}(\mathbf{p}_{1},\mathbf{p}_{2},\mathbf{p}_{1}^{^{\prime}}%
,\mathbf{p}_{2}^{^{\prime}})  &  =\left[  K(\mathbf{p}_{1})+K(\mathbf{p}%
_{2})\right]  \delta(\mathbf{p}_{1}-\mathbf{p}_{1}^{^{\prime}})\delta
(\mathbf{p}_{2}-\mathbf{p}_{2}^{^{\prime}}),K(\mathbf{p}_{i})=\frac
{\mathbf{p}_{i}^{2}}{2m},\nonumber\\
H_{12}^{^{\prime}}(\mathbf{p}_{1},\mathbf{p}_{2},\mathbf{p}_{1}^{^{\prime}%
},\mathbf{p}_{2}^{^{\prime}})  &  =\frac{1}{(2\pi\hbar)^{3}}\nu(\mathbf{p}%
_{1}-\mathbf{p}_{1}^{^{\prime}})\delta(\mathbf{p}_{1}+\mathbf{p}%
_{2}-\mathbf{p}_{1}^{^{\prime}}-\mathbf{p}_{2}^{^{\prime}}),\nonumber\\
\nu(\mathbf{p})  &  =\int\Phi(\left\vert \mathbf{r}\right\vert )\exp(-\frac
{i}{\hbar}\mathbf{pr})d\mathbf{r,}\nu(\mathbf{p})=\nu(-\mathbf{p}).
\label{58q''''}%
\end{align}

The (operator) $\delta$-function of the Hamiltonian $H$ (entering equation
(\ref{58q'''}) with a two-particle Hamiltonian $H_{12}$) can be expressed via
the imaginary part of Green's function $G(E)$:
\begin{align}
\delta(E-H)  &  =\mp\frac{1}{\pi}\operatorname{Im}G(E^{\pm}),G(E^{\pm}%
)=\frac{1}{E^{\pm}-H},\nonumber\\
E^{\pm}  &  =E\pm i\varepsilon,\varepsilon=0^{+}. \label{59q}%
\end{align}
On the other hand, there is the so called optical theorem relating
$\delta(E-H)$ (the imaginary part of Green's function $G(E^{\pm})$ for the
full Hamiltonian $H=H^{0}+H^{^{\prime}}$) and the interaction Hamiltonian
$H^{^{\prime}}$to $\delta(E-H^{0})$ (the imaginary part of Green's function
$G^{0}(E^{\pm})=(E^{\pm}-H^{0})^{-1}$) and the $T$-matrix (see, e.g.,
\cite{Ropke (1987)}):
\begin{equation}
H^{\prime}\delta(E-H)H^{^{\prime}}=T^{+}\delta(E-H^{0})T^{-}=T^{-}%
\delta(E-H^{0})T^{+}, \label{60q}%
\end{equation}
where the $T$-matrix is defined as
\begin{equation}
T^{\pm}=T(E^{\pm})=H^{^{\prime}}G(E^{\pm})G^{0^{-1}}(E^{\pm}) \label{61q}%
\end{equation}
and satisfies the equations
\begin{equation}
T^{\pm}=H^{^{\prime}}+H^{^{\prime}}G^{0}(E^{\pm})T^{\pm}=H^{^{\prime}}+T^{\pm
}G^{0}(E^{\pm})H^{^{\prime}}. \label{62q}%
\end{equation}
The relations (\ref{59q}) - (\ref{61q}) allow to convert $\Phi_{12}(E,t)$
(\ref{58q'''}) into the following form
\begin{equation}
\Phi_{12}(E,t)=\left[  H_{12}^{^{\prime-1}},T_{12}^{+}\delta(E-H_{12}%
^{0})T_{12}^{-}\left[  H_{12}^{^{\prime-1}},(1+\theta P_{12})f_{1}%
(1,t)f_{1}(2,t)\right]  T_{12}^{-}\delta(E-H_{12}^{0})T_{12}^{+}\right]  ,
\label{63q}%
\end{equation}
where $T_{12}^{\pm}$ refers to the $T$-matrix (\ref{61q}) defined for the
two-particle Hamiltonian $H_{12}=H_{12}^{0}+H_{12}^{^{\prime}}$. Using
(\ref{62q}), it is not difficult to show that
\begin{equation}
T_{12}^{-}\left[  H_{12}^{^{\prime-1}},(1+\theta P_{12})f_{1}(1,t)f_{1}%
(2,t)\right]  T_{12}^{-}=-\left[  T_{12}^{-},(1+\theta P_{12})f_{1}%
(1,t)f_{1}(2,t)\right]  , \label{64q}%
\end{equation}
where we have also taken into account that $G_{12}^{0}(E^{\pm})=(E^{\pm
}-H_{12}^{0})^{-1}$ commutes with $(1+\theta P_{12})f_{1}(1,t)f_{1}(2,t)$ for
the spatially homogeneous case (see also (\ref{34q})).

Using (\ref{64q}), (\ref{63q}), (\ref{62q}), and diagonality of $G^{0}(E^{\pm
})$ and of both sides of Eq. (\ref{58q'''}) in the momentum representation, we
can transform this equation into the following form
\begin{equation}
\frac{\partial f_{1}(1,t)}{\partial t}=-\frac{\pi n}{\hbar}\int dETr_{(2)}%
\left[  T_{12}^{+},\delta(E-H^{0})[T_{12}^{-},(1+\theta P_{12})f_{1}%
(1,t)f_{1}(2,t)]\delta(E-H^{0})\right]  . \label{65q}%
\end{equation}

To calculate the trace (integral over $\mathbf{p}_{2}$) in (\ref{65q}), one
need to have the matrix elements of $T$-matrix in the $\mathbf{p}%
$-representation. Iterating equation (\ref{62q}) and calculating the matrix
elements in the momentum representation, one can show that the $T_{12}$ matrix
(describing the particle collisions) guarantees the momentum conservation,
i.e. $T_{12}^{\pm}(\mathbf{p}_{1},\mathbf{p}_{2},\mathbf{p}_{1}^{^{\prime}%
},\mathbf{p}_{2}^{^{\prime}})\thicksim\delta(\mathbf{p}_{1}+\mathbf{p}%
_{2}-\mathbf{p}_{1}^{^{\prime}}-\mathbf{p}_{2}^{^{\prime}})$. For example, in
the first approximation in $H_{12}^{^{\prime}}$, we have $T_{12}^{\pm
(1)}(\mathbf{p}_{1},\mathbf{p}_{2},\mathbf{p}_{1}^{^{\prime}},\mathbf{p}%
_{2}^{^{\prime}})=H_{12}^{^{\prime}}(\mathbf{p}_{1},\mathbf{p}_{2}%
,\mathbf{p}_{1}^{^{\prime}},\mathbf{p}_{2}^{^{\prime}})$, where the matrix
element of the interaction Hamiltonian is given by (\ref{58q''''}). In
general, it is not difficult to obtain from (\ref{62q}) and (\ref{58q''''})
the following expression for the matrix element of $T_{12}$ in the momentum
representation:
\begin{align}
T_{12}^{\pm}(\mathbf{p}_{1},\mathbf{p}_{2},\mathbf{p}_{1}^{^{\prime}%
},\mathbf{p}_{2}^{^{\prime}})  &  =\frac{1}{(2\pi\hbar)^{3}}t^{\pm}%
(\mathbf{p}_{1},\mathbf{p}_{2},\mathbf{p}_{1}^{^{\prime}},\mathbf{p}%
_{2}^{^{\prime}})\delta(\mathbf{p}_{1}+\mathbf{p}_{2}-\mathbf{p}_{1}%
^{^{\prime}}-\mathbf{p}_{2}^{^{\prime}}),\nonumber\\
t^{\pm}(\mathbf{p}_{1},\mathbf{p}_{2},\mathbf{p}_{1}^{^{\prime}}%
,\mathbf{p}_{2}^{^{\prime}})  &  =\nu(\mathbf{p}_{1}-\mathbf{p}_{1}^{^{\prime
}})\nonumber\\
&  +\frac{1}{(2\pi\hbar)^{3}}\int d\mathbf{p}_{1}^{^{\prime\prime}}\int
d\mathbf{p}_{2}^{^{\prime\prime}}\frac{\nu(\mathbf{p}_{1}-\mathbf{p}%
_{1}^{^{\prime\prime}})\nu(\mathbf{p}_{1}^{^{\prime\prime}}-\mathbf{p}%
_{1}^{^{\prime}})}{E^{\pm}-K(\mathbf{p}_{1}^{^{\prime\prime}})-K(\mathbf{p}%
_{2}^{^{\prime\prime}})}\delta(\mathbf{p}_{1}+\mathbf{p}_{2}-\mathbf{p}%
_{1}^{^{\prime\prime}}-\mathbf{p}_{2}^{^{\prime\prime}})\nonumber\\
&  +\ldots\label{66q}%
\end{align}
This $T$-matrix obeys the following symmetry conditions
\begin{align}
T_{12}^{\pm}(\mathbf{p}_{1},\mathbf{p}_{2},\mathbf{p}_{1}^{^{\prime}%
},\mathbf{p}_{2}^{^{\prime}})  &  =T_{12}^{\pm}(\mathbf{p}_{2},\mathbf{p}%
_{1},\mathbf{p}_{2}^{^{\prime}},\mathbf{p}_{1}^{^{\prime}}),\nonumber\\
T_{12}^{\pm}(\mathbf{p}_{1},\mathbf{p}_{2},\mathbf{p}_{1}^{^{\prime}%
},\mathbf{p}_{2}^{^{\prime}})  &  =T_{12}^{\pm}(\mathbf{p}_{1}^{^{\prime}%
},\mathbf{p}_{2}^{^{\prime}},\mathbf{p}_{1},\mathbf{p}_{2}). \label{67q}%
\end{align}
Calculating the trace (integral over $\mathbf{p}_{2}$) in (\ref{65q}) with the
use of (\ref{66q}), (\ref{67q}), (\ref{40q}) and (\ref{49q'}), we obtain the
following equation for the momentum distribution function
\begin{align}
\frac{\partial w(\mathbf{p}_{1},t)}{\partial t}  &  =\frac{\pi n}{(2\pi
\hbar)^{3}\hbar}\int d\mathbf{p}_{2}d\mathbf{p}_{1}^{^{\prime}}d\mathbf{p}%
_{2}^{^{\prime}}\delta\left[  K(\mathbf{p}_{1})+K(\mathbf{p}_{2}%
)-K(\mathbf{p}_{1}^{^{\prime}})-K(\mathbf{p}_{2}^{^{\prime}})\right]
\nonumber\\
&  \times\delta(\mathbf{p}_{1}+\mathbf{p}_{2}-\mathbf{p}_{1}^{^{\prime}%
}-\mathbf{p}_{2}^{^{\prime}})\left\vert t^{+}(\mathbf{p}_{1},\mathbf{p}%
_{2},\mathbf{p}_{1}^{^{\prime}},\mathbf{p}_{2}^{^{\prime}})+\theta
t^{+}(\mathbf{p}_{1},\mathbf{p}_{2},\mathbf{p}_{2}^{^{\prime}},\mathbf{p}%
_{1}^{^{\prime}})\right\vert ^{2}\nonumber\\
&  \times\left[  w(\mathbf{p}_{1}^{^{\prime}},t)w(\mathbf{p}_{2}^{^{\prime}%
},t)-w(\mathbf{p}_{1},t)w(\mathbf{p}_{2},t)\right]  . \label{68q}%
\end{align}

Equation (\ref{68q}) is the quantum Boltzmann equation in the linear
approximation in the particle density $n$ (only the binary collisions have
been accounted for). It includes (as it should be) the quantum mechanical
processes of particle exchange at scattering (the term proportional to
$\theta$), which are due to the quantum statistics of particles. This
irreversible kinetic equation (\ref{68q}) has been obtained with no use of a
factorized initial condition (which introduces irreversibility) for the
density matrix of the system under consideration. In the suggested here
approach, the kinetic equation (\ref{68q}) follows from the evolution equation
(\ref{55q}), which is exact in the linear approximation in $n$ and describes
an evolution process on any time scale treating the initial correlations and
correlations due to collisions on the equal footing. The described procedure
of obtaining the time-irreversible Eq. (\ref{68q}) from Eq. (\ref{55q})
clearly indicates that irreversibility emerges on the macroscopic time scale
(\ref{0}) as a result of the damping of both the initial correlations caused
by interaction and correlations caused by collisions (see also \cite{Lebowitz
(1999)}). The latter can be secured by the appropriate properties of the
system's dynamics (e.g., the ergodic mixing flow in the phase space). One can,
therefore, expect that Eq. (\ref{55q}) would switch automatically from a
reversible behaviour to an irreversible one (described by the Boltzmann
equation (\ref{68q})) if all correlations vanish with time while going from
microscopic to a large enough time scale. Thus, the influence of initial
correlations on an evolution process can be revealed.

Before going to the next Section, we would again like to stress, that the
obtained above TC-HGME and TCL-HGME are linear equations for the relevant
(reduced) statistical operator because they have been derived from the linear
Liouville-von-Neumann equation by the conventional linear time-independent
projection operator technique. These equations, when applied, e.g., to the
(sub)system interacting with a thermal bath, describe the (sub)system's
evolution on all time scales (no restrictions like those determined by
(\ref{0})). But we have seen, that in order to obtain from these equations the
nonlinear evolution equations accounting for initial correlations, the second
inequality in (\ref{0}) ($t-t_{0}\ll t_{rel}$) should still be satisfied.
Therefore, we have to turn to a different approach in order to obtain from the
Liouville-von-Neumann equation the nonlinear evolution equations valid on all
time scales, i.e. to lift all restrictions imposed by inequalities (\ref{0}).

\section{NONLINEAR INHOMOGENEOUS GME}

We are considering the evolution equations for many-particle systems. It is
natural to expect that these equations should generally be nonlinear evolution
equations on any time scale due to interaction between particles. Strictly
speaking, it is impossible to obtain a nonlinear evolution equation (like the
Boltzmann equation) in the framework of the linear time-independent projection
operator formalism without additional approximations (as we have seen above).
The problem is that the conventional time-independent projection operator
(\ref{17}) (or (\ref{22q})) leads to the relevant part of an $N$-particle
distribution function (\ref{21}) (or (\ref{23q})) which is linear in the
time-dependent one-particle distribution function of interest. That is why, a
nonlinear equation for a density matrix is often simply postulated in the form
of a Lindblad-type master equation whose generator depends parametrically on
this density matrix (see, e.g., \cite{Breuer and Petruccione (2002)}).\ 

Therefore, we need a new approach which would allow to obtain, e.g., a
nonlinear relevant distribution function (statistical operator) like
$f_{r}(x_{1},\ldots,x_{N},t)=%
{\displaystyle\prod\limits_{i=1}^{N}}
F_{1}(x_{i},t)$ in (\ref{0a}).

We start again with the Liouville-von-Neumann equation (\ref{1}) describing
the evolution of a distribution function (statistical operator) $F_{N}(t)$ of
a system of $N$ ($N\gg1$) classical or quantum particles but in more general
case, when the Hamiltonian and the corresponding Liouville (evolution)
(super)operator $L(t)$, acting on $F_{N}(t)$ according to Eqs. (\ref{1''}),
(\ref{1'''}), may depend on time
\begin{equation}
\frac{\partial}{\partial t}F_{N}(t)=L(t)F_{N}(t). \label{1n}%
\end{equation}
The distribution function (or statistical operator) is subject to the
normalization condition (\ref{1'}).

The formal solution to equation (\ref{1n}) can be written as%
\begin{equation}
F_{N}(t)=U(t,t_{0})F_{N}(t_{0}),U(t,t_{0})=T\exp\left[
{\textstyle\int\limits_{t_{0}}^{t}}
dsL(s)\right]  , \label{6n}%
\end{equation}
where $T$ denotes the chronological time-ordering operator, which orders the
product of time-dependent operators such that their time arguments increase
from right to left and $F_{N}(t_{0})$ is the value of a distribution function
(statistical operator) at some initial time $t_{0}$.

We introduce the generally time-dependent operator $P(t)$ converting the
distribution function $F_{N}(t)$ into a relevant distribution function
$f_{r}(t)$, which as a rule is a vacuum (without correlations) many-particle
distribution function, i.e., a product of the reduced distribution functions
that are necessary and sufficient for describing the evolution of the
measurable values (statistical expectations). The problem is to obtain the
closed evolution equation for such a reduced distribution function in view of
the existence of correlations between the elements of the considered system
including initial correlations at the initial time $t_{0}$.

We therefore define the relevant distribution function (statistical operator)
as%
\begin{equation}
f_{r}(t)=P(t)F_{N}(t)=P(t)U(t,t_{0})F_{N}(t_{0}). \label{7n}%
\end{equation}
We note that the operator $P(t)$ is generally not a projection operator.
Moreover, the action of \ $P(t)$ on $F_{N}(t)$ is no longer a linear operation
(in contrast to applying the conventional time-independent projection operator
\cite{Zwanzig (1960)}) because $P(t)$ can depend on the time-dependent
distribution function of interest (see below). Therefore, by applying $P(t)$
to the linear Liouville equation (\ref{1n}), we generally obtain a nonlinear
equation. The operator $P(t)$ does not commute with the derivative
$\frac{\partial}{\partial t}$ (the time-independent linear projection
operators commute with $\frac{\partial}{\partial t}$) and $P(t^{^{\prime}%
})F_{N}(t)\neq f_{r}(t)$ (for $t^{^{\prime}}\neq t$).

Using Liouville-von-Neumann equation (\ref{1n}), the equation of motion for
$f_{r}(t)$ (\ref{7n}) may be written as%
\begin{equation}
\frac{\partial f_{r}(t)}{\partial t}=\left[  \frac{\partial P(t)}{\partial
t}+P(t)L(t)\right]  \left[  f_{r}(t)+f_{i}(t)\right]  , \label{8n}%
\end{equation}
where $f_{i}(t)=\left[  F_{N}(t)-f_{r}(t)\right]  $ and $\frac{\partial
P(t)}{\partial t}$ is the operator obtained by taking the derivative of the
operator $P(t)$, i.e., we assume that the operator $\frac{\partial
P(t)}{\partial t}$ exists. To make sense of $\frac{\partial P(t)}{\partial t}%
$, we can define $P(t)$ as
\begin{equation}
P(t)=C(t)D, \label{8an}%
\end{equation}
where $C(t)$ is a well defined (operator) function of time and $D$ is a
time-independent (super)operator acting on $F_{N}(t)$ and integrating over all
unnecessary variables in the spirit of the reduced description method (more
details of the specification of $P(t)$ are given below).

As it follows from (\ref{8n}), we can obtain the equation for the irrelevant
part of a distribution function $f_{i}(t)$ by using Eqs. (\ref{1n}) and
(\ref{8n}) and splitting $F_{N}(t)$ into the relevant and irrelevant parts. Thus,%

\begin{equation}
\frac{\partial f_{i}(t)}{\partial t}=\left[  Q(t)L(t)-\frac{\partial
P(t)}{\partial t}\right]  \left[  f_{i}(t)+f_{r}(t)\right]  , \label{9n}%
\end{equation}
where $Q(t)=1-P(t)$ and therefore $f_{i}(t)=Q(t)F_{N}(t)$.

The formal solution of equation (\ref{9n}) is%
\begin{equation}
f_{i}(t)=%
{\textstyle\int\limits_{t_{0}}^{t}}
dt^{^{\prime}}S(t,t^{^{\prime}})\left[  Q(t^{^{\prime}})L(t^{^{\prime}}%
)-\frac{\partial P(t^{^{\prime}})}{\partial t^{^{\prime}}}\right]
f_{r}(t^{^{\prime}})+S(t,t_{0})f_{i}(t_{0}), \label{10n}%
\end{equation}
where
\begin{equation}
S(t,t_{0})=T\exp\left\{
{\textstyle\int\limits_{t_{0}}^{t}}
ds\left[  Q(s)L(s)-\frac{\partial P(s)}{\partial s}\right]  \right\}  .
\label{11n}%
\end{equation}
We can represent the latter operator as the series%
\begin{align}
S(t,t_{0})  &  =1+%
{\textstyle\int\limits_{t_{0}}^{t}}
dt_{1}\left[  Q(t_{1})L(t_{1})-\frac{\partial P(t_{1})}{\partial t_{1}}\right]
\nonumber\\
&  +%
{\textstyle\int\limits_{t_{0}}^{t}}
dt_{1}%
{\textstyle\int\limits_{t_{1}}^{t}}
dt_{2}\left[  Q(t_{2})L(t_{2})-\frac{\partial P(t_{2})}{\partial t_{2}}\right]
\nonumber\\
&  \times\left[  Q(t_{1})L(t_{1})-\frac{\partial P(t_{1})}{\partial t_{1}%
}\right]  +\ldots. \label{11an}%
\end{align}
We note that if the operators $P$ and $L$ are independent of time, then
$S(t,t_{0})=\exp\left[  QL(t-t_{0})\right]  $.

Substituting (\ref{10n}) in (\ref{8n}), we obtain%
\begin{align}
\frac{\partial f_{r}(t)}{\partial t}  &  =\left[  P(t)L(t)+\frac{\partial
P(t)}{\partial t}\right]  \{f_{r}(t)\nonumber\\
&  +%
{\textstyle\int\limits_{t_{0}}^{t}}
dt^{^{\prime}}S(t,t^{^{\prime}})\left[  Q(t^{^{\prime}})L(t^{^{\prime}}%
)-\frac{\partial P(t^{^{\prime}})}{\partial t^{^{\prime}}}\right]
f_{r}(t^{^{\prime}})\nonumber\\
&  +S(t,t_{0})f_{i}(t_{0})\}. \label{12n}%
\end{align}

Equation (\ref{12n}) is a generalization of the Nakajima-Zwanzig
time-convolution generalized master equation (TC-GME) (see \cite{Nakajima
(1958), Zwanzig (1960), Prigogine (1962)}) for the relevant part of the
distribution function (statistical operator) to the case of time-dependent
operators $P$ and $Q$. As previously noted, $P(t)$ and $Q(t)$ are generally
not projection operators in the usual sense, and in deriving (\ref{12n}) we do
not use such a property of the projection operators as $P^{2}=P$, $Q^{2}=Q $
(for $P(t)P(t^{^{\prime}})$ and $Q(t)Q(t^{^{\prime}})$ such a property holds
only for $t=t^{^{\prime}}$, as is shown below). The obtained equation, like
the conventional TC-GME, is an exact inhomogeneous integro-differential
equation for the relevant part of the distribution function (statistical
operator) containing the irrelevant part of the distribution function (a
source) at the initial instant $f_{i}(t_{0})=F_{N}(t_{0})-f_{r}(t_{0})$. But,
Eq. (\ref{12n}) is generally equivalent to the nonlinear equation for the
reduced distribution function of interest (e.g., for the one-particle
distribution function considered in the next section) and is therefore also
convenient for studying the evolution of many-particle systems described by
nonlinear Boltzmann-type equations (in the kinetic regime) where the
nonlinearity results from particle collisions.

For time-independent $P$ and $Q$ equation, Eq. (\ref{12n}) reduces to the
conventional linear TC-GME. This type of equation is more suitable, for
example, for studying the evolution of a subsystem interacting with a large
system in thermal equilibrium (a thermal bath). Although, as we have seen
above, it is possible to obtain from the linear TC-GME the nonlinear evolution
equation but on the time scale $t-t_{0}\ll t_{rel}$, as we have seen in the
previous sections.

To simplify Eq. (\ref{12n}) and make sense of the derivative $\frac{\partial
P(t)}{\partial t}$, we consider an operator $P(t)$ of form (\ref{8an}). This
representation of $P(t)$ is suitable for studying many cases of interest (see
Sec. 9). Also, splitting a distribution function (statistical operator) into
the relevant $P(t)F_{N}(t)$ and irrelevant $Q(t)F_{N}(t)$ parts makes sense if
the operator $P(t)$ satisfies the relation \ \ \ \ \
\begin{equation}
P(t)P(t)=P(t) \label{12an}%
\end{equation}
for any $t$. In this case, if $P(t)=C(t)D$, then Eq. (\ref{12an}) implies
\begin{equation}
DC(t)=1 \label{12bn}%
\end{equation}
at any instant $t$. Equation (\ref{12bn}) generally represents the
normalization condition for the distribution function(s) making up $C(t)$ (see
(\ref{1'}) and (\ref{20})) and implies that $C(t)$ depends on the variables
that are removed from $F_{N}(t)$ by the operator $D$ ($f_{r}(t)=P(t)F_{N}(t)$
depends on the complete set of variables of the distribution function
$F_{N}(t)$ as indicated by (\ref{0b})). From condition (\ref{12bn}) we have
\begin{align}
P(t)P(t^{^{\prime}})  &  =P(t),Q(t)Q(t^{^{\prime}})=Q(t^{^{\prime}%
}),P(t)Q(t^{^{\prime}})=0,\nonumber\\
Q(t^{^{\prime}})P(t)  &  =P(t)-P(t^{^{\prime}}). \label{12cn}%
\end{align}
Condition (\ref{12bn}) also leads to the relations
\begin{equation}
D\frac{\partial P(t)}{\partial t}=D\frac{\partial C(t)}{\partial t}D=0,
\label{12dn}%
\end{equation}
for the time-derivative of $P(t)$, where we assume that the operators $D$ and
$\partial/\partial t$ commute ($D$ is independent of time).

We also introduce the reduced distribution function (statistical operator)
\begin{equation}
f_{red}(t)=DF_{N}(t)=Df_{r}(t), \label{12en}%
\end{equation}
which is actually needed for calculating the expectation values of interest in
a nonequilibrium state. By (\ref{12bn}), we have
\begin{equation}
D\frac{\partial f_{r}(t)}{\partial t}=\frac{\partial f_{red}(t)}{\partial
t},Df_{i}(t)=DQ(t)F_{N}(t)=0. \label{12fn}%
\end{equation}
Hence, applying the operator $D$ to equation (\ref{12n}) from the left, we
obtain%
\begin{align}
\frac{\partial f_{red}(t)}{\partial t}  &  =DL(t)\{f_{r}(t)+%
{\textstyle\int\limits_{t_{0}}^{t}}
dt^{^{\prime}}S(t,t^{^{\prime}})\left[  Q(t^{^{\prime}})L(t^{^{\prime}}%
)-\frac{\partial P(t^{^{\prime}})}{\partial t^{^{\prime}}}\right]
f_{r}(t^{^{\prime}})\nonumber\\
&  +S(t,t_{0})f_{i}(t_{0})\}, \label{12gn}%
\end{align}
where we use (\ref{12bn}), (\ref{12dn}), and (\ref{12fn}).

Without loss of generality, we can split the Hamiltonian or the Liouvillian of
a system into two terms as $H(t)=H^{0}(t)+H^{^{\prime}}(t)$ or $L(t)=L^{0}%
(t)+L^{^{\prime}}(t)$, where $H^{0}$ or $L^{0}$ is related to the energy of
the noninteracting particles (subsystems) and $H^{^{\prime}}$or $L^{^{\prime}%
}$ describes the interaction between particles (subsystems). Then, in addition
to satisfying condition (\ref{12bn}), the operator $D$, which selects the
reduced distribution function, should commute with $L^{0}$as%
\begin{equation}
DL^{0}(t)=L_{red}^{0}(t)D, \label{12hn}%
\end{equation}
where $L_{red}^{0}(t)$ is the reduced Liouvillian $L^{0}$ depending only on
the variables of the reduced distribution function (statistical operator)
$f_{red}(t)$. Expression (\ref{12hn}) means that both sides of this relation
act on the arbitrary function (operator) defined on the phase (Hilbert) space.
Taking into account that $C(t^{^{\prime}})L_{red}^{0}(t)=L_{red}%
^{0}(t)C(t^{^{\prime}})$ ($C(t^{^{\prime}})$ depends on the set of variables
that are removed from $F(t)$ by $D$, and $L_{red}^{0}(t)$ is independent of
them), we can write Eq. (\ref{12hn}) in the more general form%
\begin{equation}
P(t^{^{\prime}})L^{0}(t)=L_{red}^{0}(t)P(t^{^{\prime}}). \label{12h'n}%
\end{equation}
Commutation relation (\ref{12hn}) follows from a self-consistency argument
because if $L^{^{\prime}}(t)=0$, then the evolution of a "free" subsystem (not
interacting with the rest of the system) should be described in Eq.
(\ref{12gn}) only by $L_{red}^{0}(t)$ ($DL^{0}(t)f_{r}(t)=L_{red}%
^{0}(t)f_{red}(t)=L_{red}^{0}(t)Df_{r}(t)$; also see \cite{Bogoliubov
(1946)}). It can be seen from the relations%
\begin{align}
DL^{0}(t)S(t,t_{1})Q(t_{2})  &  =0,\nonumber\\
DL^{0}S(t,t_{1})\frac{\partial P(t_{2})}{\partial t_{2}}  &  =0, \label{12in}%
\end{align}
which follow from (\ref{12hn}), (\ref{11n}), (\ref{12dn}), and (\ref{12fn}),
that this is the case. Hence, all terms in (\ref{12gn}) are proportional to
$L^{^{\prime}}(t)$ except the first term, $L_{red}^{0}(t)f_{red}(t)$.

Analogously, using (\ref{12hn}), (\ref{11n}), and relations that follow from
(\ref{12bn}), we can prove the equations
\begin{align}
S(t,t_{1})Q(t_{2})  &  =\overline{U}(t,t_{1})Q(t_{2}),\nonumber\\
S(t,t_{1})\frac{\partial P(t_{2})}{\partial t_{2}}  &  =\overline{U}%
(t,t_{1})\frac{\partial P(t_{2})}{\partial t_{2}}, \label{12jn}%
\end{align}
where
\begin{equation}
\overline{U}(t,t_{1})=T\exp\left\{
{\textstyle\int\limits_{t_{1}}^{t}}
ds\left[  L^{0}(s)+Q(s)L^{^{\prime}}\right]  \right\}  . \label{12kn}%
\end{equation}
Hence, using (\ref{12in}) and (\ref{12jn}) in (\ref{12gn}), we can finally
rewrite Eq. (\ref{12n}) as the equation for the relevant part of the
distribution function (statistical operator)%
\begin{align}
\frac{\partial Df_{r}(t)}{\partial t}  &  =DL(t)f_{r}(t)\nonumber\\
&  +DL^{^{\prime}}(t)%
{\textstyle\int\limits_{t_{0}}^{t}}
dt^{^{\prime}}\overline{U}(t,t^{^{\prime}})\left[  Q(t^{^{\prime}%
})L(t^{^{\prime}})-\frac{\partial P(t^{^{\prime}})}{\partial t^{^{\prime}}%
}\right]  f_{r}(t^{^{\prime}})\nonumber\\
&  +DL^{^{\prime}}(t)\overline{U}(t,t_{0})f_{i}(t_{0}), \label{12ln}%
\end{align}
where $Df_{r}(t)=f_{red}(t)$ and (\ref{12hn}) holds.

We demonstrate that basic conditions (\ref{12bn}) and (\ref{12hn}) are
satisfied for a gas of interacting classical particles in the next section.

\section{NONLINEAR INHOMOGENEOUS EVOLUTION EQUATION FOR A NONIDEAL
INHOMOGENEOUS DILUTE GAS OF CLASSICAL PARTICLES}

In this section, we apply Eq. (\ref{12ln}) to the case of a gas of $N$
($N\gg1$) identical classical particles. The Liouville operator $L$ for such a
system can be represented in the form (\ref{23}), where $L^{0}$ corresponds to
the kinetic energy $H^{0}=%
{\textstyle\sum\limits_{i=1}^{N}}
\mathbf{p}_{i}^{2}/2m$ of particles with the momenta $\mathbf{p}_{i}$ and mass
$m$ and $L^{^{\prime}}$corresponds to the particle interaction $H^{^{\prime}}=%
{\textstyle\sum\limits_{i<j=1}^{N}}
V_{ij}$ with the pair potential $V_{ij}=V(|\mathbf{x}_{i}-\mathbf{x}_{j}|)$.
Hence, $L$ is independent of time in the considered case . We do not assume
here that the interaction is weak, but we do assume that all necessary
requirements for the properties of forces by which the particles interact are
satisfied (in particular, bound states do not form). As usual, we also assume
that all functions $\Phi(x_{1},\ldots,x_{N})$ defined on the phase space and
their derivatives vanish at the boundaries of the configuration space and at
$\mathbf{p}_{i}=\pm\infty$. These boundary conditions and the explicit form of
the Liouville operators (\ref{23}) lead to relations (\ref{24}) (see Sec. 4).

We seek an evolution equation for a one-particle distribution function
(\ref{18})
\begin{equation}
f_{1}(x_{i},t)=V\int dx_{1}\cdots\int dx_{i-1}\int dx_{i+1}\cdots\int
dx_{N}F_{N}(x_{1},\ldots,x_{N},t). \label{14n}%
\end{equation}
For this, it is convenient to define the operator $P(t)$ as%
\begin{equation}
P(t)=C(t)D,C(t)=\left[  \prod\limits_{i=2}^{N}f_{1}(x_{i},t)\right]
,D=\frac{1}{V^{N-1}}\int dx_{2}\cdots\int dx_{N} \label{15n}%
\end{equation}
(compare with (\ref{17})).

We apply the projection operator (\ref{15n}) to $f_{N}(t)=V^{N}F_{N}(t)$,
which satisfies the Liouville-von- Neumann equation (\ref{1n}), and obtain the
following relevant part of the distribution function%
\begin{equation}
f_{r}(t)=P(t)f_{N}(t)=%
{\textstyle\prod\limits_{i=1}^{N}}
f_{1}(x_{i},t). \label{18n}%
\end{equation}
From definition (\ref{18n}) we can see an advantage of the time-dependent
operator $P(t)$ (\ref{15n}): applying it to an $N$-particle distribution
function $f_{N}(t)$, we can define relevant part (\ref{18n}) of the
distribution function, which seems more natural and symmetric than (\ref{21})
used in Sec. 4 and obtained using the time-independent projection operator
(\ref{17}) (we had to use the time-independent projection operator because the
TC-HGME considered in Sec. 2 holds only for such operators). Moreover,
relevant part (\ref{18n}) has a clear physical meaning because it represents a
part of an $N$-particle distribution function without correlations which is
often of most interest.

Hence, the irrelevant part%
\begin{equation}
f_{i}(t)=Q(t)f_{N}(t)=f_{N}(t)-%
{\textstyle\prod\limits_{i=1}^{N}}
f_{1}(x_{i},t) \label{18an}%
\end{equation}
describes all interparticle correlations, in particular, the initial ones
(existing at the initial instant $t_{0}$), which compose the inhomogeneous
term in (\ref{12ln}). The irrelevant part of the $N$-particle distribution
function at $t_{0}$, $f_{i}(t_{0})=f_{N}(t_{0})-f_{r}(t_{0})$, can always be
represented (and also at any other time $t$) by the cluster expansion
(\ref{22}).

Using normalization condition (\ref{20}) for a one-particle distribution
function $\int f_{1}(x_{i},t)dx_{i}=V$, we can easily see that by applying the
integral operator $D$ to $C(t)$ (both are defined by (\ref{15n})), we obtain
$DC(t)=1$, i.e., condition (\ref{12bn}) and its consequences given by
(\ref{12cn}) and (\ref{12dn}) are satisfied. Also, in this case,
\begin{align}
f_{red}(t)  &  =Df_{r}(t)=f_{1}(x_{1},t),\nonumber\\
DQ(t)  &  =0,Df_{i}(t)=0 \label{19n}%
\end{align}
\bigskip in accordance with (\ref{12en}) and (\ref{12fn}).

Commutation rules (\ref{12hn}) and (\ref{12h'n}) also follow from (\ref{23}),
(\ref{24}), and (\ref{15n}), because%
\begin{align}
DL^{0}  &  =L_{1}^{0}D\nonumber\\
P(t)L^{0}  &  =L_{1}^{0}P(t), \label{20n}%
\end{align}
i.e., $L_{red}^{0}=L_{1}^{0}$ in the case under consideration. Therefore, the
identities (\ref{12in}) - (\ref{12kn}) are also satisfied, where $L^{0}$ and
$L^{^{\prime}}$ are independent of time in the considered case.

Using (\ref{24}), (\ref{19n}), and (\ref{20n}), we can represent Eq.
(\ref{12ln}) in the case under consideration as%

\begin{align}
\frac{\partial f_{1}(x_{1},t)}{\partial t}  &  =L_{1}^{0}f_{1}(x_{1},t)+n\int
dx_{2}L_{12}^{^{\prime}}f_{1}(x_{2},t)f_{1}(x_{1},t)\nonumber\\
&  +DL^{^{\prime}}%
{\textstyle\int\limits_{t_{0}}^{t}}
dt^{^{\prime}}\overline{U}(t,t^{^{\prime}})\left[  (L^{0}-L_{1}^{0}%
)+Q(t^{^{\prime}})L^{^{\prime}}-\frac{\partial P(t^{^{\prime}})}{\partial
t^{^{\prime}}}\right]  f_{r}(t^{^{\prime}})\nonumber\\
&  +DL^{^{\prime}}\overline{U}(t,t_{0})f_{i}(t_{0}), \label{22n}%
\end{align}
where $n=N/V$ is the density of particles and $D$, $f_{r}(t)$, and
$f_{i}(t_{0})$ are respectively defined in (\ref{15n}), (\ref{18n}), and
(\ref{22}). Evolution operator (\ref{12kn}) can be represented by the series
similar to (\ref{11an})%
\begin{align}
\overline{U}(t,t^{^{\prime}})  &  =1+%
{\textstyle\int\limits_{t^{^{\prime}}}^{t}}
dt_{1}[L^{0}+Q(t_{1})L^{^{\prime}}]\nonumber\\
&  +%
{\textstyle\int\limits_{t^{^{\prime}}}^{t}}
dt_{1}%
{\textstyle\int\limits_{t_{1}}^{t}}
dt_{2}[L^{0}+Q(t_{2})L^{^{\prime}}][L^{0}+Q(t_{1})L^{^{\prime}}]+\ldots.
\label{23n}%
\end{align}

We consider equation (\ref{22n}) in the first approximation in the particle
density $n$. The corresponding dimensionless small parameter of the
perturbation expansion is given by (\ref{16}). From the second equation in
(\ref{24}), we can easily see that all terms in expansion (\ref{23n})
containing $P(t)L^{^{\prime}}$result in expressions proportional to at least
the first power of $n$ (like the second term in the right-hand side of
(\ref{22n})). Therefore, in the first approximation in $n$, all terms of
(\ref{23n}) with $P(t)L^{^{\prime}}$ can be neglected because the terms of Eq.
(\ref{22n}) containing $\overline{U}(t,t^{^{\prime}})$ are already of the
first order in $n$ (applying $DL^{^{\prime}}$ leads to an expression of at
least the first order in $n$). In this approximation, all terms under the
integrands in (\ref{23n}) are independent of time, and the operator
$\overline{U}(t,t^{^{\prime}})$ reduces to
\begin{equation}
\overline{U}(t,t^{^{\prime}})=\exp[(L^{0}+L^{^{\prime}})(t-t^{^{\prime}})].
\label{25n}%
\end{equation}

It is now convenient to use the expansion for the exponential operator
(\ref{25n}) (see (\ref{27}))%
\begin{equation}
e^{(L^{0}+L^{^{\prime}})(t-t^{^{\prime}})}=e^{L^{0}(t-t^{^{\prime}})}%
+\int\limits_{t^{^{\prime}}}^{t}d\theta e^{L^{0}(t-\theta)}L^{^{\prime}%
}e^{(L^{0}+L^{^{\prime}})\theta}. \label{26n}%
\end{equation}
Using this expansion and relations (\ref{24}), we obtain the following
evolution equation for a one-particle distribution function in the linear
approximation in the density parameter (\ref{16}) from (\ref{22n}):%
\begin{align}
\frac{\partial f_{1}(x_{1},t)}{\partial t}  &  =L_{1}^{0}f_{1}(x_{1},t)+n\int
dx_{2}L_{12}^{^{\prime}}f_{1}(x_{2},t)f_{1}(x_{1},t)\nonumber\\
&  +n\int dx_{2}L_{12}^{^{\prime}}%
{\textstyle\int\limits_{0}^{t-t_{0}}}
dt_{1}e^{(L_{12}^{0}+L_{12}^{^{\prime}})t_{1}}[(L_{2}^{0}+L_{12}^{^{\prime}%
})f_{1}(x_{2},t-t_{1})f_{1}(x_{1},t-t_{1})\nonumber\\
&  +\frac{\partial f_{1}(x_{2},t-t_{1})}{\partial t_{1}}f_{1}(x_{1}%
,t-t_{1})]\nonumber\\
&  +n\int dx_{2}L_{12}^{^{\prime}}e^{(L_{12}^{0}+L_{12}^{^{\prime}})(t-t_{0}%
)}g_{2}(x_{1},x_{2}), \label{27n}%
\end{align}
where $L_{12}^{0}=L_{1}^{0}+L_{2}^{0}$. In obtaining (\ref{27n}), we also use
the definition (\ref{22}) for the irrelevant part of the distribution function
and take into account that each additional integration over $x_{3},\ldots$
adds an additional power of $n$ (therefore, in the linear approximation in
$n$, all formulae contain no more than one integration over the phase space).

Equation (\ref{27n}) is the main result in this section. We must stress that
(\ref{27n}) is a nonlinear\textbf{\ }inhomogeneous time-convolution
(non-Markovian) master equation for a one-particle distribution function
containing initial correlations (a source). We note, that this new equation is
rigorously derived here (the desired nonlinear master equation is usually
postulated \cite{Breuer and Petruccione (2002)}). Equation (\ref{27n}) is
exact in the linear approximation in small density parameter (\ref{16}) and
holds on any time scale and for any spatial inhomogeneity of the system under
consideration. The first term in the right-hand side is a conventional flow
term. The second term, a nonlinear Vlasov term, represents the self-consistent
field acting on the given (first) particle and determined by all particles of
the system. We note that it is impossible (without additional approximations)
to obtain this term (appearing only in a spatially inhomogeneous case) from
the GME (or HGME) containing time-independent projection operator (\ref{17})
(see Sec. 4). The term quadratic in $L_{12}^{^{\prime}}$ describes particle
collisions and can lead to dissipation in the system. The terms with
$L_{2}^{0}$ (the space derivative) and with the time derivative take the
change in space and time of the one-particle distribution function on the
respective microscopic scales (also see below) of the orders of $r_{0}$ and
$t_{cor}\sim r_{0}/\overline{v}$ into account. These two terms determine the
contribution of the pair collisions to the nondissipative characteristics
(thermodynamic functions) of the nonideal gas (see, e.g., \cite{Klimontovich
(1982)}) and are usually absent in the standard derivation (from the BBGKY
chain) of the kinetic equations in the approximation linear in $n$. Taking
these terms into account poses a certain problem because they usually appear
in the next (second) approximation in $n$ (see, e.g., \cite{Uhlenbeck and Ford
(1963)}), which is inconsistent with the dissipative three-particle collision
term appearing in this approximation. The last (irrelevant) term in
(\ref{27n}) takes initial correlations (at $t=t_{0}$) into account and is
given by the two-particle correlation function $g_{2}(x_{1},x_{2})$ in the
linear approximation in $n$.

We examine the terms in (\ref{27n}) more closely, especially their behavior in
time. The evolution in time is governed by the exact two-particle propagator
$G_{12}(t)=\exp(L_{12}t)$ ($L_{12}=L_{12}^{0}+L_{12}^{\prime}$) (this is
natural for the considered dilute gas in the lowest approximation in density),
which satisfies the integral equation (\ref{39}).The action of "free"
propagator $G^{0}(t)=\exp(L^{0}t)$ on any function defined on the phase space
is given by Eq. (\ref{28}).

If the particle dynamics have the necessary properties such as a mixing
ergodic flow in the phase space resulting from the local stochastic
instability, then the distances between particles rapidly (exponentially)
increase with time under the action of $G_{12}(t)$ (the contribution of a
"parallel motion" is negligible). It was proved that such behavior is the
case, for example, for the Boltzmann dilute gas of hard spheres and is now
viewed as common for the most real systems considered in statistical physics
(see, e.g., \cite{Klimontovich (1982)}). The mixing ergodic flow in the phase
space results in the exponential damping of correlations and is the reason for
irreversibility. Hence, if the effective particle interaction $V(\left\vert
\mathbf{r}_{i}-\mathbf{r}_{j}\right\vert )$ vanishes at a distance $\left\vert
\mathbf{r}_{i}-\mathbf{r}_{j}\right\vert >r_{0}$, then the integrand of the
integral over $t_{1}$ in (\ref{27n}) vanishes for $t-t_{0}>t_{cor}$ and the
same holds (under the action of $G_{12}(t-t_{0})$) for the initial correlation
term determined by the function $g_{2}(x_{1},x_{2})$, which also depends on
the particle interaction (we discuss this in more detail below, and it was
considered in the spatially homogeneous case in Sec. 5).

Condition (\ref{16}) implies the existence of the time hierarchy%
\begin{equation}
t_{cor}\ll t_{rel}, \label{30n}%
\end{equation}
where $t_{rel}\thicksim\gamma^{-1}t_{cor}$ is the relaxation time for a
one-particle distribution function $f_{1}$($x_{i},t$) ($1/nr_{0}^{2}$ is the
particle mean free path). In the initial evolution stage $t_{0}\leqslant
t\leqslant t_{cor}$, which is very interesting and essential for understanding
the irreversibility problem and for studying the non-Markovian processes (the
memory effects), decoherence phenomena, and the ultrafast relaxation effects,
the initial correlations can be important. But if we consider the time scale
$t-t_{0}\gg t_{cor}$, then the initial correlations are damped (as a result of
mixing flow), and the source in Eq. (\ref{27n}) containing $g_{2}(x_{1}%
,x_{2})$ can be neglected. But as Bogoliubov noted in \cite{Bogoliubov
(1946)}, this can be done only on time interval (\ref{0}), which exists
because of (\ref{30n}). The latter conclusion follows because the corrections
to the solution of (\ref{27n}) due to initial correlations turn out to be
proportional to $\left\vert t-t_{0}\right\vert $ (secular terms) and are small
only for $\left\vert t-t_{0}\right\vert \ll t_{rel}$. Therefore, the expansion
of the irrelevant initial condition term in $n$ (as in passing from
(\ref{22n}) to (\ref{27n})) is ineffective on the time scale of interest in
the kinetic theory $t-t_{0}\gtrsim t_{rel}$.

To get rid of the undesirable initial condition terms, it is often assumed
that $f_{i}(t_{0})=0$, which corresponds to RPA and is incorrect in principle
(see \cite{van Kampen (2004)}). Therefore (as mentioned in Sec. 1), to study
the kinetic evolution stage $t\gtrsim t_{rel}$, a more sophisticated approach
like that based on the principle of weakening of initial correlations
(Bogoliubov's ansatz) \cite{Bogoliubov (1946)} is needed. But the Bogoliubov
approach does not allow considering the initial evolution stage $t_{0}%
\leqslant t\leqslant t_{cor}$ and the situations where the large-scale
correlations with $t_{cor}\gtrsim t_{rel}$ play an essential role.

If $f_{1}(x_{i},t)$ changes significantly on the macroscopic (hydrodynamic)
length $l_{h}$ with the corresponding time scale $t_{h}=l_{h}/\overline{v}$
(for the spatially homogeneous gas, there are only the characteristic scales
$l$ and $t_{rel}$ on which $f_{1}(x_{i},t)$ changes), then the terms with
$L_{2}^{0}$ and $\frac{\partial F_{1}(x_{2},t-t^{^{\prime}})}{\partial
t^{^{\prime}}}$ in Eq. (\ref{27n}) can be neglected because they are both at
least of the order of $\gamma$, i.e., their contribution is of the second
order in $n$ (but these terms should not be omitted if we wish to take the
nondissipative characteristics of the nonideal gas into account
self-consistently). On the kinetic time scale $t-t_{0}\gtrsim t_{rel}$, the
upper limit of integration over $t_{1}$ can be extended to infinity
($t_{0}\rightarrow-\infty$) because of the damping of correlations resulting
from collisions (as noted above). If we also accept the RPA (or principle of
weakening of initial correlations) for $t_{0}\rightarrow-\infty$, i.e.,
neglect the initial condition term with $g_{2}(x_{1},x_{2})$, then Eq.
(\ref{27n}) reduces to%
\begin{align}
\frac{\partial f_{1}(x_{1},t)}{\partial t}  &  =L_{1}^{0}f_{1}(x_{1},t)+n\int
dx_{2}L_{12}^{^{\prime}}f_{1}(x_{2},t)f_{1}(x_{1},t)\nonumber\\
&  +n\int dx_{2}L_{12}^{^{\prime}}%
{\textstyle\int\limits_{0}^{\infty}}
dt_{1}e^{(L_{12}^{0}+L_{12}^{^{\prime}})t_{1}}L_{12}^{^{\prime}}f_{1}%
(x_{2},t)f_{1}(x_{1},t), \label{31n}%
\end{align}
where we take into account that for $t\gtrsim t_{rel}\gg t_{cor}$ in the
adopted approximation, we can replace $F_{1}(x_{i},t-t_{1})$ with $F_{1}%
(x_{i},t)$ (the integration over $t_{1}$ in (\ref{27n}) gives an essential
contribution only up to $t_{1}\thicksim$ $t_{cor}$). In the last (collision)
term in (\ref{31n}), the distribution functions of the colliding particles can
be taken at the same space point $\mathbf{r}_{1}=\mathbf{r}_{2} $ with the
adopted accuracy $\thicksim r_{0}/r_{h}\ll1$ (or $r_{0}/l\ll1$).

If the last integral in Eq. (\ref{31n}) exists (see \cite{Prigogine and
Resibois (1961)}), then Eq. (\ref{31n}) represents the nonlinear Markovian
kinetic equation irreversible in time. In the case of a weak particle
interaction, this equation in the first approximation in the small interaction
parameter $\varepsilon$ gives the reversible Vlasov equation (Eq. (\ref{31n})
with neglect of the third collision term in the right-hand side, which is of
the second order in $\varepsilon$). In the second approximation in
$\varepsilon$, it coincides with the Vlasov-Landau kinetic equation (see,
e.g., \cite{Balescu (1975)}). In the spatially homogeneous case, where a
one-particle distribution function is independent of the particle coordinate,
$f_{1}(x_{j},t)=f_{1}(\mathbf{p}_{j},t)$ ($\int f_{1}(\mathbf{p,}%
t)d\mathbf{p=}1$), the second (Vlasov) term in (\ref{31n}) vanishes because we
consider a potential $V_{ij}$ given by (\ref{23}) that depends on the particle
coordinate difference. In this case, Eq. (\ref{31n}) is equivalent to the
nonlinear Boltzmann equation (see, e.g., \cite{Balescu (1975)}). We stress
that the nonlinearity of (\ref{31n}) appears quite naturally in this approach
(based on the time-dependent $P(t)$ given by (\ref{15n})) in contrast to the
time-independent projection operator approach presented in Secs. 2-7. In the
latter case, we must approximate $f_{1}(x_{2},t_{0})$ by $f_{1}(x_{2},t)$
(which is possible on the time scale given by (\ref{0})) in the evolution
equation in order to obtain a nonlinear equation from the linear GME or HGME
(in the case of the BBGKY chain approach, the nonlinearity is introduced by
principle of weakening of initial correlations (\ref{0a})).

\section{NONLINEAR HOMOGENEOUS GME}

In Sec. 8, we obtained the nonlinear GME for the relevant part of a
distribution function (statistical operator). But Eq. (\ref{12n}) is not
closed in the sense that it is inhomogeneous and contains an irrelevant term
generally containing all multiparticle correlations. As noted above, to
eliminate this irrelevant part and obtain a homogeneous equation (like
(\ref{31n})), for example, for a one-particle distribution function, the
Bogoliubov principle of weakening of initial correlations or the RPA is
usually used.

To take initial correlations into account and to obtain the evolution
equations applicable on all time scales, we use here a method for converting
conventional linear inhomogeneous GMEs into the homogeneous GMEs presented in
Secs. 2 and 3. We now apply this approach to equation (\ref{12ln}). Our goal
is to convert Eq. (\ref{12ln}) into a homogeneous equation containing the
irrelevant part of the distribution function (statistical operator)
$f_{i}(t_{0})$ in the "mass" (super)operator acting on the relevant part of
the distribution function (statistical operator) $f_{r}(t)$. Such an equation
allows treating the initial correlations (contained in $f_{i}(t_{0}) $) on an
equal footing with all other correlations (collisions) and is applicable on
any time scale including the initial evolution stage when the initial
correlations are certainly essential. This is so because there is no need for
restrictions given by inequalities (\ref{0}) that allow neglecting the initial
correlations and avoiding the "secular terms" in the solution of the evolution
equation (the perturbation expansion of the "mass" (super)operator does not
lead to such terms on the large (kinetic) time scale). In contrast to the HGME
obtained in Sec. 2, the equation derived below is nonlinear and is also
applicable to the nonconservative systems (with time-dependent Hamiltonians).

Following Sec. 2, we identically represent the irrelevant part of the
distribution function (statistical operator) in the form%
\begin{align}
f_{i}(t_{0})  &  =F_{N}(t_{0})-f_{r}(t_{0})=Q(t_{0})F_{N}(t_{0})\nonumber\\
&  =\left[  Q(t_{0})F_{N}(t_{0})\right]  F_{N}^{-1}(t_{0})U^{-1}%
(t,t_{0})[P(t)+Q(t)]U(t,t_{0})F_{N}(t_{0})\nonumber\\
&  =C_{0}U^{-1}(t,t_{0})[f_{r}(t)+f_{i}(t)],\nonumber\\
C_{0}  &  =\left[  Q(t_{0})F_{N}(t_{0})\right]  F_{N}^{-1}(t_{0}%
),U^{-1}(t,t_{0})=T_{-}\exp[-%
{\textstyle\int\limits_{t_{0}}^{t}}
dsL(s)], \label{32n}%
\end{align}
where $U^{-1}(t,t_{0})$ is the backward-in-time evolution operator for the
density matrix $F_{N}(t)$ (compare with (\ref{6n})), $U^{-1}(t,t_{0}%
)U(t,t_{0})=1$, $T_{-}$ is the antichronological time-ordering operator
arranging the time-dependent operators $L(s)$ such that the time arguments
increase from left to right, $F_{N}^{-1}(t_{0})$ is the inverse of
$F_{N}(t_{0})$, $F_{N}^{-1}(t_{0})F_{N}(t_{0})=1$, $P(t)+Q(t)=1$. Hence,
additional identity (\ref{32n}) is obtained by multiplying the irrelevant part
by the unity $F_{N}^{-1}(t_{0})F_{N}(t_{0})$ (we assume that $F_{N}^{-1}%
(t_{0})$ exists) and inserting the unities $U^{-1}(t,t_{0})U(t,t_{0})=1$ and
$P(t)+Q(t)=1$. Therefore, neither a divergence (due to $F_{N}(t_{0})$ possible
vanishing ) nor an indeterminacy of the $0/0$ type (the behaviors of the
numerator and denominator in $F_{N}(t_{0})/F_{N}(t_{0})=1$ are similar) can
occur. This holds for all further (identical) manipulations (see below).

In (\ref{32n}) we introduced the parameter of initial correlations
\begin{align}
C_{0}  &  =[Q(t_{0})F_{N}(t_{0})]F_{N}^{-1}(t_{0})=f_{i}(t_{0})[f_{r}%
(t_{0})+f_{i}(t_{0})]^{-1}\nonumber\\
&  =f_{i}(t_{0})f_{r}^{-1}(t_{0})[1+f_{i}(t_{0})f_{r}^{-1}(t_{0}%
)]^{-1}\nonumber\\
&  =(1-C_{0})f_{i}(t_{0})f_{r}^{-1}(t_{0}). \label{33n}%
\end{align}
It is important that operator $Q(t_{0})$ acts only on $F_{N}(t_{0})$ in
(\ref{32n}) and (\ref{33n}), which is indicated by enclosing $Q(t_{0}%
)F_{N}(t_{0}) $ in brackets. This follows because $f_{r}(t)$ and $f_{i}%
(t_{0})$ are the basic quantities we are dealing with in Eq. (\ref{12ln}). All
functions of dynamical variables, whose average values we can calculate using
$f_{r}(t)$ (or $f_{red}(t)$) by multiplying equation (\ref{12ln}) by the
corresponding functions (operators) from the right and taking an average value
(a trace), depend only on the variables that are not integrated off by $D$
($D$ integrates over all excess variables in $F_{N}(t)$). Therefore, if we
correspondingly represent $f_{r}(t)$ and $f_{i}(t_{0})$ in (\ref{12ln}) as
$f_{r}(t)=P(t)F_{N}(t)$ and $f_{i\text{ }}(t_{0})=Q(t_{0})F_{N}(t_{0})$, then
the projection operators $P(t)=C(t)D$ and $Q(t)=1-P(t)$ in these expressions
act only on $F_{N}(t)$ but not on the functions (if any) to the right of them.
This is the essence of the reduced description method: to calculate the
average values of the functions that depend on a much smaller number of
variables than the whole distribution function $F_{N}(t)$ does, we in fact
need only the reduced distribution function (density matrix) $f_{red}(t)$.

As can be seen from (\ref{33n}), the correlation parameter is a series in
$f_{i}(t_{0})f_{r}^{-1}(t_{0})$, and we may therefore only need the formal
existence of the function (operator) $f_{r}^{-1}(t_{0})$, which is the inverse
of the relevant distribution function (statistical operator) chosen using the
appropriate operator $P(t_{0})$. It seems plausible that the inverse of the
relevant part of the distribution function (statistical operator) defined in
the sense noted above (uncorrelated part) can be constructed (see Secs. 4,6,
and the next section).

We now have two equations, (\ref{14n}) and (\ref{32n}), relating $f_{i}(t)$
and $f_{i}(t_{0})$. Using relations (\ref{12jn}), finding $f_{i}(t_{0})$ from
the indicated equations as a function of $f_{r}(t)$, and substituting it in
(\ref{12ln}), we obtain the equation
\begin{align}
\frac{\partial Df_{r}(t)}{\partial t}  &  =DL^{0}(t)f_{r}(t)+DL^{^{\prime}%
}(t)R(t,t_{0})f_{r}(t)\nonumber\\
&  +DL^{^{\prime}}(t)R(t,t_{0})%
{\textstyle\int\limits_{t_{0}}^{t}}
dt^{^{\prime}}\overline{U}(t,t^{^{\prime}})\left[  Q(t^{^{\prime}%
})L(t^{^{\prime}})-\frac{\partial P(t^{^{\prime}})}{\partial t^{^{\prime}}%
}\right]  f_{r}(t^{^{\prime}}), \label{34n}%
\end{align}
where the operator $R(t,t_{0})$ is defined as%
\begin{align}
R(t,t_{0})  &  =1+C(t,t_{0}),\nonumber\\
C(t,t_{0})  &  =\overline{U}(t,t_{0})\left[  1-C_{0}(t,t_{0})\right]
^{-1}C_{0}U^{-1}(t,t_{0}),\nonumber\\
C_{0}(t,t_{0})  &  =C_{0}U^{-1}(t,t_{0})\overline{U}(t,t_{0}), \label{35n}%
\end{align}
$Df_{r}(t)=f_{red}(t)$ and relation (\ref{12hn}) holds.

Equation (\ref{34n}) is the central result of this section. We have derived
the desired homogeneous generalized evolution equation for the relevant part
of the distribution function (statistical operator). This equation differs
from the linear TC-HGME obtained in Sec. 2 from the conventional
Nakajima-Zwanzig TC-GME using the time-independent projection operator
technique. Equation (\ref{34n}) is generally nonlinear and holds for systems
with a time-dependent Hamiltonian. It holds in both the classical and quantum
physics cases if the symbols are properly redefined and all (super)operators
exist (we address the latter question below). We have not made any
approximations in deriving Eq. (\ref{34n}), and, it is therefore an exact
integro-differential equation that takes the initial correlations and their
dynamics into account via the modified (super)operator (memory kernel) of
(\ref{12ln}) acting on the relevant part of the distribution function
(statistical operator) $f_{r}(t)$.

The obtained exact kernel of (\ref{34n}) can serve as a starting point for
effective perturbation expansions. In many cases, such expansions of
homogeneous equations (like (\ref{34n})) have a much broader applicability
domain than those of the inhomogeneous equations (like (\ref{12ln})) when the
expansions of the functions ($f_{r}$, $f_{i}$), rather than equations, are
involved (see also \cite{Bogoliubov (1946)}). Particularly, we can expect that
the divergencies appearing in the higher-order terms in the gas density
expansion \cite{Dorfman and Cohen (1967)} are tackled more conveniently using
the "mass" (super)operator expansion. As can be seen from (\ref{34n}), the
time evolution of initial correlations ($C(t,t_{0})$) influences the collision
integral (and also the second term) and therefore leads to an additional
dependence on the density $n$ (when the expansion in $n$ is used). It is
known, that initial correlations in the higher orders in $n$ become more
long-lived and are damped only at $t-t_{0}\thicksim t_{rel}$ because of
successive pair collisions (see \cite{Dorfman and Cohen (1967), Ferziger and
Kaper (1972)}). While the initial correlations in the lowest approximation in
$n$ (pair collision approximation) are typically the short-scale ones with the
characteristic $t_{cor}\thicksim r_{0}/\overline{v}\ll t_{rel}$ and
$l_{cor}\thicksim\overline{v}t_{cor}\ll l$, they are large-scale ones in the
next approximations in $n$ and therefore cannot be neglected on the kinetic
time scale. Hence, the initial correlations in the higher orders terms (than
linear) of the expansion of Eq. (\ref{34}) in $n$ may result in the cut-off of
the discussed divergencies of the kinetic equation at $t\thicksim t_{rel}$
(this was noted in \cite{Klimontovich (1975)}). The foregoing means that the
expansion of the "mass" (super)operator governing the evolution of $f_{r}(t)$
in (\ref{34n}) provides a new possibility for expanding the kinetic
coefficients in $n$.

Equation (\ref{34n}) reduces to the linear TC-HGME obtained in Sec. 2 if the
(projection) operator $P$ and the Hamiltonian $H$ are independent of time and
relations (\ref{12hn}) and (\ref{12h'n}) hold.

As in Sec. 2, the problem of the existence (convergence) of $R(t,t_{0})$ can
be raised. The function $R(t,t_{0})$ behaves properly at all times. Moreover,
expanding the kernel of (\ref{34n}) may result in canceling the pole in the
function $R(t,t_{0})$. For the linear TC-HGME, this correspondingly shown in
Secs. 4 and 6 in the linear approximation in the small density for a spatially
homogeneous dilute gas of classical and quantum particles. In such cases there
is no problem with the existence of $R(t,t_{0})$. Examining (\ref{33n}) (from
which the relation $(1-C_{0})^{-1}C_{0}=f_{i}(t_{0})/f_{r}(t_{0})$ follows),
we can expect that the same is true for $R(t,t_{0})$ given by (\ref{35n}) (see below).

\section{NONLINEAR EVOLUTION EQUATION FOR A NONIDEAL INHOMOGENEOUS DILUTE GAS
OF CLASSICAL PARTICLES WITH INITIAL CORRELATIONS TAKEN INTO ACCOUNT}

We now apply Eq. (\ref{34n}) to a system of $N$ ($N\gg1$) identical
interacting classical particles described by the Liouvillian (\ref{23}) and
considered in Sec. 4. To obtain an equation for the one-particle distribution
function $f_{1}(x_{i},t)$ ($i=1,\ldots,N$), we use the operator $P(t)$\ given
by (\ref{15n}). Hence, taking (\ref{24}), (\ref{19n}), and (\ref{20n}) into
account, we obtain the equation for the one-particle distribution function
from (\ref{34n}):%
\begin{align}
\frac{\partial f_{1}(x_{1},t)}{\partial t}  &  =L_{1}^{0}f_{1}(x_{1},t)+n\int
dx_{2}L_{12}^{^{\prime}}f_{1}(x_{2},t)f_{1}(x_{1},t)+DL^{^{\prime}}%
C(t,t_{0})f_{r}(t)\nonumber\\
&  +DL^{^{\prime}}%
{\textstyle\int\limits_{t_{0}}^{t}}
dt^{^{\prime}}R(t,t_{0})\overline{U}(t,t^{^{\prime}})\left[  (L^{0}-L_{1}%
^{0})+Q(t^{^{\prime}})L^{^{\prime}}-\frac{\partial P(t^{^{\prime}})}{\partial
t^{^{\prime}}}\right]  f_{r}(t^{^{\prime}}). \label{38n}%
\end{align}
We consider Eq. (\ref{38n}) in the linear approximation in density parameter
(\ref{16}). Given that the action of $P(t)$ or $D$ results in expressions
proportional to at least the first power in the density $n$ and that all terms
on the right-hand side of (\ref{38n}) (except the first flow term) are already
proportional to $n$ (or $D$), in the first approximation in $n$, we can
neglect all terms with $P(t)$ in (\ref{38n}) including such terms in
$\overline{U}(t,t^{^{\prime}})$. Then $\overline{U}(t,t^{^{\prime}%
})=U(t,t^{^{\prime}})$ in this approximation and is given by (\ref{25n}).
Therefore, we can approximate $U^{-1}(t,t_{0})\overline{U}(t,t_{0})$ in
(\ref{35n}) by unity and $C_{0}(t,t_{0})$ by $C_{0}$. Hence, in the adopted
approximation, the correlation parameter is
\begin{equation}
C(t,t_{0})=U(t,t_{0})[f_{i}(t_{0})f_{r}^{-1}(t_{0})]U^{-1}(t,t_{0}),
\label{40n}%
\end{equation}
as follows from (\ref{33n}) and (\ref{35n}). To calculate contribution
(\ref{40n}) of initial correlations to evolution equation (\ref{38n}), we
should use definitions (\ref{18n}) and (\ref{22}) of $f_{r}(t_{0})$ and
$f_{i}(t_{0})$.

Proceeding as in Sec. 9, i.e., using (\ref{24}), (\ref{19n}), (\ref{26n}), and
the fact (noted above) that each additional integration over the phase space
leads to an additional power of $n$, we can represent Eq. (\ref{38n}) in the
first approximation in $n$ as%
\begin{align}
\frac{\partial f_{1}(x_{1},t)}{\partial t}  &  =L_{1}^{0}f_{1}(x_{1},t)+n\int
dx_{2}L_{12}^{^{\prime}}[1+C_{12}(t-t_{0})]f_{1}(x_{2},t)f_{1}(x_{1}%
,t)\nonumber\\
&  +n\int dx_{2}L_{12}^{^{\prime}}[1+C_{12}(t-t_{0})]%
{\textstyle\int\limits_{0}^{t-t_{0}}}
dt_{1}e^{L_{12}t_{1}}[(L_{2}^{0}+L_{12}^{^{\prime}})f_{1}(x_{2},t-t_{1}%
)f_{1}(x_{1},t-t_{1})\nonumber\\
&  +\frac{\partial f_{1}(x_{2},t-t_{1})}{\partial t_{1}}f_{1}(x_{1},t-t_{1})],
\label{41n}%
\end{align}
where%
\begin{equation}
C_{12}(t-t_{0})=e^{L_{12}(t-t_{0})}\frac{g_{2}(x_{1},x_{2})}{f_{1}(x_{1}%
)f_{1}(x_{2})}e^{-L_{12}(t-t_{0})}, \label{42n}%
\end{equation}
$L_{12}=L_{12}^{0}+L_{12}^{^{\prime}}$, $L_{12}^{0}=L_{1}^{0}+L_{2}^{0}$ and
$f_{1}(x_{i})=f_{1}(x_{i},t_{0})$.

Homogeneous nonlinear equation (\ref{41n}) is the central result of this
section. This is a new equation that (in the linear approximation in $n$)
exactly describes the evolution of a one-particle distribution function of a
nonideal classical gas of particles with arbitrary spatial inhomogeneity and
on any time scale. This equation takes initial correlations into account
exactly (in the adopted approximation in $n$) and treats them on an equal
footing with the particle collisions and other processes. The initial
correlations enter the memory kernel (mass operator) governing the evolution
of the one-particle distribution function and do not result in any undesirable
inhomogeneous term in evolution equation (\ref{41n}) (cf. (\ref{27n})). The
contribution of initial correlations to the evolution process is given by
time-dependent correlation parameter (\ref{42n}), which describes the
evolution in time of the normalized two-particle correlation function
$\frac{g_{2}(x_{1},x_{2})}{f_{1}(x_{1})f_{1}(x_{2})}$. The time-evolution of
all terms in (\ref{41n}) is determined by the exact two-particle propagator
$G_{12}(t)=\exp(L_{12}t)$.

We consider the terms in the right-hand side of (\ref{41n}) and their
evolution in time in more detail. The first term is a conventional flow term
resulting from the gas inhomogeneity. The second term represents the Vlasov
self-consistent field modified by initial correlations. If $C_{12}(t-t_{0})$
vanishes (under the action of $G_{12}(t-t_{0})$) with time for $t-t_{0}\gtrsim
t_{cor}$ (see below), then we obtain the conventional Vlasov term. The third
term in the right-hand side of (\ref{41n}) describes the particle collisions
and the influence on the collisions of the change in the one-particle
distribution function on the microscopic scales in space and time . All these
processes are modified by initial correlations given by $C_{12}(t-t_{0})$.
Equation (\ref{41n}) thus takes the influence of all pair collisions and
correlations on the dissipative and nondissipative characteristics of the
nonideal spatially inhomogeneous gas of particles into account (this is not
the case in the conventional approaches \cite{Klimontovich (1975)}).

Noting that the term with $L_{2}^{0}$ is of the order of $r_{0}/r_{h}$ and the
term with the time derivative of $f_{1}(x_{2},t)$ is $\thicksim t_{cor}%
/t_{rel}$ ($\thicksim\gamma$), we can neglect these terms (if we are not
interested in the nondissipative corrections to the nonideal gas dynamics) in
the adopted approximation (linear in $n$). Equation (\ref{41n}) then takes the
simpler form%
\begin{align}
\frac{\partial f_{1}(x_{1},t)}{\partial t}  &  =L_{1}^{0}f_{1}(x_{1},t)+n\int
dx_{2}L_{12}^{^{\prime}}[1+C_{12}(t-t_{0})]f_{1}(x_{2},t)f_{1}(x_{1}%
,t)\nonumber\\
&  +n\int dx_{2}L_{12}^{^{\prime}}[1+C_{12}(t-t_{0})]\nonumber\\
&  \times%
{\textstyle\int\limits_{0}^{t-t_{0}}}
dt_{1}e^{L_{12}t_{1}}L_{12}^{^{\prime}}f_{1}(x_{2},t-t_{1})f_{1}(x_{1}%
,t-t_{1}), \label{43n}%
\end{align}
where the distribution functions of the colliding particles in the last
(collision) term can be taken at the same space point $\mathbf{r}%
_{1}=\mathbf{r}_{2}$ with the adopted accuracy $\thicksim r_{0}/r_{h}\ll1$ (in
the case of plasma, it is not good to do this in the second term because of
the long-range Coulomb interaction).

As an example, we consider the evolution in time of the terms in (\ref{43n})
originating from the initial correlations (terms with $C_{12}(t-t_{0})$) when
the dependence of the initial correlation function and the interparticle
interaction on the distance between the particles is given by
\begin{align}
g_{2}(x_{1},x_{2})  &  =g_{0}\exp(-\frac{r^{2}}{r_{cor}^{2}})\phi(x_{1}%
,x_{2})\text{ ,}\nonumber\\
V(r)  &  =V_{0}\exp(-\frac{r^{2}}{r_{0}^{2}})\text{,} \label{43an}%
\end{align}
where $\mathbf{r}=\mathbf{x}_{1}-\mathbf{x}_{2}$ is the distance between
particles 1 and 2, $r_{cor}\thicksim r_{0}$, $g_{0}$ and $V_{0}$ are constant
parameters, and $\phi(x_{1},x_{2})$ is a properly normalized function of
$x_{1}$ and $x_{2}$. We estimate the time dependence of the terms in
(\ref{43n}) determined by initial correlations in the case of a weak
interparticle interaction when the time evolution is governed by the "free"
propagator $G_{12}^{0}(t)=\exp(L_{12}^{0}t)$. Then we have (see (\ref{28}))%
\begin{equation}
G_{12}^{0}(t)\frac{g_{2}(x_{1},x_{2})}{F_{1}(x_{1})F_{1}(x_{2})}=g_{0}%
\exp(-\frac{|\mathbf{r}-\mathbf{g}t|^{2}}{r_{cor}^{2}})\frac{\phi
(\mathbf{x}_{1}-\mathbf{v}_{1}t,\mathbf{p}_{1},\mathbf{x}_{2}-\mathbf{v}%
_{2}t,\mathbf{p}_{2})}{F_{1}(\mathbf{x}_{1}-\mathbf{v}_{1}t,\mathbf{p}%
_{1})F_{1}(\mathbf{x}_{2}-\mathbf{v}_{2}t,\mathbf{p}_{2})},\text{ }
\label{43bn}%
\end{equation}
where $\mathbf{g=v}_{1}-\mathbf{v}_{2}$ is the relative velocity. Function
(\ref{43bn}) tends to zero at $t\rightarrow\infty$ with any fixed distance
$\mathbf{r}$ and velocity $\mathbf{g}$. Using (\ref{23}), (\ref{43an}), and
(\ref{43bn}), we can represent the term determining the contribution of
initial correlation to the second term in the right-hand side of (\ref{43n})
as%
\begin{align}
&  n\int d\mathbf{p}_{2}\int d\mathbf{r}\left[  \frac{\partial}{\partial
\mathbf{r}}V_{0}\exp(-\frac{r^{2}}{r_{0}^{2}})\right]  (\frac{\partial
}{\partial\mathbf{p}_{1}}-\frac{\partial}{\partial\mathbf{p}_{2}})g_{0}%
\exp(-\frac{|\mathbf{r}-\mathbf{g}t|^{2}}{r_{cor}^{2}})\nonumber\\
&  \times\frac{\phi(\mathbf{x}_{1}-\mathbf{v}_{1}t,\mathbf{p}_{1}%
,\mathbf{x}_{1}-\mathbf{r}-\mathbf{v}_{2}t,\mathbf{p}_{2})}{F_{1}%
(\mathbf{x}_{1}-\mathbf{v}_{1}t,\mathbf{p}_{1})F_{1}(\mathbf{x}_{1}%
-\mathbf{r}-\mathbf{v}_{2}t,\mathbf{p}_{2})}F_{1}(\mathbf{x}_{1}%
-\mathbf{r,p}_{2},t)F_{1}(\mathbf{x}_{1},\mathbf{p}_{1},t)\text{ }
\label{43cn}%
\end{align}
(where we set $t_{0}=0$ for brevity). It is easy to see that the integral over
$\mathbf{r}$ in (\ref{43cn}) is nonzero only if $t<t_{cor}$, where
$t_{cor}\backsim r_{cor}/\bar{v}\thicksim r_{0}/\bar{v}$. For $t\gg t_{cor}$,
the integral (\ref{43cn}) practically vanishes because of the finite
interparticle interaction range $r_{0}$. Physically, this means that the
initial correlations propagate outward from the region of interest determined
by $r_{0}$ and do not return in the approximation linear in $n$ under
consideration (also see \cite{Balescu (1975)}). Of course, such behavior is
possible if the contribution to (\ref{43cn}) of the \textquotedblright
parallel motion\textquotedblright\ with small $\mathbf{g}$ is negligible.
Therefore, the mixing flow in the phase space is necessary to ensure that the
initial correlation term vanishes. The same behavior of the initial
correlation term in the master equation was found in \cite{Tasaki et
al1,Tasaki et al2} in the case of a quantum mechanical system interacting with
a thermal bath with the mixing property in the van Hove limit. The term
related to initial correlations and contributing to the third term in
right-hand side of Eq. (\ref{43n}) (collision integral) displays the same
behavior with time. In the considered example, the terms with initial
correlations in Eq. (\ref{43n}) thus vanish for $t\gg t_{cor}$ (in the
considered first approximation in $n$).

If we pass to the kinetic time scale $t-t_{0}\gtrsim t_{rel}\gg t_{cor}$ in
Eq. (\ref{43n}), then the initial correlation terms determined by the
two-particle correlation function $C_{12}(t-t_{0})$ given by (\ref{42n})
vanish on this time scale because of the mixing flow in the phase space (as in
the example considered above), and we can rewrite Eq. (\ref{43n}) in the form
of the irreversible Markovian kinetic equation%
\begin{align}
\frac{\partial f_{1}(x_{1},t)}{\partial t}  &  =L_{1}^{0}f_{1}(x_{1},t)+n\int
dx_{2}L_{12}^{^{\prime}}f_{1}(x_{2},t)f_{1}(x_{1},t)\nonumber\\
&  +n\int dx_{2}L_{12}^{^{\prime}}%
{\textstyle\int\limits_{0}^{\infty}}
dt_{1}e^{L_{12}t_{1}}L_{12}^{^{\prime}}f_{1}(x_{2},t)f_{1}(x_{1},t).
\label{44n}%
\end{align}
For an interparticle interaction with the small parameter $\varepsilon\ll1$,
the two-particle propagator $\exp(L_{12}t)$ in the collision term in
(\ref{44n}) can be replaced with the "free" propagator $\exp(L_{12}^{0}t)$
(see (\ref{39})), and Eq. (\ref{44n}) reduces to the Vlasov-Landau equation in
the second approximation in $\varepsilon$ (see, e.g., \cite{Balescu (1975)}).
In the spatially homogeneous case, Eq. (\ref{44n}) coincides with the
nonlinear Boltzmann equation for the momentum distribution function
$f_{1}(\mathbf{p}_{1},t)$.

We note, that all stages of the evolution of the system under consideration
can be followed (in the adopted approximation) using Eq. (\ref{41n}) beginning
from the initial reversible regime $t_{0}\leqslant t\leqslant t_{cor}$. If the
reversible terms in (\ref{41n}), related to the initial correlations
$C_{12}(t-t_{0})$, vanish with time and the last integral $%
{\textstyle\int\limits_{0}^{\infty}}
$ in (\ref{44n}) exists (see Secs. 5 and 7), then the evolution of the system
can switch automatically from the reversible to the irreversible stage
described by kinetic equation (\ref{44n}).

\section{CONCLUSION}

In the preceding sections we have introduced a new type of evolution
equations: the homogeneous generalized master equations (HGMEs) and nonlinear
GMEs. The idea behind this is that we tried to obtain an evolution equations
which are capable of describing all stages of the (sub)system of interest
evolution in time beginning from the initial stage dependent of the initial
system state (initial correlations). Although the BBGKY chain or conventional
Nakajima-Zwanzig TC-GME and TCL-GME are formally valid for all $t>t_{0}$, the
actual consideration of the transient (from reversible initial $0<t-t_{0}%
<t_{cor}$ to irreversible kinetic $t_{cor}<(t-t_{0})\thicksim t_{rel}$) regime
and of the influence of initial correlations on the evolution process poses
essential difficulties (see, e.g., \cite{Zubarev (1997)} and references therein).

In the framework of the conventional time-independent projection operator
approach, we have derived the exact time-convolution homogeneous generalized
master equation (TC-HGME) (\ref{8}) and the exact time-convolutionless
generalized master equation (TCL-HGME) (\ref{14}) for the relevant part of a
distribution function (statistical operator). In the derivation we have not
used any approximation (like a factorizing initial condition or RPA) or
principle (like the Bogoliubov principle of weakening of initial
correlations). These equations have several advantages as compared to the
conventional generalized master equations (GMEs). The HGMEs contain the
parameter of initial correlations depending on time in the \textquotedblright
mass\textquotedblright\ (super)operator acting on the relevant part of a
distribution function (statistical operator). These equations allow treating
the initial correlations consistently and on an equal footing with the
collision integral by expanding the \textquotedblright mass\textquotedblright%
\ (super)operator into the series in the appropriate small parameter. The
obtained equations are valid on any time scale, in particular on the initial
stage of evolution $t_{0}\leqslant t-t_{0}\leqslant t_{cor}$, which can be
important for studying the irreversibility problem, the non-Markovian and the
ultrafast relaxation processes. The HGMEs enable the consideration of the
entire evolution process of the relevant part of the distribution function
(statistical operator) and of the influence of initial correlations on this
process. These equations can automatically switch, in principle, from the
initial (reversible) regime into the kinetic (irreversible) one if the
particle dynamics is characterized by the ergodic mixing flow in the phase
space leading to the damping of initial correlations and correlations caused
by collisions.

By appropriate selection of the projection operator, one can rewrite the
obtained HGMEs for the system under consideration. We have considered a
spatially homogeneous dilute gas of classical and quantum particles with an
arbitrary repulsive inter-particle interaction and obtained in the linear
approximation in the small density parameter (\ref{16}) the closed equation
for a one-particle distribution function (\ref{38}) and a one-particle density
matrix (\ref{55q}) retaining initial correlations. In this approximation, the
evolution equation for a one-particle momentum distribution function contains
only binary collisions and a two-particle time-dependent (via only
two-particle dynamics) correlation function in the parameter accounting for
initial correlations. We have shown as on the macroscopic time-scale (\ref{0})
these equations may become equivalent to the Boltzmann classical and quantum
equations if all correlations caused by inter-particle interaction vanish on
this time scale.

We have also derived (not postulated as it is usually done) the nonlinear GMEs
describing the evolution of the relevant part of a distribution function
(statistical operator). The new approach leading to these equations is based
on using the nonlinear time-dependent operator $P(t)$ determining the relevant
part of the distribution function (statistical operator). This operator is
generally not a projection operator. The obtained inhomogeneous nonlinear GME
(\ref{12n}) (and (\ref{12ln})) can be viewed as a generalization of the
Nakajima-Zwanzig linear TC-GME. In the case of time-independent projection
operators $P$ and $Q$, the obtained equation reduces to the conventional
TC-GME. This inhomogeneous nonlinear GME is equally useful for deriving both
the nonlinear and linear evolution equations for the reduced distribution
function (statistical operator) of interest in contrast to the linear TC-GME,
which is naturally more convenient for deriving linear evolution equations,
for example, for a subsystem interacting with a thermal bath. Using the
obtained equation, we derived new inhomogeneous nonlinear equation (\ref{27n})
for a one-particle distribution function in the linear approximation in the
small density parameter of a gas of classical particles. This equation, which
holds for an arbitrary spatial inhomogeneity, contains the space and time
changes of the one-particle distribution function on the microscopic time
scale (contributing to the nondissipative characteristics of a nonideal gas
and thus eliminating the inconsistency in the standard approach) and an
irrelevant part (a source) given by the initial two-particle correlation
function. Although the obtained inhomogeneous nonlinear GME allows deriving
the closed nonlinear equation for a one particle distribution function, an
additional assumption like the RPA or the Bogoliubov principle of weakening of
initial correlations is still needed in order to obtain either the
Vlasov-Landau or the Boltzmann kinetic equation.

To take initial correlations into account, we used the method suggested in
Sec. 2 to convert the inhomogeneous nonlinear GME into the homogeneous form.
The obtained exact homogeneous nonlinear GME (\ref{34n}) for the relevant part
of the distribution function (statistical operator) describes all evolution
stages of the (sub)system of interest including the initial stage where the
initial correlations play a role. In deriving this equation, which can be used
to obtain both the nonlinear and linear evolution equations for the reduced
distribution functions, we used no approximation like the RPA or the principle
of weakening of initial correlations. The initial correlations are treated in
this equation on an equal footing with collisions via the modified memory
kernel, which is a starting point for effective perturbation expansions. To
test this equation, we used it for a spatially inhomogeneous nonideal gas of
classical particles and obtained new homogeneous nonlinear evolution equation
(\ref{41n}) for a one-particle distribution function retaining initial
correlations in the memory kernel in the approximation linear in the density.
Equation (\ref{41n}) describes all evolution stages, takes all two-particle
correlations (collisions) into account, and converts into the nonlinear
Boltzmann or Vlasov-Landau equation on the appropriate time scale when all
initial correlations (and the ones due to collisions) vanish. It is important
to note, that in order to obtain the nonlinear evolution (kinetic) equation,
we are not now restricted by the second inequality (\ref{0}) ($t-t_{0}\ll
t_{rel}$) as it was in the case of linear TC-HGME. The irreversible Boltzmann
and Vlasov-Landau equations can be thus obtained from the Liouville equation
without conventional additional approximations (like the RPA or principle of
weakening of initial correlations) if the system dynamics have the necessary
properties (such as an ergodic mixing flow in the phase space).

\end{document}